\documentclass[aps, prx, twocolumn, superscriptaddress]{revtex4-2}

\pdfoutput=1

\usepackage[english]{babel}
\usepackage[utf8]{inputenc}
\usepackage{newtxtext, newtxmath}
\usepackage[scaled=0.8]{beramono}
\usepackage[T1]{fontenc}

\usepackage{acronym}
\usepackage{bm}
\usepackage{dsfont}
\usepackage{graphicx}
\usepackage{mathtools}
\usepackage{microtype}
\usepackage{tikz}

\usetikzlibrary{decorations.markings}
\usetikzlibrary{decorations.pathmorphing}

\definecolor{mauve}{RGB}{94, 60, 153}
\definecolor{orange}{RGB}{230, 97, 1}

\usepackage[absolute]{textpos}

\usepackage[colorlinks, allcolors=mauve]{hyperref}
\usepackage[figure, figure*]{hypcap}

\let\Re\relax
\let\Im\relax
\DeclareMathOperator\Re{Re}
\DeclareMathOperator\Im{Im}
\DeclareMathOperator\Tr{Tr}

\def\D{\mathrm d}
\def\E{\mathrm e}
\def\I{\mathrm i}

\def\abs#1{|#1|}
\def\bra#1{\langle#1|}
\def\ket#1{|#1\rangle}
\def\av#1{\langle#1\rangle}
\def\sub#1{_{\textnormal{#1}}}
\def\super#1{^{\textnormal{#1}}}

\def\inv{^{-1}}
\def\ps{^{\vphantom0}}

\let\Pi\varPi
\let\op\hat
\let\pow\textsuperscript
\let\s\textsubscript
\let\vec\mathbf

\tikzset{%
    baseline=-0.43ex,
    el/.style={
        line cap=round,
        decoration={markings, mark=at position 1/2 with {\arrow[xshift=1.5pt]>}},
        preaction={decorate},
        thick,
        },
    ph/.style={
        line cap=rect,
        decoration={coil, segment length=1mm, pre length=1mm, post length=1mm},
        decorate,
        thick,
        },
    elel/.style={
        line cap=rect,
        decoration={snake, aspect=0, pre length=0.3mm, post length=0.3mm},
        decorate,
        thick,
        },
    bubble/.pic={
        \fill[white] (0, 0) to[out=-45, in=-135] (2, 0) to[out=135, in=45] cycle;
        \draw[el] (0, 0) to[out=-45, in=-135] (2, 0);
        \draw[el] (2, 0) to[out=+135, in=+45] (0, 0);
        },
      chi/.pic={\pic{bubble}; \node at (1, 0.05) {$\chi \super b (T, \omega)$};},
     chi0/.pic={\pic{bubble}; \node at (1, 0.05) {$\chi \super b (\sigma, 0)$};},
     chiR/.pic={\pic{bubble}; \node at (1, 0.05) {$\chi \super{b,R} (T, \omega)$};},
     chiA/.pic={\pic{bubble}; \node at (1, 0.05) {$\chi \super{b,A} (T, \omega)$};},
    chiA0/.pic={\pic{bubble}; \node at (1, 0.05) {$\chi \super{b,A} (\sigma, 0)$};},
    chis/.pic={
        \draw[el] (0, 0) to[out=-70, in=-110] (1, 0);
        \draw[el] (1, 0) to[out=+110, in=+70] (0, 0);
        },
    Gb/.pic={
        \draw[ph] (0, 0) -- (1, 0);
        \useasboundingbox (-0.15, 0) -- (1.15, 0);
        \node[below=1mm] at (0.5, 0) {$G \super b$};
        },
    Gp/.pic={
        \draw[ph, double] (0, 0) -- (1, 0);
        \useasboundingbox (-0.15, 0) -- (1.15, 0);
        \node[below=1mm] at (0.5, 0) {$G \super p (T, \omega)$};
        },
    G/.pic={
        \draw[ph, ultra thick] (0, 0) -- (1, 0);
        \useasboundingbox (-0.15, 0) -- (1.15, 0);
        \node[below=1mm] at (0.5, 0) {$G(T, \omega)$};
        },
    v/.pic={
        \draw[elel] (0, 0) -- (1, 0);
        \node[below=2pt] at (0.5, 0) {$v$};
        \useasboundingbox (-0.15, 0) -- (1.15, 0);
        },
    U/.pic={
        \draw[elel, double] (0, 0) -- (1, 0);
        \useasboundingbox (-0.15, 0) -- (1.15, 0);
        \node[below=2pt] at (0.5, 0) {$U(T, \omega)$};
        },
    U0/.pic={
        \draw[elel, double] (0, 0) -- (1, 0);
        \useasboundingbox (-0.15, 0) -- (1.15, 0);
        \node[below=2pt] at (0.5, 0) {$U(\sigma, 0)$};
        },
    W/.pic={
        \draw[elel, ultra thick] (0, 0) -- (1, 0);
        \useasboundingbox (-0.15, 0) -- (1.15, 0);
        \node[below=2pt] at (0.5, 0) {$W(T, \omega)$};
        },
      ob/.pic={\fill (0, 0) circle (2pt);},
      op/.pic={\draw[thick, fill=white] (0, 0) circle (2pt);},
       o/.pic={\fill (0, 0) circle (3pt);},
      gb/.pic={\pic{ob}; \node[above=2pt] {$g\mathrlap{\super b}$};},
      gp/.pic={\pic{op}; \node[above=2pt] {$g \super p (T, \omega)$};},
     gpl/.pic={\pic{op}; \node[above=2pt] {$\mathllap{g \super p (T, {}}\omega\mathrlap)$};},
    gpl0/.pic={\pic{op}; \node[above=2pt] {$\mathllap{g \super p (\sigma, {}}0\mathrlap)$};},
     gpr/.pic={\pic{op}; \node[above=2pt] {$g\mathrlap{\super p (T, \omega)}$};},
       g/.pic={\pic {o}; \node[above=2pt] {$g(T, \omega)$};},
      gl/.pic={\pic {o}; \node[above=2pt] {$\mathllap{g(T, {}}\omega\mathrlap)$};},
     gl0/.pic={\pic {o}; \node[above=2pt] {$\mathllap{g(\sigma, {}}0\mathrlap)$};},
      gr/.pic={\pic {o}; \node[above=2pt] {$g\mathrlap{(T, \omega)}$};},
     gr0/.pic={\pic {o}; \node[above=2pt] {$g\mathrlap{(\sigma, 0)}$};},
    grT0/.pic={\pic {o}; \node[above=2pt] {$g\mathrlap{(T, 0)}$};},
    gbl0/.pic={\pic {o}; \node[below=2pt] {$\mathllap{g(\sigma, {}}0\mathrlap)$};},
    gbr0/.pic={\pic {o}; \node[below=2pt] {$g\mathrlap{(\sigma, 0)}$};},
      gs/.pic={\pic {o}; \node[above=2pt] {$g$};},
    }

\newacro{BZ}{Brillouin zone}
\newacro{cDFPT}{constrained density-functional perturbation theory}
\newacro{CDW}{charge-density wave}
\newacro{cRPA}{constrained random-phase approximation}
\newacro{DFG}{Deutsche Forschungsgemeinschaft}
\newacro{DFPT}{density-functional perturbation theory}
\newacro{DFT}{density-functional theory}
\newacro{HLRN}{North-German Supercomputing Alliance}
\newacro{PBE}{Perdew-Burke-Ernzerhof}
\newacro{RPA}{random-phase approximation}
\newacro{SC}{semicore}
\newacro{SNSF}{Swiss National Science Foundation}

\def\Bremen{%
    U~Bremen Excellence Chair,
    Bremen Center for Computational Materials Science,
    and MAPEX Center for Materials and Processes,
    University of Bremen,
    28359~Bremen,
    Germany}

\def\Zagreb{%
    Institute of Physics,
    10000~Zagreb,
    Croatia}

\def\Hamburg{%
    I. Institute of Theoretical Physics,
    University of Hamburg,
    22607~Hamburg,
    Germany}

\def\CUI{%
    The Hamburg Centre for Ultrafast Imaging,
    22761~Hamburg,
    Germany}

\def\Lausanne{%
    Theory and Simulation of Materials~(THEOS),
    and National Centre for Computational Design and Discovery of Novel Materials~(MARVEL),
    \'Ecole Polytechnique F\'ed\'erale de Lausanne,
    1015~Lausanne,
    Switzerland}

\def\Louvain{%
    European Theoretical Spectroscopy Facility,
    Institute of Condensed Matter and Nanosciences,
    Universit\'e catholique de Louvain,
    1348~Louvain-la-Neuve,
    Belgium}

\begin{document}

\title{Phonon Self-Energy Corrections: To Screen, or Not to Screen}

\author{Jan Berges}
\affiliation\Bremen

\author{Nina Girotto}
\affiliation\Zagreb

\author{Tim Wehling}
\affiliation\Hamburg
\affiliation\CUI

\author{Nicola Marzari}
\affiliation\Lausanne
\affiliation\Bremen

\author{Samuel Ponc\'e}
\affiliation\Louvain
\affiliation\Lausanne

\begin{abstract}
First-principles calculations of phonons are often based on the adiabatic approximation and on Brillouin-zone samplings that might not always be sufficient to capture the subtleties of Kohn anomalies.
These shortcomings can be addressed through corrections to the phonon self-energy arising from the low-energy electrons.
The exact self-energy involves a product of a bare and a screened electron-phonon vertex~\nocite{Giustino2017}[\href{https://doi.org/10.1103/RevModPhys.89.015003}{Rev.\@ Mod.\@ Phys.~\textbf{89}, 015003 (2017)}]; still, calculations often employ two adiabatically screened vertices, which have been proposed as a reliable approximation for self-energy differences~\nocite{Calandra2010}[\href{https://doi.org/10.1103/PhysRevB.82.165111}{Phys.\@ Rev.~B~\textbf{82}, 165111 (2010)}].
We assess the accuracy of both approaches in estimating the phonon spectral functions of model Hamiltonians and the adiabatic low-temperature phonon dispersions of monolayer TaS\s2 and doped MoS\s2.
We find that the approximate method yields excellent corrections at low computational cost, due to its designed error cancellation to first order, while using a bare vertex could in principle improve these results but is challenging in practice.
We offer an alternative strategy based on downfolding to partially screened phonons and interactions~\nocite{Nomura2015}[\href{https://doi.org/10.1103/PhysRevB.92.245108}{Phys.\@ Rev.~B~\textbf{92}, 245108 (2015)}].
This is a natural scheme to include electron-electron interactions and tackle phonons in strongly correlated materials and the frequency dependence of the electron-phonon vertex.
\end{abstract}

\maketitle

\begin{textblock*}{\paperwidth}(0mm, \paperheight-18mm)
    \centering
    \small
    Published in \href{https://doi.org/10.1103/PhysRevX.13.041009}{Phys.\@ Rev.\@ X \textbf{13}, 041009 (2023)}
\end{textblock*}

\section{Introduction}

The interplay between interacting electrons and lattice vibrations in solids gives rise to diverse phenomena ranging from the quantitative, such as changes in the electric~\cite{Ponce2020} or thermal conductivities~\cite{Knoop2020}, to the qualitative, such as instabilities toward charge~\cite{Wilson1974} and superconducting order~\cite{Revolinsky1963}.
Accordingly, the simulation of lattice dynamics is an exceptionally well established branch of condensed-matter physics~\cite{Baroni2001}.
Readily accessible total energies from \ac{DFT}~\cite{Hohenberg1964, Kohn1965} and the development of \ac{DFPT}~\cite{Zein1984, Gonze1997, Baroni2001} have significantly advanced this field.
Nowadays, phonon frequencies and normal modes are obtained routinely and at reasonable computational cost, enabling even high-throughput studies~\cite{Petretto2018, Mounet2018}.
For a wide range of materials, from semiconductors~\cite{Baroni1987, Giannozzi1991} to metals~\cite{deGironcoli1995}, the agreement between theory and experiment is remarkable.

Nevertheless, \ac{DFPT} still depends on crucial approximations:
First, the adiabatic or Born-Oppenheimer approximation~\cite{Born1927} implicitly assumes that the dynamics of electrons and ions happens on two well separated energy scales.
However, in some materials the relevant energies of the electrons are similar to or even smaller than those of the phonons, leading to nonadiabatic effects~\cite{Maksimov1996, Lazzeri2006, Pisana2007, Bauer2009, Calandra2010, Ponce2015, Caruso2017, Verdi2017, Miglio2020, Girotto2023}.
Second, calculations are in practice limited to a sparse sampling of the \ac{BZ} or, equivalently, short-range interatomic force constants.
This poses a problem for materials close to a lattice instability, signaled by Kohn anomalies~\cite{Kohn1959} driven by a strong long-range electronic response.
Furthermore, the exchange-correlation energy is described by an approximate functional.
Thus, \ac{DFT}-based methods will inevitably fail for strongly correlated materials where the single-electron picture breaks down~\cite{Hubbard1963}.

A popular approach in cases where the abovementioned approximations cannot be applied is to amend the results through suitable corrections to the self-energy of the electrons~\cite{Geilikman1971, Kocer2020} and phonons~\cite{Brovman1967, Ipatova1974, Falter1981, Falter1988, Calandra2010, Giustino2017}.
Here, one usually faces a difficulty referred to as ``double counting'' or ``overscreening''~\cite{Paleari2021a, Paleari2021b, Marini2023a}:
An effect that is accounted for in \ac{DFT} must be removed before a more elaborate description of the same effect can be applied.
This is closely related to the concept of ``downfolding''~\cite{Aryasetiawan2004, vanLoon2021a, vanLoon2021b}:
The full problem is mapped to an \emph{ab initio} low-energy effective system with a significantly reduced number of degrees of freedom~\cite{Imada2010}.
Established downfolding methods for the electron-electron and electron-phonon interactions~\cite{Giovannetti2014} are the \ac{cRPA}~\cite{Aryasetiawan2004, Aryasetiawan2006} and \ac{cDFPT}~\cite{Nomura2015}, respectively.
While for the former several implementations in popular simulation software exist~\cite{Friedrich2010, Sasioglu2013, Amadon2014, Kaltak2015, Nakamura2021}, a general workflow for the latter is at an earlier development stage~\cite{Nomura2015, Berges2020a, Novko2020a}.

In this work, we revisit the problem of obtaining adiabatic and nonadiabatic phonons at a low electronic temperature that are converged with respect to the sampling of the \ac{BZ}, originally addressed in Ref.~\citenum{Calandra2010}.
In doing so, we also react to the recently revived controversy about the correct screening of the electron-phonon vertices in the phonon self-energy:
It is universally acknowledged that both perturbative~\cite{Reichardt2018, Paleari2021a, Paleari2021b, Marini2023a} and nonperturbative treatments of the problem based on the Hedin-Baym equations~\cite{Giustino2017} yield a phonon self-energy with one bare and one screened vertex.
Still, there are strong arguments~\cite{Calandra2010} that the traditional choice~\cite{Allen1972, Lazzeri2006, Pisana2007, Giustino2007a, Novko2018, Novko2020b} of two screened vertices can not only be ascribed to the fact that these are readily available from \ac{DFPT}, but is indeed preferable for phonon self-energy \emph{differences} as long as the screened vertex is approximated at the \ac{DFT} level.

Here, by comparing the two approaches and supporting them with numerical results for monolayer TaS\s2 and n-doped MoS\s2, we will detail the following.
(i)~Working with one bare vertex is exact and thus allows for systematic improvements.
However, the quality of the result depends on the achievable precision of the screened vertex.
The computational cost is increased by the necessity to sum over many electronic bands, but a possible way around this problem has recently been published~\cite{Lihm2021}.
Also, the bare vertex is not routinely obtained in a pseudopotential framework.
In practice, some authors have approximated the bare vertex by \emph{unscreening} the \ac{DFPT} vertex with model dielectric functions~\cite{Caruso2017, PrasadKafle2020}, but their accuracy is limited by the validity of these models.
(ii)~The established approach with two screened vertices~\cite{Calandra2010} yields excellent results at low cost for a wide parameter range.
Its robustness can be explained by a designed cancellation of errors to first order.

We also illustrate a third option~(iii) based on downfolding via \ac{cDFPT}~\cite{Nomura2015}, which can be shown to be equivalent to (i) with a computational cost similar to (ii).
By splitting the electronic transitions into an ``active subspace'' and its complement, the ``rest subspace\rlap,'' we can gradually switch between the bare and the screened vertex and settle for the optimum.
Whenever we deal with strongly correlated materials~\cite{Savrasov2003, Kocer2020} or wish to also describe the frequency dependence of the electron-phonon vertex~\cite{Giustino2017, Reichardt2018}, which is beyond the scope of this work, it becomes necessary to explicitly consider the electron-electron interaction.
We believe that in these cases downfolding via a combination of \ac{cRPA} and \ac{cDFPT}, as opposed to (i), is a viable option because it provides a complexity reduction without introducing a double-counting problem.
In fact, within such a framework, the ideas of (ii) can still be applied~%
\footnote{Francesco Mauri (private communication).}.

Importantly, not only the bare but also the partially screened quantities are, in good approximation, adiabatic---in an Engelsberg-Schrieffer sense~\cite{Engelsberg1963, Saitta2008}---and independent of the electronic temperature since the rest system is gapped.
As a consequence, they are also smooth in large parts of reciprocal space, which facilitates interpolation and ensures convergence already at coarse \ac{BZ} sampling.
Nevertheless, there is also a drawback since the absence of these screening effects is accompanied by the emergence of long-range Fr\"ohlich terms.
Since the associated discontinuities in the derivatives of phonon dispersion and electron-phonon coupling render a straightforward Fourier interpolation impossible, an accurate description of these long-range terms, which allows us to subtract and add them before and after interpolation, is needed.

We implemented approaches~(i), (ii), and (iii) in \textsc{Quantum ESPRESSO}~\cite{Giannozzi2009, Giannozzi2017, Giannozzi2020} and the \textsc{EPW} code~\cite{Giustino2007b, Noffsinger2010, Ponce2016}, see Supplemental Material~%
\footnote{See Supplemental Material at \url{https://arxiv.org/src/2212.11806/anc} for the source code and data associated with this work.
It is also available in the Materials Cloud Archive at \url{https://doi.org/10.24435/materialscloud:he-pv}.},
independently of an existing realization of (ii)~\cite{Marini2023c}.
On this basis, we can efficiently study the fine features of Kohn anomalies~\cite{He2020}, which arise from the active subspace, at the required (and otherwise prohibitive) ultradense \ac{BZ} sampling and with frequency dependence.

This paper is organized as follows.
In Sec.~\ref{sec:theory}, we present the relevant theoretical concepts and formulas, which are illustrated using simple models in Sec.~\ref{sec:models}.
Subsequently, in Sec.~\ref{sec:implementation}, we describe our \emph{ab initio} implementation, results from which are shown in Sec.~\ref{sec:results}.
We finish by summarizing our work and discussing possible generalization of the scheme in Sec.~\ref{sec:conclusions}.

\section{Theory}
\label{sec:theory}

In this section, we review the theoretical background, including the phonon Green's function (Sec.~\ref{sec:green}), downfolding (Sec.~\ref{sec:downfolding}), the diagrams describing the screening of phonons and interactions (Sec.~\ref{sec:rpa}), approximate phonon self-energies (Sec.~\ref{sec:approx}), their symmetrization (Sec.~\ref{sec:sym}), Friedel and Fr\"ohlich long-range effects (Sec.~\ref{sec:lr}), and the basics of Wannier-Fourier interpolation (Sec.~\ref{sec:wannier}).
We employ Hartree atomic units where $m \sub e = e = \hbar = 4 \pi \epsilon_0 = 1$.

\subsection{Phonon Green's function and dynamical matrix}
\label{sec:green}

The lattice dynamics of a material can be described by the phonon Green's function or, more precisely, the retarded displacement-displacement correlation function~%
\footnote{Strictly speaking, the phonon Green's function is the correlation function of phonon ladder operators instead of displacement operators~\cite{Giustino2017} and thus differs from Eq.~\eqref{eq:green_t} by a factor of $2 \omega$, which thus appears in Eq.~\eqref{eq:specfun} for the phonon spectral function.}:
\begin{multline}
    \label{eq:green_t}
    G_{\vec R - \vec R' \kappa \alpha \kappa' \beta}(T, t - t')
    = -\I \Theta(t - t') \sqrt{M_\kappa M_{\kappa'}}
    \\
    \times
    \av{[\op u_{\vec R \kappa \alpha}^{\vphantom\dagger}(t), \op u_{\vec R' \kappa' \beta}^\dagger(t')]}_T,
\end{multline}
where the Heaviside function $\Theta$ ensures that a displacement introduced at time $t'$ can only propagate forward in time $t > t'$, $u$ and $M$ are atomic displacements and masses, $\vec R, \vec R'$ are Bravais lattice vectors, $\kappa, \kappa'$ and $\alpha, \beta$ enumerate basis atoms and Cartesian directions, $\av \cdots_T$ denotes an ensemble average at electronic temperature $T$, and $[{\cdots}, {\cdots}]$ is the commutator.
Knowledge of the correlations between any two atomic displacements at different times and positions as in Eq.~\eqref{eq:green_t} completely characterizes the lattice dynamics.

Because of translational invariance, Eq.~\eqref{eq:green_t} depends on \emph{differences} of times $t - t'$ and lattice vectors $\vec R - \vec R\mathrlap'$, which allows for a Fourier transform to frequency $\omega$ and phonon momentum $\vec q$,
\begin{equation}
    \label{eq:green_omega}
    G_{\vec q}(T, \omega)
    = \frac{\mathds 1}{\omega^2 \mathds 1 - D_{\vec q}(T, \omega)},
\end{equation}
with the screened dynamical matrix $D$.
Here, $\omega$ is defined on the whole complex plane and the \emph{retarded} phonon Green's function is obtained at $\omega + \I 0^+$ with $\omega$ real and $0^+$ a positive infinitesimal.
The right-hand side of Eq.~\eqref{eq:green_omega} is a matrix inversion in the basis of displacements $\kappa \alpha, \kappa' \beta$, and $\mathds 1$ is the corresponding identity matrix.

The link to experiments such as inelastic neutron- or x-ray-scattering spectroscopy~\cite{Caruso2017} is the phonon spectral function.
It assigns an intensity to each combination of energies $\omega$ and crystal momenta $\vec q$ and thus provides the quasiparticle band structure, where applicable, including many-body effects such as broadening and satellites.
We compute the phonon spectral function as~\cite{Abrikosov1963}
\begin{equation}
    \label{eq:specfun}
    A_{\vec q}(T, \omega)
    = -\frac {2 \omega} \pi \Tr \Im G_{\vec q}(T, \omega + \I 0^+).
\end{equation}

Instead, the static ($\omega = 0$) adiabatic phonon frequencies $\omega_{\vec q \nu}$ as obtained from \ac{DFPT}, usually at high electronic smearing $\sigma$, follow from the eigenvalue equation for the dynamical matrix:
\begin{equation}
    \label{eq:eigen}
    D_{\vec q}(\sigma, 0) \vec e_{\vec q \nu} \ps
    = \omega^2_{\vec q \nu} \vec e_{\vec q \nu} \ps,
\end{equation}
where $ \vec e_{\vec q \nu} $ are the phonon eigenvectors with mode index $\nu$.

\subsection{Screened, partially screened, and bare quantities}
\label{sec:downfolding}

In a typical \emph{ab initio} calculation, all electronic states are treated equally.
However, the relevant physics often takes place in a small subset of these states, namely the low-energy states close to the Fermi level, which we refer to as active states.
In particular, if an \emph{ab initio} calculation for chosen parameters and approximations fails to converge or capture the effects of interest, this is likely due to processes within this subset of active states alone, while the rest might already be properly described.
Before resorting to more elaborate treatments, it is thus instructive to consider the different subsets of states and corresponding energy scales separately~\cite{Falter1981, Falter1988}.

While pure \emph{ab initio} approaches always address the full system, the downfolding approach~\cite{Aryasetiawan2004, Aryasetiawan2006, Giovannetti2014, Nomura2015} uses \emph{ab initio} calculations to construct tractable low-energy systems with active states only, thus reducing the number of degrees of freedom significantly.
This approach is exact, provided the complexity reduction is properly compensated by the partial screening of the system parameters.
While the full system consists of simple \emph{bare} elementary particles and interactions, the parameters of the downfolded system usually acquire dependences on the involved quantum numbers and are in general also frequency dependent~\cite{Aryasetiawan2004, Falter2006}.
The chosen active states must thus span an energy window large enough that low-energy effects and related dependences can be safely neglected and at the same time small enough to keep the computational cost affordable.

\begin{figure}
    \includegraphics{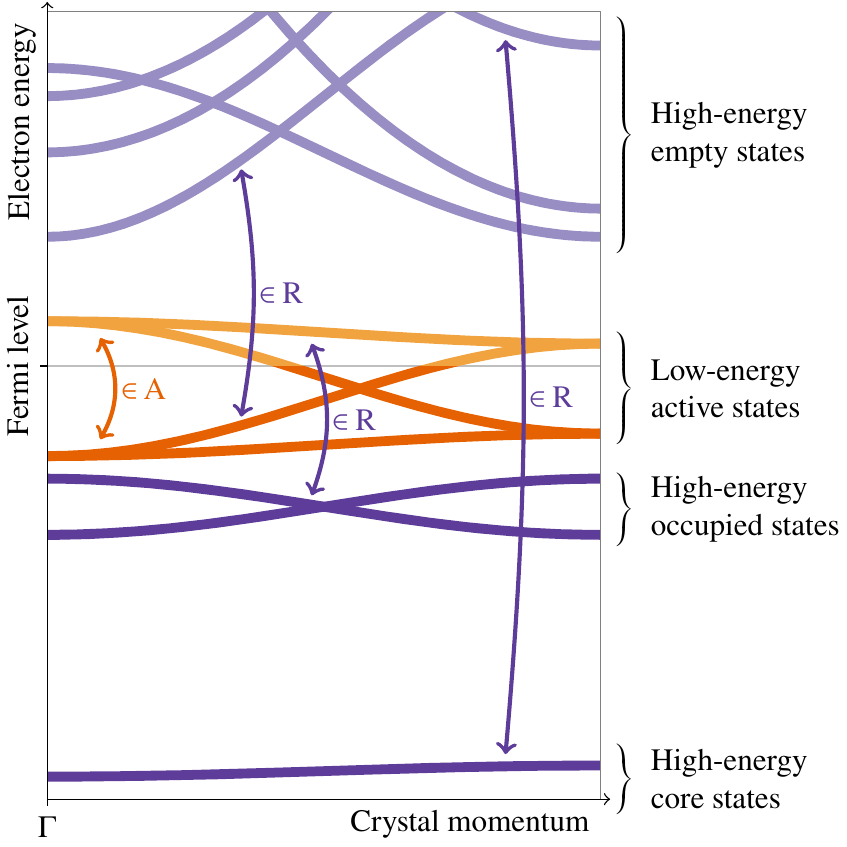}
    \caption{Visualization of active (A) and rest (R) subspaces of electronic transitions based on a generic band structure.
    A only includes transitions between a chosen set of low-energy states highlighted in orange, R all remaining transitions.
    Note that there is in general an infinite number of empty states beyond the shown energy range.}
    \label{fig:downfolding}
\end{figure}

A typical choice of active and rest states is sketched in Fig.~\ref{fig:downfolding}.
Note that here, for convenience but departing from what is usually done, we define the active subspace not as the set of active \emph{states} but as the set of \emph{transitions} between them.

The full system with bare parameters would be recovered if \emph{all} states were counted among the active states.
Hence, special care has to be taken in the context of pseudopotentials.
Since quasiparticle energies and interactions decrease with screening, the bare quantities in an all-electron calculation are larger than those in a pseudopotential framework with less core states.

\subsection{Random-phase approximation (RPA)}
\label{sec:rpa}

\begin{table*}
    \caption{List of symbols used in this paper.
    Subscript momenta and band indices have been omitted for brevity.}
    \label{tab:symbols}
    \medskip
    \begin{tabular}[t]{*{11}l}
        \multicolumn2l{\bfseries Phonon Green's function}
        &
        \multicolumn2l{\bfseries Electron-phonon coupling}
        &
        \multicolumn5l{\bfseries Bare electron susceptibility}
        &
        \multicolumn2l{\bfseries Miscellaneous}
        \\
        $G \super b$ & bare
        &
        $g \super b$ & bare
        &
        $\chi \super b (T, \omega)$ & \multicolumn4l{full}
        &
        $\omega$ & frequency argument
        \\
        $G \super p (T, \omega)$ & partially screened
        &
        $g \super p (T, \omega)$ & partially screened
        &
        $\chi \super{b,R} (T, \omega)$ & \multicolumn4l{rest subspace}
        &
        $T$ & electronic temperature
        \\
        $G(T, \omega)$ & screened
        &
        $g(T, \omega)$ & screened
        &
        $\chi \super{b,A} (T, \omega)$ & \multicolumn4l{active subspace}
        &
        $\sigma$ & \emph{ab initio} smearing
        \\[1mm]
        \multicolumn2l{\bfseries Dynamical matrix}
        &
        \multicolumn2l{\bfseries Electron-electron interaction}
        &
        \multicolumn5l{\bfseries Phonon self-energy}
        &
        $\vec R$ & Bravais lattice
        \\
        $D \super b$ & bare
        &
        $v$ & bare
        &
        $\Pi(T, \omega)$ & $=$ & $g \super b$ & $\chi \super b (T, \omega)$ & $g(T, \omega)$
        &
        $\vec G$ & reciprocal lattice
        \\
        $D \super p (T, \omega)$ & partially screened
        &
        $U(T, \omega)$ & partially screened
        &
        $\Pi \super R (T, \omega)$ & $=$ & $g \super b$ & $\chi \super{b,R} (T, \omega)$ & $g \super p (T, \omega)$
        &
        $\vec k$ & fermionic momentum
        \\
        $D(T, \omega)$ & screened
        &
        $W(T, \omega)$ & screened
        &
        $\Pi \super A (T, \omega)$ & $=$ & $g \super p (T, \omega)$ & $\chi \super{b,A}(T, \omega)$ & $g(T, \omega)$
        &
        $\vec q$ & bosonic momentum
        \\[1mm]
        \multicolumn2l{\bfseries Basis indices}
        &
        \multicolumn2l{\bfseries Long-range electrostatics}
        &
        $\Pi \super{00} (T, \omega)$ & $=$ & $g(\sigma, 0)$ & $\chi \super{b,A}(T, \omega)$ & $g(\sigma, 0)$
        &
        $\varepsilon$ & electronic energy
        \\
        $s, p$ & electronic orbitals
        &
        $\epsilon$ & dielectric constant
        &
        $\Pi \super{p0} (T, \omega)$ & $=$ & $g \super p (\sigma, 0)$ & $\chi \super{b,A}(T, \omega)$ & $g(\sigma, 0)$
        &
        $f$ & electronic occupation
        \\
        $m, n$ & electronic bands
        &
        $\vec Z^*$ & Born effective charge
        &
        $\Pi \super{p$T$} (T, \omega)$ & $=$ & $g \super p (\sigma, 0)$ & $\chi \super{b,A}(T, \omega)$ & $g(T, 0)$
        &
        $M$ & atomic mass
        \\
        $\kappa, \kappa'$ & basis atoms
        &
        $Q$ & quadrupole tensor
        &
        $\Pi \super{b0} (T, \omega)$ & $=$ & $g \super b$ & $\chi \super{b,A}(T, \omega)$ & $g(\sigma, 0)$
        &
        $\bm \uptau$ & atomic position
        \\
        $\alpha, \beta$ & Cartesian directions
        &
        $L$ & separation parameter
        &
        $\Pi \super{b$T$} (T, \omega)$ & $=$ & $g \super b$ & $\chi \super{b,A}(T, \omega)$ & $g(T, 0)$
        &
        $\vec u$ & atomic displacement
    \end{tabular}
\end{table*}
In this section, we review the formulas that describe the electronic screening of phonons and interactions in the framework of the \ac{RPA}~\cite{Bohm1951, GellMann1957, Ren2012, vanLoon2021b}.
In this approximation, the electronic response is determined by the bare polarizability of the system.
In terms of Feynman diagrams, we consider all possible diagrams consisting of bare phonons, the bare electron-phonon interaction, the bare electron-electron interaction, and electron-hole ``bubbles\rlap.''
In the static case ($\omega = 0$), this is equivalent to the screening in \ac{DFPT} as long as the electrons are given by the adiabatically screened Kohn-Sham states and the exchange-correlation kernel is included in the bare electron-electron interaction~\cite{Nomura2015, Giustino2017}.
A summary of the most relevant symbols used in the following is provided in Table~\ref{tab:symbols}.

\subsubsection{Electron-electron interaction}
\label{sec:rpa_elel}

We start with the screening of the electron-electron interaction, for which the \ac{RPA} has been originally derived~\cite{Bohm1951}.
In the basis of electronic eigenstates, used throughout the paper, the bare Coulomb interaction can be written as
\begin{equation}
    \label{eq:coulomb}
    v_{\vec q \vec k m n \vec k' m' n'}
    = \bra{\vec k {+} \vec q m; \vec k' n'} \op v \ket{\vec k n; \vec k' {+} \vec q m'},
\end{equation}
with electron momentum $\vec k$, band indices $m, n$, and the virtual photon momentum transfer $\vec q$.
We purposely used the symbol $\vec q$ as in Eq.~\eqref{eq:green_omega} to highlight the fact that in \ac{RPA} the phonon and photon momentum transfers have to be the same---this is no longer true beyond \ac{RPA}.
More precisely, Eq.~\eqref{eq:coulomb} quantifies the scattering of two electrons from single-particle states $\ket{\vec k n}$ and $\ket{\vec k' {+} \vec q m'}$ into states $\ket{\vec k {+} \vec q m}$ and $\ket{\vec k' n'}$, respectively.
In the basis of electronic positions, it has the well-known diagonal representation $\bra{\vec r; \vec r'} \op v \ket{\vec r; \vec r'} = 1 / \abs{\vec r - \vec r'}$.

The interaction between two electrons is, however, screened by the polarizability of all other electrons.
Taking the formation of any number of electron-hole pairs into account, the screened electron-electron interaction $W(T, \omega)$ is related to the bare Coulomb interaction $v$ via
\begin{equation}
    \label{eq:v2w}
    \begin{tikzpicture}
        \pic at (0, 0) {W};
        \node at (1.25, 0) {$=$};
        \pic at (1.5, 0) {v};
        \node at (2.75, 0) {$+$};
        \pic at (3, 0) {v};
        \pic at (6, 0) {W};
        \pic at (4, 0) {chi};
    \end{tikzpicture},
\end{equation}
where $\vec q$ is transferred from the right to the left.
As a formula,
\begin{multline}
    \label{eq:v2w_form}
    W_{\vec q \vec k m n \vec k' m' n'}(T, \omega)
    = v_{\vec q \vec k m n \vec k' m' n'}
    + \sum_{\vec k'' m'' n''}
    v_{\vec q \vec k m n \vec k'' m'' n''}
    \\
    \times
    \chi_{\vec q \vec k'' m'' n''} \super b (T, \omega)
    W_{\vec q \vec k'' m'' n'' \vec k' m' n'} \ps (T, \omega),
\end{multline}
where the summation is over \emph{all} pairs of band indices.
$W(T, \omega)$ as defined here is also used in the $G W$ approximation~\cite{Hedin1965, Golze2019, Li2019}.
The summation over an infinite number of bands can be circumvented via the Sternheimer approach~\cite{Sternheimer1954, Giustino2010, Schlipf2020, Lihm2021}.

The bare electronic susceptibility or ``polarizability\rlap,'' from which any $T$ or $\omega$~dependence originates, reads
\begin{equation}
    \label{eq:susc}
    \chi_{\vec q \vec k m n} \super b (T, \omega)
    = \frac 2 N \frac
        {f(\varepsilon_{\vec k n} / T) - f(\varepsilon_{\vec k + \vec q m} / T)}
        {\varepsilon_{\vec k n} - \varepsilon_{\vec k + \vec q m} + \omega},
\end{equation}
where $\varepsilon$ and $f$ are the electronic energies and occupations, the factor 2 accounts for the spin, and $N$ is the number of $\vec k$~points.

The most important contributions in Eq.~\eqref{eq:susc} come from transitions between occupied and empty states with similar energies, i.e., from the low-energy electrons near the Fermi level.
This suggests to split the screening in Eq.~\eqref{eq:v2w} into two steps, namely the \emph{downfolding} to a low-energy system and the \emph{renormalization} to recover physical results.
We label the subset of transitions between properly chosen low-energy active states as A and the remaining transitions as the rest R (cf.\@ Fig.~\ref{fig:downfolding}).
The bare susceptibility can then be decomposed as
\begin{equation}
    \label{eq:split}
    \chi_{\vec q \vec k m n} \super b (T, \omega)
    = \chi_{\vec q \vec k m n} \super{b,A} (T, \omega)
    + \chi_{\vec q \vec k m n} \super{b,R} (T, \omega),
\end{equation}
where the first and second term can only be nonzero if the transition from $\ket{\vec k n}$ to $\ket{\vec k {+} \vec q m}$ is part of A and R, respectively.

First, in the downfolding step, the partially screened electron-electron interaction $U(T, \omega)$ is calculated from the bare Coulomb interaction $v$:
\begin{equation}
    \label{eq:v2u}
    \begin{tikzpicture}
        \pic at (0, 0) {U};
        \node at (1.25, 0) {$=$};
        \pic at (1.5, 0) {v};
        \node at (2.75, 0) {$+$};
        \pic at (3, 0) {v};
        \pic at (6, 0) {U};
        \pic at (4, 0) {chiR};
    \end{tikzpicture}.
\end{equation}
This is known as the \ac{cRPA}~\cite{Aryasetiawan2004, Aryasetiawan2006}.
The corresponding formula is equivalent to Eq.~\eqref{eq:v2w_form}, except that the summation is constrained to the transitions in R.
In general, the partially screened $U(T, \omega)$ depends on $T$ and $\omega$~\cite{Aryasetiawan2004, Falter2006}.
However, excluding low-energy transitions makes the system effectively gapped and the dependence on $T$ and $\omega$ can be controlled via the size of the active subspace~\cite{Miyake2009}.

Second, in the renormalization step, the screened $W(T, \omega)$ can be recovered from the partially screened $U(T, \omega)$~\cite{Aryasetiawan2004, Falter2006},
\begin{equation}
    \label{eq:u2w}
    \begin{tikzpicture}
        \pic at (0, 0) {W};
        \node at (1.25, 0) {$=$};
        \pic at (1.5, 0) {U};
        \node at (2.75, 0) {$+$};
        \pic at (3, 0) {U};
        \pic at (6, 0) {W};
        \pic at (4, 0) {chiA};
    \end{tikzpicture}.
\end{equation}
The corresponding formula is again equivalent to Eq.~\eqref{eq:v2w_form}, but now the summation is constrained to the finite number of active bands.
This allows us to evaluate the low-energy response, which requires a dense \ac{BZ} sampling at low electronic temperature, at an affordable computational cost.

\subsubsection{Electron-phonon interaction}
\label{sec:rpa_elph}

The change of the electronic energies upon ionic displacements is also screened by the surrounding electrons.
Without this screening, the bare electron-phonon coupling in the electronic eigenbasis reads
\begin{equation}
    \label{eq:gb}
    g_{\vec q \kappa \alpha \vec k m n} \super b
    = \frac 1 {\sqrt{M_\kappa}}
    \bra{\vec k {+} \vec q m}
        \frac{\partial \op V \super b}{\partial u_{\vec q \kappa \alpha}}
    \ket{\vec k n},
\end{equation}
with the bare external potential $V \super b$ acting on an electron amid the ensemble of ions.
In the position representation, we have $\bra{\vec r} \op V \super b \ket{\vec r} = -\sum_{\vec R \kappa} Z_\kappa / \abs{\vec R + \bm \uptau_\kappa + \vec u_{\vec R \kappa} - \vec r}$, where $Z_\kappa$ and $\bm \uptau$ are ionic charges and equilibrium positions within the unit cell.

The screened electron-phonon interaction can be written as $g = \epsilon \inv g \super b$ with the dielectric function $\epsilon = 1 - v \chi \super b$, similarly to the screened electron-electron interaction $W = \epsilon \inv v$ of Eq.~\eqref{eq:v2w}, or, using diagrams, as (cf.\@ Appendix~I of Ref.~\citenum{Geilikman1975})
\begin{equation}
    \label{eq:gb2g}
    \begin{tikzpicture}
        \pic at (0, 0) {g};
        \node at (0.5, 0) {$=$};
        \pic at (1, 0) {gb};
        \node at (1.5, 0) {$+$};
        \pic at (2, 0) {v};
        \pic at (3, 0) {chi};
        \pic at (5, 0) {gr};
    \end{tikzpicture}.\qquad
\end{equation}
Translated to a formula, it reads
\begin{multline}
    \label{eq:gb2g_form}
    g_{\vec q \kappa \alpha \vec k m n} \ps (T, \omega)
    = g_{\vec q \kappa \alpha \vec k m n} \super b
    + \sum_{\vec k' m' n'}
    v_{\vec q \vec k m n \vec k' m' n'} \ps
    \\
    \times
    \chi_{\vec q \vec k' m' n'} \super b (T, \omega)
    g_{\vec q \kappa \alpha \vec k' m' n'} \ps (T, \omega).
\end{multline}

Also the renormalization of the electron-phonon coupling can be split into two steps:
The partially screened $g \super p (T, \omega)$ is obtained from the bare $g \super b$,
\begin{equation}
    \label{eq:gb2gp}
    \begin{tikzpicture}
        \pic at (0, 0) {gp};
        \node at (0.5, 0) {$=$};
        \pic at (1, 0) {gb};
        \node at (1.5, 0) {$+$};
        \pic at (2, 0) {v};
        \pic at (3, 0) {chiR};
        \pic at (5, 0) {gpr};
    \end{tikzpicture},\qquad
\end{equation}
which can be accomplished using \ac{cDFPT}~\cite{Nomura2015}.
The screened $g(T, \omega)$ follows from the partially screened $g \super p (T, \omega)$,
\begin{equation}
    \label{eq:gp2g}
    \begin{tikzpicture}
        \pic at (0, 0) {g};
        \node at (0.5, 0) {$=$};
        \pic at (1, 0) {gpr};
        \node at (1.5, 0) {$+$};
        \pic at (2, 0) {U};
        \pic at (3, 0) {chiA};
        \pic at (5, 0) {gr};
    \end{tikzpicture}.\qquad
\end{equation}
The corresponding formulas are equivalent to Eq.~\eqref{eq:gb2g_form}.

Using the alternative expression for the inverse dielectric function $\epsilon \inv = 1 + W \chi \super b$ and the corresponding formulas for the active and rest subspace, we can also write Eqs.~\eqref{eq:gb2g}, \eqref{eq:gb2gp}, and \eqref{eq:gp2g} as
\begin{align}
    \label{eq:gb2g2}
    \begin{tikzpicture}
        \pic at (0, 0) {g};
        \node at (0.5, 0) {$=$};
        \pic at (1, 0) {gb};
        \node at (1.5, 0) {$+$};
        \pic at (2, 0) {W};
        \pic at (3, 0) {chi};
        \pic at (5, 0) {gb};
    \end{tikzpicture},\qquad
    \\
    \label{eq:gb2gp2}
    \begin{tikzpicture}
        \pic at (0, 0) {gp};
        \node at (0.5, 0) {$=$};
        \pic at (1, 0) {gb};
        \node at (1.5, 0) {$+$};
        \pic at (2, 0) {U};
        \pic at (3, 0) {chiR};
        \pic at (5, 0) {gb};
    \end{tikzpicture},\qquad
    \\
    \label{eq:gp2g2}
    \begin{tikzpicture}
        \pic at (0, 0) {g};
        \node at (0.5, 0) {$=$};
        \pic at (1, 0) {gpr};
        \node at (1.5, 0) {$+$};
        \pic at (2, 0) {W};
        \pic at (3, 0) {chiA};
        \pic at (5, 0) {gpr};
    \end{tikzpicture}.\qquad
\end{align}
In practice, these explicit definitions can be more convenient if $U(T, \omega)$ and $W(T, \omega)$ are known.

\subsubsection{Phonons}
\label{sec:rpa_ph}

Finally, we consider how electronic screening affects the phonons.
Without any electronic response, the bare interatomic force constants read~\cite{Baroni2001}
\begin{equation}
    C_{\vec R - \vec R' \kappa \alpha \kappa' \beta} \super b
    = \int \D^3 r \frac
        {\partial^2 V \super b (\vec r)}
        {\partial u_{\vec R \kappa \alpha} \partial u_{\vec R' \kappa' \beta}}
    n(\vec r)
    + \frac
        {\partial^2 \varPhi}
        {\partial u_{\vec R \kappa \alpha} \partial u_{\vec R' \kappa' \beta}},
\end{equation}
with the electron density $n$ and the classical electrostatic energy $\varPhi = 1/2 \sum_{\vec R \kappa \neq \vec R' \kappa'} Z_\kappa Z_{\kappa'} / \abs{\vec R + \bm \uptau_\kappa + \vec u_{\vec R \kappa} - \vec R' - \bm \uptau_{\kappa'} - \vec u_{\vec R' \kappa'}}$ of the ensemble of ions, which is often called Ewald energy.
The corresponding bare phonon Green's function $G \super b$ follows from the bare dynamical matrix $D \super b = C \super b / \sqrt{M_\kappa M_{\kappa'}}$ and Eq.~\eqref{eq:green_omega}.

As derived step by step in Sec.~5.1 of Ref.~\citenum{Berges2020b}, the screened phonon Green's function $G(T, \omega)$ satisfies
\begin{equation}
    \label{eq:db2d}
    \begin{tikzpicture}
        \pic at (0, 0) {G};
        \node at (1.25, 0) {$=$};
        \pic at (1.5, 0) {Gb};
        \node at (2.75, 0) {$+$};
        \pic at (3, 0) {Gb};
        \pic at (6, 0) {G};
        \pic at (4, 0) {chi};
        \pic at (4, 0) {gb};
        \pic at (6, 0) {gr};
    \end{tikzpicture},
\end{equation}
or, using matrices in the basis of ionic displacements,
\begin{equation}
    \label{eq:db2d_form}
    G_{\vec q}(T, \omega)
    = G_{\vec q} \super b
    + G_{\vec q} \super b
    \cdot \Pi_{\vec q} \ps (T, \omega)
    \cdot G_{\vec q} \ps (T, \omega),
\end{equation}
where we have defined the phonon self-energy as an electron-hole bubble connected to a bare---to avoid double counting the Coulomb interaction---and a screened electron-phonon vertex:
\begin{multline}
    \label{eq:selfen}
    \Pi_{\vec q \kappa \alpha \kappa' \beta}(T, \omega)
    \\
    = \sum_{\vec k m n}
    \bar g_{\vec q \kappa \alpha \vec k m n} \super b
    \chi_{\vec q \vec k m n} \super b (T, \omega)
    g_{\vec q \kappa' \beta \vec k m n} \ps (T, \omega).
\end{multline}
The bar denotes the complex conjugate.
If we multiply Eq.~\eqref{eq:db2d_form} with the matrix inverses $(G \super b) \inv$ and $G \inv (T, \omega)$ from the left and right, respectively, and insert Eq.~\eqref{eq:green_omega}, we are left with a simple additive formula for the screened dynamical matrix:
\begin{equation}
    \label{eq:d}
    D_{\vec q}(T, \omega)
    = D_{\vec q} \super b
    + \Pi_{\vec q} \ps (T, \omega).
\end{equation}

Eventually, also the screening of the phonons can be split into two steps.
First, the partially screened $G \super p (T, \omega)$ is derived from the bare $G \super b$:
\begin{equation}
    \label{eq:db2dp}
    \begin{tikzpicture}
        \pic at (0, 0) {Gp};
        \node at (1.25, 0) {$=$};
        \pic at (1.5, 0) {Gb};
        \node at (2.75, 0) {$+$};
        \pic at (3, 0) {Gb};
        \pic at (6, 0) {Gp};
        \pic at (4, 0) {chiR};
        \pic at (4, 0) {gb};
        \pic at (6, 0) {gpr};
    \end{tikzpicture}.
\end{equation}
Just like Eq.~\eqref{eq:db2d}, Eq.~\eqref{eq:db2dp} can be rewritten as an additive equation for the dynamical matrix:
\begin{equation}
    D_{\vec q} \super p (T, \omega)
    = D_{\vec q} \super b
    + \Pi_{\vec q} \super R (T, \omega).
\end{equation}
In the second step, the screened $G(T, \omega)$ is retrieved from the partially screened $G \super p (T, \omega)$:
\begin{equation}
    \label{eq:dp2d}
    \begin{tikzpicture}
        \pic at (0, 0) {G};
        \node at (1.25, 0) {$=$};
        \pic at (1.5, 0) {Gp};
        \node at (2.75, 0) {$+$};
        \pic at (3, 0) {Gp};
        \pic at (6, 0) {G};
        \pic at (4, 0) {chiA};
        \pic at (4, 0) {gpl};
        \pic at (6, 0) {gr};
    \end{tikzpicture}.
\end{equation}
Again, this translates into a simple addition of dynamical matrix and phonon self-energy:
\begin{equation}
    D_{\vec q}(T, \omega)
    = D_{\vec q} \super p (T, \omega)
    + \Pi_{\vec q} \super A (T, \omega).
\end{equation}
The phonon self-energy can thus be decomposed in the same way as the bare susceptibility in Eq.~\eqref{eq:split} [cf.\@ Eq.~(57) of Ref.~\citenum{Falter1981} or Eq.~(4.29) of Ref.~\citenum{Falter1988}]:
\begin{equation}
    \Pi_{\vec q}(T, \omega)
    = \Pi_{\vec q} \super A (T, \omega)
    + \Pi_{\vec q} \super R (T, \omega).
\end{equation}
In the following, we will focus on corrections to the first term, which reflects the relevant physics of the active subspace.

\subsection{Approximations to the phonon self-energy}
\label{sec:approx}

As outlined in the previous section, the exact phonon self-energy [Eq.~\eqref{eq:selfen}] is calculated using one bare and one screened electron-phonon vertex.
However, expressions with two screened vertices are often used in practice.
In Ref.~\citenum{Calandra2010}, the connection between these two formulations is established:
Using Eq.~\eqref{eq:gb2g}, and leaving $T$ and $\omega$~dependences understood for brevity, we can recast the phonon self-energy as
\begin{align}
    \label{eq:nonstationary}
    \begin{tikzpicture}
        \pic at (0, 0) {chis};
        \pic at (0, 0) {gb};
        \pic at (1, 0) {gs};
        \node at (1.5, 0) {$=$};
        \pic at (2, 0) {chis};
        \pic at (2, 0) {gb};
        \node at (3, 0) {$\,\Big[$};
        \pic at (3.25, 0) {gb};
        \node at (3.5, 0) {$+$};
        \pic at (3.75, 0) {v};
        \pic[mauve] at (4.75, 0) {chis};
        \pic[mauve] at (5.75, 0) {gs};
        \node at (6, 0) {$\Big]$};
    \end{tikzpicture}
    \\
    \label{eq:stationary}
    \hspace*{-3mm}%
    \begin{tikzpicture}
        \node at (-1.5, 1) {$=$};
        \pic at (2, 1) {chis};
        \node at (-1, 1) {$\Big[$};
        \pic at (0.25, 1) {v};
        \pic[mauve] at (-0.75, 1) {chis};
        \pic[mauve] at (-0.75, 1) {gs};
        \node at (1.5, 1) {$+$};
        \pic at (1.75, 1) {gb};
        \node at (2, 1) {$\Big]\,$};
        \node at (3, 1) {$\,\Big[$};
        \pic at (3.25, 1) {gb};
        \node at (3.5, 1) {$+$};
        \pic at (3.75, 1) {v};
        \pic[mauve] at (4.75, 1) {chis};
        \pic[mauve] at (5.75, 1) {gs};
        \node at (6, 1) {$\Big]$};
        \node at (2.25, 0) {$-$};
        \pic at (3.75, 0) {v};
        \pic[mauve] at (2.75, 0) {chis};
        \pic[mauve] at (4.75, 0) {chis};
        \pic[mauve] at (2.75, 0) {gs};
        \pic[mauve] at (5.75, 0) {gs};
    \end{tikzpicture}
    \\
    \begin{tikzpicture}
        \node at (0.25, 0) {$=$};
        \pic at (0.75, 0) {chis};
        \pic at (0.75, 0) {gs};
        \pic at (1.75, 0) {gs};
        \node at (2.25, 0) {$-$};
        \pic at (3.75, 0) {v};
        \pic at (2.75, 0) {chis};
        \pic at (4.75, 0) {chis};
        \pic at (2.75, 0) {gs};
        \pic at (5.75, 0) {gs};
        \node at (6, 0) {$\phantom{\Big]}$};
        \useasboundingbox;
        \node at (6, 0) {$,$};
    \end{tikzpicture}
\end{align}
i.e., as a phonon self-energy with two screened vertices less a double-counting term [cf.\@ Eq.~(4.23) of Ref.~\citenum{Allen1980} and Eq.~(4.5) of Ref.~\citenum{Maksimov2008}].
All of the above expressions evaluate to the same result as long as all their constituents are exact.
If however---for reasons that will become evident soon---each occurrence of the electron response $g \chi \super b$ or $\chi \super b g$ in Eqs.~\eqref{eq:nonstationary} and \eqref{eq:stationary} (\emph{only} the mauve parts) is replaced by an approximation to it, their values will differ.
Here, by construction Eq.~\eqref{eq:stationary} will deviate least from the exact value.
This is because its \emph{partial} functional derivative with respect to $\chi \super b g$ vanishes:
\begin{equation}
    \frac
        {\delta \eqref{eq:stationary}}
        {\delta \begin{tikzpicture}
            \draw[el] (0, 0) to[out=-70, in=-110] (0.5, 0);
            \draw[el] (0.5, 0) to[out=+110, in=+70] (0, 0);
            \fill (0.5, 0) circle (3pt);
        \end{tikzpicture}}
    =
    \begin{tikzpicture}
        \pic at (1, 0) {v};
        \pic at (0, 0) {chis};
        \pic at (0, 0) {gs};
    \end{tikzpicture}
    -
    \begin{tikzpicture}
        \pic at (1, 0) {v};
        \pic at (0, 0) {chis};
        \pic at (0, 0) {gs};
    \end{tikzpicture}
    = 0.
\end{equation}
The same holds true for the derivative with respect to $g \chi \super b$.
This is not the case for the right-hand side of Eq.~\eqref{eq:nonstationary}:
\begin{equation}
    \frac
        {\delta \eqref{eq:nonstationary}}
        {\delta \begin{tikzpicture}
            \draw[el] (0, 0) to[out=-70, in=-110] (0.5, 0);
            \draw[el] (0.5, 0) to[out=+110, in=+70] (0, 0);
            \fill (0.5, 0) circle (3pt);
        \end{tikzpicture}}
    =
    \begin{tikzpicture}
        \pic at (1, 0) {v};
        \pic at (0, 0) {chis};
        \pic at (0, 0) {gb};
    \end{tikzpicture}
    \neq 0.
\end{equation}
Equation~\eqref{eq:stationary} is a stationary functional of the electron response~\cite{Calandra2010}.
Consequently, an approximate electron response yields errors only at second order.
Hence, it appears to be a reasonable approximation to replace the electron response in Eq.~\eqref{eq:stationary} by the static ($\omega = 0$) and high-smearing ($T = \sigma$) response we obtain from an \emph{ab initio} calculation using \ac{DFPT}:
\begin{equation}
    \begin{tikzpicture}
        \pic at (0, 1.3) {chi};
        \pic at (0, 1.3) {gb};
        \pic at (2, 1.3) {gr};
        \node at (2.6, 1.3) {$\approx$};
        \pic at (3.2, 1.3) {chi};
        \pic at (3.2, 1.3) {gbl0};
        \pic at (5.2, 1.3) {gbr0};
        \node at (0.4, 0) {$-$};
        \pic at (3, 0) {v};
        \pic at (1, 0) {chi0};
        \pic at (4, 0) {chi0};
        \pic at (1, 0) {gl0};
        \pic at (6, 0) {gr0};
    \end{tikzpicture}.
\end{equation}
Since the approximate double-counting term does not depend on $T$ or $\omega$, we only have to focus on the first term when correcting \emph{ab initio} phonon self-energies.

Using Eq.~\eqref{eq:gp2g} and the fact that the partially screened $U(T, \omega)$ and $g \super p (T, \omega)$ exclude low-energy screening and are thus only weakly $T$ and $\omega$~dependent (in the phononic energy range), a corresponding expression can be derived for the active-subspace phonon self-energy $\Pi \super A(T, \omega)$:
\begin{equation}
    \label{eq:approx}
    \begin{tikzpicture}
        \pic at (0, 1.3) {chiA};
        \pic at (0, 1.3) {gpl};
        \pic at (2, 1.3) {gr};
        \node at (2.6, 1.3) {$\approx$};
        \pic at (3.2, 1.3) {chiA};
        \pic at (3.2, 1.3) {gbl0};
        \pic at (5.2, 1.3) {gbr0};
        \node at (0.4, 0) {$-$};
        \pic at (3, 0) {U0};
        \pic at (1, 0) {chiA0};
        \pic at (4, 0) {chiA0};
        \pic at (1, 0) {gl0};
        \pic at (6, 0) {gr0};
    \end{tikzpicture}.
\end{equation}

For later analysis, we define the following five approximate active-subspace phonon self-energies:
\begin{align}
    \label{eq:pi00}
    \Pi \super{00} (T, \omega)
    \equiv &&
    \begin{tikzpicture}
        \pic at (0, 0) {chiA};
        \pic at (0, 0) {gl0};
        \pic at (2, 0) {gr0};
    \end{tikzpicture}, &&
    \\
    \label{eq:pip0}
    \Pi \super{p0} (T, \omega)
    \equiv &&
    \begin{tikzpicture}
        \pic at (0, 0) {chiA};
        \pic at (0, 0) {gpl0};
        \pic at (2, 0) {gr0};
    \end{tikzpicture}, &&
    \\
    \label{eq:pipt}
    \Pi \super{p$T$} (T, \omega)
    \equiv &&
    \begin{tikzpicture}
        \pic at (0, 0) {chiA};
        \pic at (0, 0) {gpl0};
        \pic at (2, 0) {grT0};
    \end{tikzpicture}, &&
    \\
    \label{eq:pib0}
    \Pi \super{b0} (T, \omega)
    \equiv &&
    \begin{tikzpicture}
        \pic at (0, 0) {chiA};
        \pic at (0, 0) {gb};
        \pic at (2, 0) {gr0};
    \end{tikzpicture}, &&
    \\
    \label{eq:pibt}
    \Pi \super{b$T$} (T, \omega)
    \equiv &&
    \begin{tikzpicture}
        \pic at (0, 0) {chiA};
        \pic at (0, 0) {gb};
        \pic at (2, 0) {grT0};
    \end{tikzpicture}. &&
\end{align}

With Eq.~\eqref{eq:pi00}, the approach of Ref.~\citenum{Calandra2010} to approximate converged low-temperature nonadiabatic phonons based on adiabatic high-smearing calculations can be formulated as
\begin{align}
    \label{eq:d00}
    D_{\vec q}(T, \omega)
    \approx D_{\vec q} \super{00} (T, \omega)
    &\equiv D_{\vec q} \super u(\sigma, 0)
    + \Pi_{\vec q} \super{00} (T, \omega),
    \\
    \label{eq:du}
    D_{\vec q} \super u(\sigma, 0)
    &\equiv D_{\vec q}(\sigma, 0)
    - \Pi_{\vec q} \super{00} (\sigma, 0),
\end{align}
where we have defined the ``unscreened'' dynamical matrix $D \super u (\sigma, 0)$.
Note that in the original publication, a slightly different approach via phonons at the high electronic temperature $T_\infty$, chosen such that they can be safely interpolated, is proposed.
Equation~\eqref{eq:d00} is equivalent for $\sigma \approx T_\infty$.

With Eq.~\eqref{eq:pip0}, the corresponding \ac{cDFPT}-based formula is
\begin{align}
    \label{eq:dp0}
    D_{\vec q}(T, \omega)
    \approx D_{\vec q} \super{p0} (T, \omega)
    &\equiv D_{\vec q} \super p (\sigma, 0)
    + \Pi_{\vec q} \super{p0} (T, \omega),
    \\
    \label{eq:dp}
    D_{\vec q} \super p (\sigma, 0)
    &\equiv D_{\vec q}(\sigma, 0)
    - \Pi_{\vec q} \super{p0} (\sigma, 0),
\end{align}
where the partially screened dynamical matrix $D \super p (\sigma, 0)$ is calculated via unscreening $D(\sigma, 0)$ from \ac{DFPT} but could also be directly obtained from a self-consistent \ac{cDFPT} calculation.
We note that the choice between unscreening and constraining is also relevant in other contexts~\cite{Nomura2012}.

A shortcoming of Eq.~\eqref{eq:pip0} is that the bare susceptibility is generally calculated at a different electronic temperature than the screened vertex.
Taking the temperature dependence of the screened vertex into account yields Eq.~\eqref{eq:pipt} and
\begin{equation}
    \label{eq:dpt}
    D_{\vec q}(T, \omega)
    \approx D_{\vec q} \super{p$T$} (T, \omega)
    \equiv D_{\vec q} \super p (\sigma, 0)
    + \Pi_{\vec q} \super{p$T$} (T, \omega).
\end{equation}
This will be insightful but is useless in practice since $g(T, 0)$ is as computationally expensive as $D(T, 0)$.

In the limit of an infinitely large active subspace, where the partially screened vertex becomes bare, and for a single electronic temperature $T = \sigma$, Eqs.~\eqref{eq:dp0} and \eqref{eq:dp} are equivalent to Eq.~(145) of Ref.~\citenum{Giustino2017}.
In practice, the phonon self-energy with one bare vertex might also be calculated for a finite number of active bands as in Eq.~\eqref{eq:pib0}, which leads to
\begin{align}
    \label{eq:db0}
    D_{\vec q}(T, \omega)
    \approx D_{\vec q} \super{b0} (T, \omega)
    &\equiv D_{\vec q} \super{ub} (\sigma, 0)
    + \Pi_{\vec q} \super{b0} (T, \omega),
    \\
    \label{eq:dub}
    D_{\vec q} \super{ub} (\sigma, 0)
    &\equiv D_{\vec q}(\sigma, 0)
    - \Pi_{\vec q} \super{b0} (\sigma, 0).
\end{align}
Note that the unscreened $D \super{ub} \neq D \super b$ for finite active subspaces.
Also here, a variant with temperature-corrected screened vertex can be studied.
With Eq.~\eqref{eq:pibt},
\begin{equation}
    \label{eq:dbt}
    D_{\vec q}(T, \omega)
    \approx D_{\vec q} \super{b$T$} (T, \omega)
    \equiv D_{\vec q} \super{ub} (\sigma, 0)
    + \Pi_{\vec q} \super{b$T$} (T, \omega).
\end{equation}

Importantly, the quality of Eqs.~\eqref{eq:d00}--\eqref{eq:dbt} is not determined by how close the approximate phonon self-energies defined in Eqs.~\eqref{eq:pi00}--\eqref{eq:pibt} are to the exact phonon self-energy---in fact, the absolute values can be quite different.
Instead, what matters is the difference $\Pi \super A (T, \omega) - \Pi \super A (\sigma, 0)$, the deviations from which we can quantify.

For the asymmetric formulations using one bare or partially screened vertex we can readily recognize that
\begin{multline}
    \label{eq:linear}
    \Pi_{\vec q} \super{b0/p0} (T, \omega) - \Pi_{\vec q} \super{b0/p0} (\sigma, 0)
    \approx \Pi_{\vec q} \super A (T, \omega) - \Pi_{\vec q} \super A (\sigma, 0)
    \\
    + \smash[b]{\sum_{\vec k m n}}
    \bar g_{\vec q \kappa \alpha \vec k m n} \super{b/p} (\sigma, 0)
    \chi_{\vec q \vec k m n} \super{b,A} (T, \omega)
    \\
    \times
    [g(\sigma, 0) - g(T, \omega)]_{\vec q \kappa' \beta \vec k m n},
\end{multline}
where the error in the phonon self-energy correction is \emph{linear} both in the bare or partially screened vertex and in the error of the screened vertex.
Equation~\eqref{eq:linear} becomes exact for $g \super p (T, \omega) = g \super p (\sigma, 0)$, i.e., for a large enough active subspace.
Since the strength of the partially screened vertex increases with the size of the active subspace (asymptotically approaching the bare vertex), we expect larger errors in Eq.~\eqref{eq:db0} than in Eq.~\eqref{eq:dp0}, especially for small low-energy active subspaces.

For the symmetric formulation using two screened vertices, via the straightforward derivation in Appendix~\ref{app:error}, we can analogously show that
\begin{multline}
    \label{eq:quadratic}
    \Pi_{\vec q} \super{00} (T, \omega) - \Pi_{\vec q} \super{00} (\sigma, 0)
    \approx \Pi_{\vec q} \super A (T, \omega) - \Pi_{\vec q} \super A (\sigma, 0)
    \\
    - 1/N^2 \smash[b]{\sum_{\mathclap{\vec k m n \vec k' m' n'}}}
    [\overline{g(\sigma, 0) - g(T, -\bar \omega)}]_{\vec q \kappa \alpha \vec k m n} \ps
    W_{\vec q \vec k m n \vec k' m' n'} \inv (T, \omega)
    \\
    \times
    [g(\sigma, 0) - g(T, \omega)]_{\vec q \kappa' \beta \vec k' m' n'},
\end{multline}
where the error is only \emph{quadratic} in the error of the screened vertex and where $W \inv = v \inv \epsilon$ is the inverse screened electron-electron interaction [cf.\@ Eq.~\eqref{eq:v2w}].
Equation~\eqref{eq:quadratic} becomes exact for $g \super p (T, \omega) = g \super p (\sigma, 0)$ and $U(T, \omega) = U(\sigma, 0)$ [cf.\@ Eq.~\eqref{eq:approx}].

Finally, we remark that throughout the paper by the word ``static'' we mean $\omega = 0$ and by ``dynamical'' that a quantity has a frequency dependence $\omega$.
With the word ``nonadiabatic'' we refer to an Engelsberg-Schrieffer~\cite{Engelsberg1963} type of nonadiabaticity, i.e., a nonadiabatic electronic renormalization of \emph{adiabatic} phonons~\cite{Saitta2008, Calandra2010}.
However, we still neglect adiabatic effects beyond the Migdal theorem~\cite{Migdal1958} arising from vertex corrections, which are of the order $\sqrt{m \sub e \smash/ M}$ and thus usually much smaller.

\subsection{Symmetrization of the phonon self-energy}
\label{sec:sym}

The approximate phonon self-energies $\Pi \super{p0}$ and $\Pi \super{b0}$ have been defined such that the screened vertex is situated on the right of the electron-hole bubble.
An equivalent definition with the screened vertex on the left is possible.
While for the exact $\Pi$ both choices yield the same result, this is no longer true when the bubbles contained in $g$ through Eq.~\eqref{eq:gb2g} or Eq.~\eqref{eq:gp2g} are approximated (cf.\@ Fig.~3 of Ref.~\citenum{Marini2023a}).
To ensure that the static $\Pi \super{b0/p0}(T, 0)$ are Hermitian~\cite{Berges2020a} and the dynamical $\Pi \super{b0/p0}(T, \omega)$ fulfill the fluctuation-dissipation theorem~\cite{Marini2023a}, for the rest of the paper we replace $\smash{\bar g_{\vec q \kappa \alpha \vec k m n} \super{b/p}} (\sigma, 0) g_{\vec q \kappa' \beta \vec k m n} \ps (\sigma, 0)$ by
\begin{multline}
    \label{eq:g2}
    \mathcal G_{\vec q \kappa \alpha \kappa' \beta \vec k m n} \equiv
    \frac 1 2 \bigl[
        \bar g_{\vec q \kappa \alpha \vec k m n} \super{b/p} (\sigma, 0)
        g_{\vec q \kappa' \beta \vec k m n} \ps (\sigma, 0)
        \\
        + \bar g_{\vec q \kappa \alpha \vec k m n} \ps (\sigma, 0)
        g_{\vec q \kappa' \beta \vec k m n} \super{b/p} (\sigma, 0)
    \bigr]
\end{multline}
in Eqs.~\eqref{eq:pip0} and \eqref{eq:pib0} and similarly in Eqs.~\eqref{eq:pipt} and \eqref{eq:pibt}.
Note that complex conjugation of the coupling inverts the scattering process in the static case where the bare susceptibility is real and invariant under $\vec q \leftrightarrow -\vec q$.
In the dynamical case, care has to be taken to keep the frequency argument in the correct quadrant of the complex plane [cf.~Eq.~\eqref{eq:quadratic}].

By construction, $\mathcal G$ is a Hermitian matrix in the basis of displacements $\kappa \alpha, \kappa' \beta$ for each $\vec q, \vec k$ and $m, n$.
It has at most two nonzero eigenvalues $\lambda^{(1)} > \lambda^{(2)}$ of opposite sign belonging to the normalized eigenvectors $\vec v^{(1, 2)}$, i.e., a spectral representation
\begin{equation}
    \mathcal G_{\kappa \alpha \kappa' \beta}
    = \bar g^{(1)}_{\kappa \alpha \vphantom\beta} g^{(1)}_{\kappa' \beta}
    - \bar g^{(2)}_{\kappa \alpha \vphantom\beta} g^{(2)}_{\kappa' \beta},
\end{equation}
with $\vec g^{(1, 2)} = \sqrt{\abs{\lambda^{(1, 2)}}} \vec v^{(1, 2)}$.
As long as $g \super{b/p}$ and $g$ mostly differ in magnitude but not in direction, the negative eigenvalue is negligibly small and the phonon self-energy can be rewritten with the same effective vertex $g^{(1)}$ on both sides.
This applies if the dielectric function is largely independent of or diagonal in the electronic indices.
For small $\vec q$, long-range effects that will be discussed in Sec.~\ref{sec:froehlich} enhance the orthogonal component of $g \super{b/p}$ and thus the negative eigenvalue.

\subsection{Long-range effects}
\label{sec:lr}

\subsubsection{Friedel long-rangedness}
\label{sec:friedel}

In a metal, the bare electronic susceptibility usually has sharp features at finite momenta in reciprocal space or, equivalently, it is long-range in real space.
This can be referred to as Friedel~\cite{Friedel1958}, Peierls~\cite{Peierls1955}, or Kohn~\cite{Kohn1959} long-rangedness.
These features are generated by the low-energy system; i.e., they affect $\chi \super{b,A} (T, \omega)$ while $\chi \super{b,R} (T, \omega)$ [cf.\@ Eq.~\eqref{eq:split}] can safely be assumed to be smooth in reciprocal space or short-range in real space.
This long-rangedness is inherited by the other screened quantities discussed in Sec.~\ref{sec:rpa}, i.e., $G(T, \omega)$ or $D(T, \omega)$, $g(T, \omega)$, and $W(T, \omega)$.
All remaining quantities are bare or partially screened and thus smooth, which allows us not only to calculate them using coarse integration meshes but also to interpolate them easily.
We show an example of Friedel long-range effects for a one-dimensional model in Fig.~\ref{fig:models}\,(a).

\subsubsection{Fr\"ohlich long-rangedness}
\label{sec:froehlich}

The bare and partially screened phonons and electron-phonon coupling do not show any Friedel long-rangedness.
However, as soon as the metallic screening is removed, we deal with another type of long-range phenomenon usually observed in insulators and semiconductors, which we will refer to as Fr\"ohlich or Coulomb long-rangedness, most clearly recognizable in the bare Coulomb interaction $v(\vec r, \vec r') = 1 / \abs{\vec r - \vec r'}$ itself.
In the absence of metallic screening, this propagates into the bare or partially screened quantities and manifests as divergences or discontinuities at the center of the \ac{BZ}, which would lead to Gibbs oscillations when performing a naive Fourier interpolation.
However, analytical models for these effects exist~\cite{Brovman1967, Giannozzi1991, Gonze1994, Verdi2015, Sjakste2015, Sohier2016, Royo2020, Brunin2020, Jhalani2020, Ponce2021, Royo2021, Macheda2022, Macheda2023, Sio2022}, allowing us to split the electron-phonon coupling and the dynamical matrix into a long-range ($\mathcal L$) and a short-range part ($\mathcal S$), the latter of which can be easily interpolated.

Here, we will apply a recent approach to model the long-range electrostatics of two-dimensional materials~\cite{Royo2021, Ponce2023a, Ponce2023b}, including dipolar and quadrupolar terms, where we neglect the effect of the out-of-plane polarizability on the dynamical matrix.
We decompose the bare or partially screened dynamical matrix and electron-phonon coupling as
\begin{align}
    D_{\vec q \kappa \alpha \kappa' \beta} \super{b/p}
    &= D_{\vec q \kappa \alpha \kappa' \beta}^{\mathcal S}
    + D_{\vec q \kappa \alpha \kappa' \beta}^{\mathcal L},
    \\
    g_{\vec q \kappa \alpha \vec R s p} \super{b/p}
    &= g_{\vec q \kappa \alpha \vec R s p}^{\mathcal S}
    + g_{\vec q \kappa \alpha}^{\mathcal L}
    \delta_{\smash{\vec R 0}}^{\vphantom{\mathcal L}}
    \delta_{s p}^{\vphantom{\mathcal L}},
\end{align}
where $s$ and $p$ label electronic orbitals located in the unit cells at the origin and $\vec R$, respectively.
The long-range parts read
\begin{align}
    D_{\vec q \kappa \alpha \kappa' \beta}^{\mathcal L}
    &= \widetilde D_{\vec q \kappa \alpha \kappa' \beta}^{\mathcal L}
    - \delta_{\kappa \kappa'} \sum_{\kappa''}
    \widetilde D_{0 \kappa \alpha \kappa'' \beta}^{\mathcal L},
    \\
    \widetilde D_{\vec q \kappa \alpha \kappa' \beta}^{\mathcal L}
    &= \sum_{\mathclap{\vec G \neq -\vec q}}
    a_{\vec q + \vec G}
    \bar b_{\vec q + \vec G \kappa \alpha}
    b_{\vec q + \vec G \kappa' \beta},
    \\
    g_{\vec q \kappa \alpha}^{\mathcal L}
    &= \sum_{\mathclap{\vec G \neq -\vec q}}
    a_{\vec q + \vec G}
    b_{\vec q + \vec G \kappa \alpha},
\end{align}
where the summations over \emph{in plane} reciprocal lattice vectors $\vec G$ converge fast.
The scalar part (independent of $\kappa$ and $\alpha$) is
\begin{equation}
    \label{eq:lr_scalar}
    a_{\vec q}
    = \frac{2 \pi f_L(\vec q)}{A \abs{\vec q}}
    \bigl[
        1 + \frac{c f_L(\vec q)}{2 \abs{\vec q}}
        \vec q^T (\epsilon - \mathds 1) \vec q
    \bigr] \inv,
\end{equation}
with the unit cell area $A$ and height $c$, the dielectric constant $\epsilon$, and the cutoff function $f_L(\vec q) = 1 - \tanh(\abs{\vec q} L / 2)$.
The range-separation parameter $L$ is chosen such that the real-space force constants are minimized.
The vectorial part is given by
\begin{equation}
    \label{eq:lr_vector}
    b_{\vec q \kappa \alpha}
    = \frac{\E^{-\I \vec q \bm \uptau_\kappa}}{\sqrt{M_\kappa}}
    \bigl[
        \I \vec Z^*_{\kappa \alpha} \vec q
        + \frac 1 2 \vec q^T Q_{\kappa \alpha} \vec q
    \bigr],
\end{equation}
including the Born effective charges $Z^*_{\kappa \alpha \beta}$ and the quadrupole tensors $Q_{\kappa \alpha \beta \beta'}$.
Note that neither phonons nor effective charges from \ac{cDFPT} calculations fulfill the acoustic sum rule.
In the bare system, $\epsilon = \mathds 1$ and $\vec Z^* = Z \mathds 1$ with nuclear charge $Z$.

\subsection{Wannierization and Fourier interpolation}
\label{sec:wannier}

In practice, calculations of the electronic energies and especially the dynamical matrices and electron-phonon couplings are limited to a coarse grid of $\vec k$ and $\vec q$~points in the \ac{BZ}.
The points in between are usually obtained via Fourier interpolation, i.e., by a discrete Fourier transform into a localized representation and the smoothest possible back transform to arbitrary points.
The phononic degrees of freedom already have a natural localized basis, namely the Cartesian ionic displacement directions; for the electrons, the basis of Wannier functions~\cite{Marzari2012}, i.e., localized orthogonal orbitals, is used.

For instance, the interpolant of the short-range part of the electron-phonon coupling~\cite{Giustino2007b} as used in the \textsc{EPW} code~\cite{Giustino2007b, Noffsinger2010, Ponce2016} reads
\begin{equation}
    \label{eq:epw}
    g_{\vec q \nu \vec k m n}^{\mathcal S}
    = \sum_{\mathclap{\vec R \kappa \alpha \vec R' s p}}
    e_{\vec q \kappa \alpha \nu}^{\vphantom{\mathcal S}}
    \bar \psi_{\vec k + \vec q s m}^{\vphantom{\mathcal S}}
    g_{\vec R \kappa \alpha \vec R' s p}^{\mathcal S}
    \psi_{\vec k p n}^{\vphantom{\mathcal S}}
    \E^{\I (\vec q \vec R + \vec k \vec R')},
\end{equation}
where $e$ and $\psi$ are the eigenvectors of dynamical matrix and Wannier Hamiltonian, for which equivalent formulas hold.

Having handled the Fr\"ohlich long-rangedness (Sec.~\ref{sec:froehlich}), we can safely interpolate all quantities except those related to $\chi \super{b,A} (T, \omega)$ at low $T$, which has to be evaluated on a dense mesh because of the inherent Friedel long-rangedness (Sec.~\ref{sec:friedel}).

\section{Model results}
\label{sec:models}

Inspired by Ref.~\citenum{Marini2023a}, to illustrate selected formulas given in the previous section, we will now apply them to minimal models for which the dynamical phonon self-energy $\Pi(T, \omega)$ in \ac{RPA} as defined in Eq.~\eqref{eq:selfen} is accessible.
First, in Sec.~\ref{sec:peierls}, we consider a one-dimensional chain with nearest-neighbor hopping parameter, force constant, and electron-phonon coupling, similar to what has been used by Peierls~\cite{Peierls1955} and Su \emph{et al.}~\cite{Su1979}.
Second, in Sec.~\ref{sec:heg}, we consider the lowest band of a periodic homogeneous electron gas with Fr\"ohlich electron-phonon coupling.
We complement both models with suitable electron-electron interactions to realize different screening levels of the electron-phonon coupling.
As there is only one electronic band, which is considered as active, no distinction between bare and partially screened quantities is made here.

\subsection{Generalized one-dimensional Peierls model}
\label{sec:peierls}

\begin{figure*}
    \includegraphics{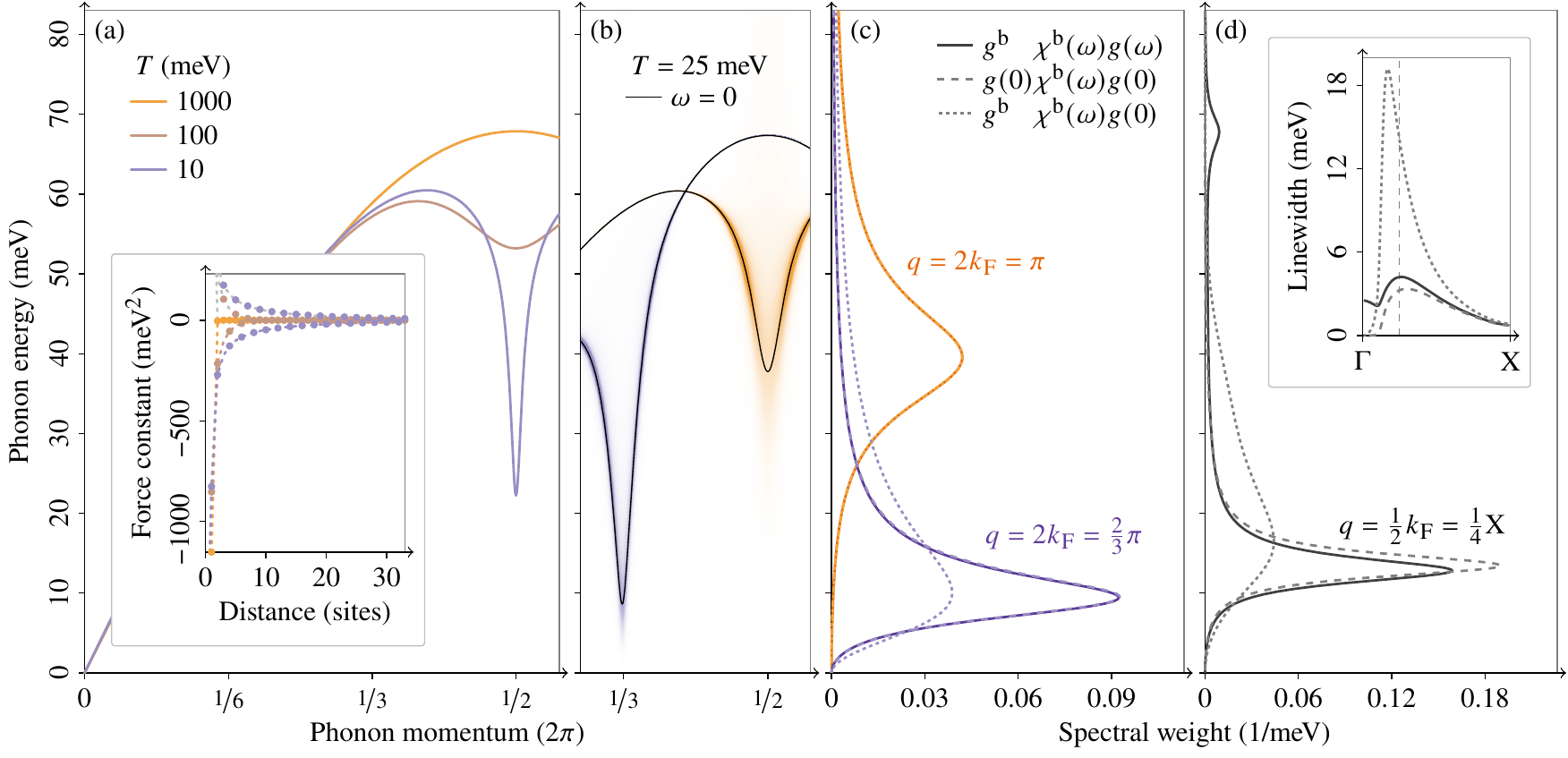}
    \caption{Results for the generalized (a--c)~one-dimensional Peierls and (d)~three-dimensional Fr\"ohlich models.
    (a)~Adiabatically screened phonon dispersion in \ac{RPA} with Kohn anomaly for different electronic temperatures $T$ at half filling.
    The inset depicts the corresponding interatomic force constants, where dashed lines are guides for the eye.
    (b)~Phonon spectral function in \ac{RPA} together with adiabatic phonon dispersion at $T = 25$~meV at half filling (orange) and one-third filling (mauve).
    (c)~Cross section through the spectral function at the Kohn anomaly according to Eq.~\eqref{eq:d} (\ac{RPA}), Eq.~\eqref{eq:d00} (``screened-screened''), and Eq.~\eqref{eq:db0} (``bare-screened'').
    (d)~Corresponding figure for the Fr\"ohlich model for a selected Fermi wave vector $k \sub F$ and phonon momentum $q$.
    The inset depicts the linewidth according to Eq.~\eqref{eq:gamma} as a function of $q$.}
    \label{fig:models}
\end{figure*}

We model the electron and phonon dispersion as well as the electron-phonon and electron-electron interaction as
\begin{align}
    \varepsilon_k &= -t \cos(k),
    \\
    D_q \super b &= \omega_0^2 [1 - \cos(q)],
    \\
    g_{q k} (T, 0) &= \I g_0 [\sin(k) - \sin(k + q)],
    \\
    U_q &= U_0 - V_0 \log[\sin(\abs q / 2)],
\end{align}
where we choose $t = 1$~eV, $\omega_0 = 50$~meV, $g_0 = 0.02$~eV\pow{3/2}, a local $U_0 = 1$~eV, and a nonlocal $V_0 = 0.5$~eV.
At half filling, the Fermi wave vector is $k \sub F = \pi / 2$ and the chemical potential zero.

Note that while $D \super b$ and $U$ are bare, we define an adiabatically screened $g(T, 0)$, from which the bare $g \super b$ follows through unscreening using Eq.~\eqref{eq:gp2g}.
The screened $W(T, \omega)$, $g(T, \omega)$, and $D(T, \omega)$ are then obtained via Eqs.~\eqref{eq:v2w}, \eqref{eq:gb2g2}, and \eqref{eq:db2d}.
Our choice for the coupling has the disadvantage that $g \super b$ rather than $g$ depends on electronic temperature $T$ and the filling $k \sub F$ via a $k$-independent ``on site'' term.
This however ensures a physically sound $g(T, 0)$---describing the displacement-induced change of the hopping parameter $t$---and does not invalidate our central findings, displayed in Fig.~\ref{fig:models}\,(a--c).

In Fig.~\ref{fig:models}\,(a), we show the relation between static low-energy electronic screening and long-range force constants at half filling, as discussed in Sec.~\ref{sec:friedel}.
The force constants are obtained from $D(T, 0)$ via a discrete Fourier transform.
At the high electronic temperature of $T = 1$~eV, the phonon dispersion is smooth and the corresponding force constants decay to nearly zero already after the first neighboring site.
As soon as the electronic temperature is lowered, a Kohn anomaly~\cite{Kohn1959} in the phonon dispersion emerges at the \ac{BZ} boundary $q = 2 k \sub F = \pi$, and the force constants become long-range exhibiting Friedel oscillations~\cite{Friedel1958}, whereby the sign changes from one site to the other.
For the chosen parameters, the phonons become soft (imaginary frequencies) below the \ac{CDW} temperature $T \sub{CDW} = 6.16$~meV, indicating the onset of the dimerization.

We now turn to the nonadiabatic case.
Figure~\ref{fig:models}\,(b) shows the phonon spectral function in \ac{RPA} using Eq.~\eqref{eq:specfun} at $T = 25$~meV for $k \sub F = \pi / 2$ (orange shade) and $k \sub F = \pi / 3$ (mauve shade), together with the corresponding adiabatic phonon frequencies (solid lines).
In practice, we replace the infinitesimal $0^+$ by a small value of $0.4$~meV and sample the \ac{BZ} using 55\,440 $k$~points.
There are no noteworthy nonadiabatic effects in the vicinity of the $\Gamma$~point, toward which the coupling vanishes.
However, when approaching the Kohn anomaly, whose position $q = 2 k \sub F$ moves away from the \ac{BZ} boundary upon doping, the branch starts to broaden significantly.

In Fig.~\ref{fig:models}\,(c), we show cross sections through the phonon spectral function at the Kohn anomaly.
We compare the results using $\Pi(T, \omega)$ in \ac{RPA} [Eq.~\eqref{eq:selfen}] (solid lines) with corresponding data from $\Pi \super{00} (T, \omega)$ [Eq.~\eqref{eq:pi00}] (dashed lines) and $\Pi \super{b0} (T, \omega)$ [Eq.~\eqref{eq:pib0}] (dotted lines).
Here, we first calculate $D(T, 0)$ in static \ac{RPA}, to which we add phonon self-energy \emph{differences} according to Eqs.~\eqref{eq:d00} and \eqref{eq:db0}.
As we do not change the electronic temperature, no distinction between low $\sigma$ and high $T$ is made.
While at half filling all approaches agree since $g = g \super b$ by symmetry, away from it they start to deviate.
The approach using two screened vertices remains close to the \ac{RPA} result, while the combination of bare and approximately screened vertices overestimates the linewidth.
This is consistent with the linear and quadratic errors quantified in Sec.~\ref{sec:approx}.
Indeed, subtracting the second and third lines of Eqs.~\eqref{eq:linear} and \eqref{eq:quadratic}, accessible for this model, from the approximate $D \super{b0} (T, \omega)$ and $D \super{00} (T, \omega)$, respectively, we recover the \ac{RPA} result.
The quadratic error simplifies to $\av{g_{q k}(T, 0) - g_{q k}(T, \omega)}_k^2 / W_q(T, \omega)$, where $\av \cdots_k$ denotes the $k$ average.

\subsection{Generalized three-dimensional Fr\"ohlich model}
\label{sec:heg}

As a second model, similar to the one considered in Sec.~V\,C of Ref.~\citenum{Marini2023a}, we now consider the lowest band of a periodic homogeneous electron gas,
\begin{equation}
    \varepsilon_{\vec k}
    = \frac 1 {2 m^*} \min_{\vec G}
    \abs{\vec k + \vec G}^2,
\end{equation}
with an effective mass $m^* = 2 m \sub e$ in a cell of lattice constant $a = 10$~\AA{} (defining the reciprocal lattice vectors $\vec G$), a constant Einstein phonon of frequency
\begin{equation}
    \omega_{\vec q} \super b = \omega_0,
\end{equation}
with $\omega_0 = 25$~meV, the Fr\"ohlich electron-phonon coupling
\begin{equation}
    g_{\vec q} \super b
    = \I g_0
    \max_{\vec G} \frac 1 {\abs{\vec q + \vec G}},
\end{equation}
with $g_0 = 55$~meV\pow{3/2}/\AA, and the Coulomb interaction
\begin{equation}
    U_{\vec q} = U_0 \max_{\vec G} \frac 1 {\abs{\vec q + \vec G}^2},
\end{equation}
with $U_0 = 5$~meV/\AA\pow2.
We choose $k \sub F = \pi / 2 a$.

Our model differs from the one of Ref.~\citenum{Marini2023a} in the choice of parameters, which in our case do not describe MgB\s2.
Also, Ref.~\citenum{Marini2023a} considers the full $\vec k$ and $\vec q$ dependence of the extended system.
We obtain the screened quantities as described in the previous section, Sec.~\ref{sec:peierls}, this time using $T = \eta = 3$~meV in combination with $240^3$ $\vec k$~points.
In addition, we approximate the phonon linewidth as~\cite{Marini2023a}
\begin{equation}
    \label{eq:gamma}
    \gamma_{\vec q} \super{(00/b0)} = -\frac 1 {\omega_0} \Im \Pi_{\vec q} \super{(00/b0)} (\omega_0 + \I \eta).
\end{equation}

The results are shown in Fig.~\ref{fig:models}\,(d).
Interestingly, for the selected phonon momentum $q \equiv q_x = k \sub F / 2$, the phonon self-energy in \ac{RPA} yields a nontrivial phonon line shape with a satellite at about 68~meV (solid line).
This feature is not captured by the screened-screened approach (dashed line), which means that it is encoded in the infinite series of dynamical electron-hole bubbles included in $g(T, \omega)$.
However, the main phonon peak at about 13~meV is relatively well reproduced.
Again, the bare-screened approach (dotted line) overestimates the linewidth.
This can also be seen in the inset, where $\gamma$ from Eq.~\eqref{eq:gamma} is shown as a function of $q$.
As the coupling in this model does not depend on $\vec k$, the quadratic error of $D \super{00} (T, \omega)$ simplifies even more and reads $[g_{\vec q}(T, 0) - g_{\vec q}(T, \omega)]^2 / W_{\vec q}(T, \omega)$.

The findings presented above suggest that the approach of Ref.~\citenum{Calandra2010} can indeed be used to estimate the frequency-dependent dynamical matrix based on adiabatic results.
It is however not clear to what extent the conclusions drawn from these models can be transferred to realistic materials simulated \emph{ab initio}.
This question will be addressed in the following sections.
Since here, different from the model case, nonadiabatic reference data are missing, we will not consider the transition from the static to the dynamical case but rather from high to low electronic temperature.

\section{Implementation}
\label{sec:implementation}

We implemented routines to perform \ac{cDFPT} calculations, based on existing code provided by the authors of Ref.~\citenum{Nomura2015}, and to renormalize the phonons according to Eqs.~\eqref{eq:d00}--\eqref{eq:dbt} in the \textsc{PHonon} and \textsc{EPW} codes~\cite{Giustino2007b, Noffsinger2010, Ponce2016}, which are part of the \textsc{Quantum ESPRESSO} distribution~\cite{Giannozzi2009, Giannozzi2017, Giannozzi2020}.
The corresponding patch is provided in Supplemental Material~\cite{Note2}.

The implementation of constrained theories such as \ac{cRPA} and \ac{cDFPT} on top of existing programs to perform unconstrained \ac{RPA} or \ac{DFPT} calculations is straightforward and requires only minor modifications of the source code~\cite{Nomura2015}.
In fact, the most difficult aspect is the definition of suitable electronic active subspaces, i.e., the identification of the band indices (usually sorted by energy in \emph{ab initio} codes) that belong to the active bands for each $\vec k$~point.
In fortunate cases, an appropriate low-energy subspace is isolated from all other bands~\cite{Berges2020a}.
In the general case, however, the active bands will be entangled with other bands.

For simplicity, we define the \ac{cDFPT} active subspace via an energy window, but a selection via band indices or orbital projections is also possible~\cite{Berges2017}.
We use a slightly modified version of the \textsc{PHonon} code with additional input parameters \texttt{cdfpt\_min} and \texttt{cdfpt\_max}, which define the lower and upper bounds of the energy window, as well as \texttt{bare} to suppress the electronic response.

The calculation and interpolation of the electron-phonon coupling as well as the phonon renormalization are done with the \textsc{EPW} code~\cite{Giustino2007b, Noffsinger2010, Ponce2016}.
Usually, the \textsc{EPW} code reads a single directory \texttt{dvscf\_dir} including the dynamical matrices $D$ and the change of the self-consistent potential $\partial V$ from \ac{DFPT} as calculated with the \textsc{PHonon} code.
We define a second input parameter \texttt{cdfpt\_dir} pointing to analogous data $D \super{b/p}$ and $\partial V \super{b/p}$ from a bare or \ac{cDFPT} calculation.
This directory also contains the values of \texttt{cdfpt\_min} and \texttt{cdfpt\_max}, which for convenience are used to set the default ``frozen window'' for the generation of Wannier functions~\cite{Pizzi2020}.
If both \texttt{dvscf\_dir} and \texttt{cdfpt\_dir} are specified, the modified code performs the calculation of the electron-matrix elements $g(\sigma, 0)$ and $g \super{b/p} (\sigma, 0)$ and the Fourier interpolation of the dynamical matrices $D(\sigma, 0)$ and $D \super{b/p} (\sigma, 0)$ and the matrix elements in the same way.
The fact that identical basis transforms are employed on both the \ac{DFPT} and the \ac{cDFPT} data ensures a consistent gauge.
Finally, we evaluate phonon self-energies and spectral functions for arbitrary $\vec q$~points using dense $\vec k$~meshes and small electronic temperatures.

Besides the above, we introduced additional inputs for the \textsc{EPW} code:
Since the \ac{cDFPT} quantities do not always fulfill the acoustic sum rule~\cite{vanLoon2021a}, enforcing it can be disabled using \texttt{asr\_typ = \textquotesingle none\textquotesingle}.
To properly handle the long-range terms in \ac{cDFPT}, we define \texttt{lpolarc} and read the file \texttt{quadrupolec.fmt} in addition to the existing \texttt{lpolar} and \texttt{quadrupole.fmt}.
Using \texttt{unscreen\_fine}, $D \super u (\sigma, 0)$ and $D \super{ub/p} (\sigma, 0)$ can be calculated on the dense instead of coarse \ac{BZ} meshes [cf.\@ Eqs.~\eqref{eq:du}, \eqref{eq:dp}, and \eqref{eq:dub}].
We select $T$ and the corresponding smearing function $f$ via \texttt{temps} and \texttt{types}.
Finally, the $\I 0^+$ appearing in the phonon spectral function in Eq.~\eqref{eq:specfun} is set in practice via two smearings, \texttt{degaussw} and \texttt{degaussq}, using Eq.~\eqref{eq:double_smearing} defined later.

\section{\emph{Ab initio} results}
\label{sec:results}

In this section, we apply the above to monolayer TaS\s2, for which we calculate screened, partially screened, and bare phonons (Sec.~\ref{sec:phonons}), the corresponding electron-phonon coupling (Sec.~\ref{sec:coupling}), renormalized phonons using the different approaches (Secs.~\ref{sec:renorm} and \ref{sec:corr}), and the phonon spectral function (Sec.~\ref{sec:specfun}).
To test for general validity, we perform additional calculations for n-doped MoS\s2; see Appendix~\ref{app:mos2}.

The trigonal-prismatic transition-metal dichalcogenide TaS\s2 is long known to be a showplace of competing \acp{CDW}~\cite{Tidman1974} and superconductivity~\cite{Nagata1992, Vano2023}, which are suppressed and enhanced, respectively, when reducing the material thickness to the monolayer~\cite{NavarroMoratalla2016, Yang2018}.
Based on \ac{cDFPT} results~\cite{Berges2020a}, the lattice instability and associated Kohn anomalies are exclusively due to low-energy electronic screening from an isolated half-filled band at the Fermi level.
These signs of significant electron-phonon coupling and well separable electronic energy scales make monolayer TaS\s2 an ideal system to test the discussed methods and to study the different levels of electronic screening.

\begin{figure*}
    \includegraphics{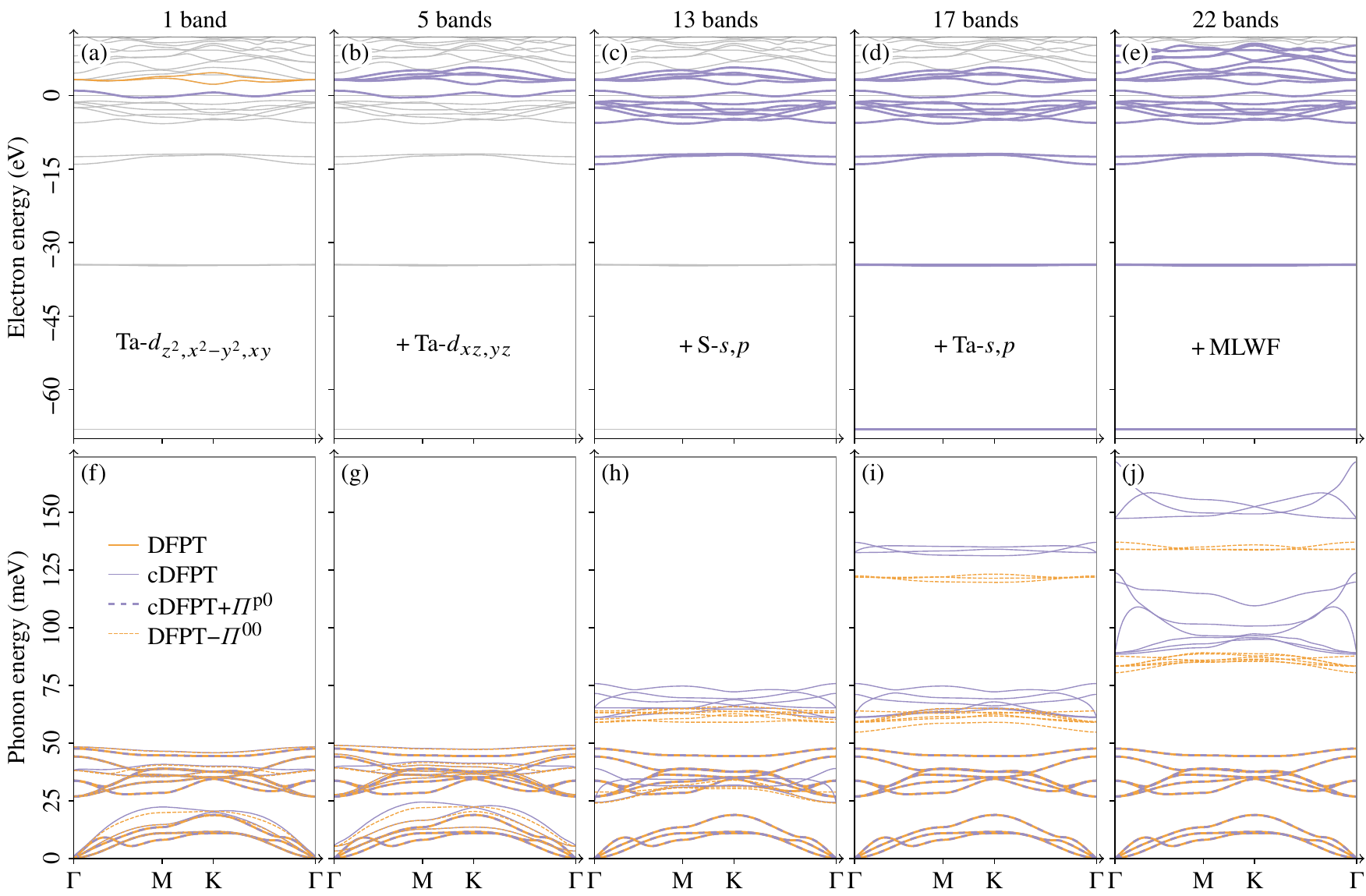}
    \caption{(a--e)~Electronic band structure of monolayer TaS\s2 from \ac{DFT}.
    Possible choices for sets of active bands of increasing size are shown using thick mauve lines.
    All colored bands (mauve or orange) have been Wannierized based on projections onto the indicated atomic orbitals.
    (f--j)~Corresponding phonon dispersions for a high Marzari-Vanderbilt smearing of $\sigma = 20$~mRy.
    Screened and partially screened phonons from \ac{DFPT} and \ac{cDFPT} are shown using solid lines, renormalized (screened \ac{cDFPT} and unscreened \ac{DFPT}) phonons using dashed lines.}
    \label{fig:subspaces}
\end{figure*}

In the \emph{ab initio} calculations, we apply the \ac{PBE} functional~\cite{Perdew1996} and corresponding norm-conserving pseudopotentials with nonlinear core correction and without \ac{SC} states from the \textsc{PseudoDojo} table~\cite{Hamann2013, vanSetten2018} at an energy cutoff of 100~Ry.
We separate periodic images of the layer using a unit-cell height of 15~\AA{} together with a truncation of the Coulomb interaction in this direction~\cite{Sohier2017}.
The relaxed lattice constant is $a = 3.34$~\AA.
For the high-smearing starting point, we use a Marzari-Vanderbilt smearing~\cite{Marzari1999} of $\sigma = 20$~mRy in combination with uniform $12 \times 12$ $\vec k$- and $\vec q$-point meshes (including $\Gamma$), sufficient for this smearing.
Reference low-temperature data are generated using a Fermi-Dirac smearing of $T = 1.9$~mRy (300~K) and $48 \times 48$ $\vec k$~points.
Minimizing forces to below 1~\textmu Ry/bohr yields a layer thickness (sulfur-sulfur distance) of $d = 3.13$~\AA, which is recomputed for each considered smearing and $\vec k$~mesh but does not change significantly.
For the Fourier interpolation, we use one-shot Wannier functions obtained from projections onto atomic orbitals (cf.\@ Fig.~\ref{fig:subspaces}) to ensure perfect symmetry, with the exception of the 22-bands case where we use maximally localized Wannier functions (MLWF)~\cite{Marzari2012} with (i)~a Ta-$2s$ orbital, (ii)~a $1s$~orbital vertically centered between two S atoms, and (iii)~another three $1s$~orbitals halfway between (i) and (ii) as initial projections for the additional five conduction bands.

\subsection{From screened to bare phonons}
\label{sec:phonons}

The Kohn-Sham band structure of monolayer TaS\s2 from \ac{DFT} is shown in Fig.~\ref{fig:subspaces}\,(a--e).
At the Fermi level, there is a half-filled isolated band of Ta-$d_{z^2}$, -$d_{x^2 - y^2}$, and -$d_{x y}$ orbital character.
These orbitals also span the two lower empty bands, which partially overlap but do not hybridize with the two higher empty bands of Ta-$d_{x z}$ and -$d_{y z}$ character.
The occupied bands are formed by four isolated blocks of (in order of decreasing energy) six S-$p$, two S-$s$, three Ta-$p$, and one Ta-$s$ band.

To trace the transition from screened to bare phonons and interactions, we perform \ac{cDFPT} calculations for active subspaces of different size, including zero (\ac{DFPT}), one, five, 13, 17, and 22 bands.
Note that even if all bands explicitly calculated in \ac{DFT} were considered active, the infinite number of empty bands accounted for via the Sternheimer approach as well as the core bands hidden into the pseudopotential would still contribute to the screening (cf.\@ Appendix~\ref{app:bare}).

The corresponding Fourier-interpolated phonon dispersions are shown in Fig.~\ref{fig:subspaces}\,(f--j).
The \ac{DFPT} phonon dispersion, obtained from $D(\sigma, 0)$ via Eq.~\eqref{eq:eigen}, is reproduced as a reference in each panel using solid orange lines.
It features a softening of the longitudinal-acoustic branch, most pronounced at $\vec q = 2/3\,\mathrm M$, signaling the tendency toward the experimentally observed $3 \times 3$ \ac{CDW}~\cite{Tidman1974}.
Interestingly, at the high Marzari-Vanderbilt smearing of $\sigma = 20$~mRy, the system is dynamically stable with no imaginary frequencies.

The partially screened phonons, obtained from $D \super p (\sigma, 0)$, corresponding to the different choices of active bands, are shown using solid mauve lines.
Excluding electronic screening from within the isolated band only [Fig.~\ref{fig:subspaces}\,(f)] already removes all $\vec q$-dependent softening of the longitudinal branch, which is now highest in energy among the acoustic branches.
For five active bands [Fig.~\ref{fig:subspaces}\,(g)], the situation is similar, except that the acoustic sum rule is no longer fulfilled because we freeze long-wavelength dipole-allowed transitions~\cite{vanLoon2021a}, and the acoustic phonons acquire a finite energy at $\Gamma$ (cf.\@ Appendix~\ref{app:bare}).
This effect is even more pronounced for 13 [Fig.~\ref{fig:subspaces}\,(h)], 17 [Fig.~\ref{fig:subspaces}\,(i)], and 22 [Fig.~\ref{fig:subspaces}\,(j)] active bands, where the partially screened dispersions have shifted to much higher energies, the originally acoustic phonons reaching about 175~meV in the latter case.
Note that some of the branches feature a finite slope near $\Gamma$, which is due to the lack of metallic screening and corresponds to long-range interactions in real space.
We used the electrostatic model introduced in Sec.~\ref{sec:froehlich} to properly interpolate these phonons, the details of which are found in the next section.

If we renormalize the partially screened phonons evaluating Eq.~\eqref{eq:dp0} at $\omega = 0$ using exactly the same smearing $T = \sigma$ and \ac{BZ} sampling as in the \emph{ab initio} calculation, we obtain the dashed mauve lines in Fig.~\ref{fig:subspaces}\,(f--j).
They coincide with the \ac{DFPT} result for all sizes of the active subspace, showing that the phonon self-energy with one partially screened and one screened vertex is exact as long as all involved quantities are exact too.
Note that all quantities entering Eq.~\eqref{eq:dp0} have been directly computed using \ac{DFPT} and \ac{cDFPT} on the coarse grid and only the resulting $D \super{p0} (\sigma, 0)$ has been Fourier interpolated along the high-symmetry lines shown in Fig.~\ref{fig:subspaces}\,(f--j).

Finally, in Fig.~\ref{fig:subspaces}\,(f--j) we also show the unscreened phonons from $D \super u (\sigma, 0)$ according to Eq.~\eqref{eq:du}, again calculated on the original coarse mesh and interpolated only in the end, using dashed orange lines.
Since $g(\sigma, 0)$ is smaller than $g \super p (\sigma, 0)$, $\Pi \super{00} (\sigma, 0)$ is smaller than $\Pi \super{p0} (\sigma, 0)$.
As a consequence, also the unscreened phonon frequencies are lower than (or at most equal to) the \ac{cDFPT} ones; some $\vec q$-dependent softening is still present albeit hardly discernible because of the large smearing chosen.
A practical advantage is the absence of long-range terms~\cite{Pickett1976, Allen1980}; the slope of all branches vanishes toward $\Gamma$.

\subsection{From screened to bare electron-phonon coupling}
\label{sec:coupling}

\begin{figure*}
    \includegraphics{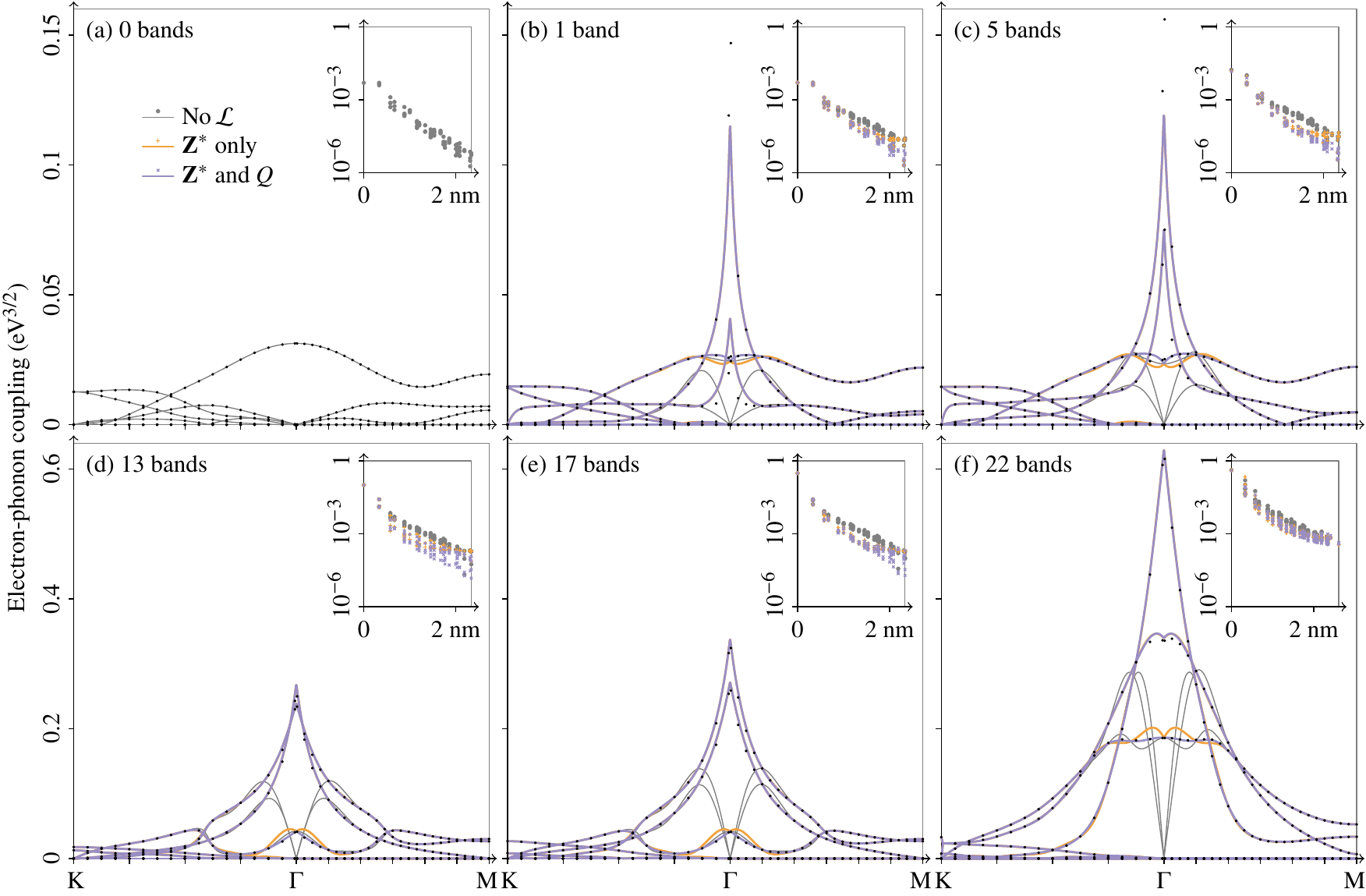}
    \caption{Electron-phonon coupling of monolayer TaS\s2 from (a)~DFPT and (b--f)~cDFPT for different sizes of the active subspace at a Marzari-Vanderbilt smearing of $\sigma = 20$~mRy.
    We show the absolute value of the coupling to the isolated low-energy electronic band as a function of $\vec q$ with $\vec k = 0$ for all phononic eigenmodes and using different approaches to handle the long-range part.
    The black dots indicate direct \ac{DFPT} and \ac{cDFPT} results and serve as a reference.
    The insets show the decay of the maximum absolute value of the short-range part ($\max \smash{\abs{g_{\vec R}^{\mathcal S}}}$) with the length of the lattice vector $\abs{\vec R}$ [cf.\@ Eq.~\eqref{eq:epw}].}
    \label{fig:coupling}
\end{figure*}

Now we will discuss the screened and partially screened electron-phonon coupling $g(\sigma, 0)$ and $g \super p (\sigma, 0)$ corresponding to the different active subspaces.
In Fig.~\ref{fig:coupling}, the absolute value of the interpolated coupling with the isolated electronic band at the Fermi level is shown as a function of $\vec q$ with $\vec k = 0$ for all nine phononic eigenmodes.
Reference data from direct \ac{DFPT} and \ac{cDFPT} calculations are shown using black dots.
Tick marks have been placed at those $\vec q$~points that are part of the $12 \times 12$ mesh on which the interpolation is based.
The interpolated quantities are guaranteed to match the reference data at these original $\vec q$~points by construction.

In the case of the screened $g(\sigma, 0)$ from \ac{DFPT} shown in Fig.~\ref{fig:coupling}\,(a), the interpolated coupling matches the reference coupling everywhere.
Here the system is metallic such that the coupling is continuous and smooth at the zone center, and due to the high electronic smearing there are no sharp features away from $\Gamma$ either.
Thus, the coupling is short-ranged (inset).

However, as soon as low-energy screening is excluded, peaks at $\Gamma$ emerge whose magnitude increases with the number of active bands, and the overall magnitude of the coupling increases too, as seen in Fig.~\ref{fig:coupling}\,(b--f).
A naive Fourier interpolation of these data yields the gray curves, which by definition match the reference at all original $\vec q$~points but are wrong in the vicinity of $\Gamma$.
This is because the peaks belong to the long-range part, which has to be subtracted before interpolation and added back afterward, as detailed in Sec.~\ref{sec:froehlich}.
The insets show the decay in real space with (gray) and without (mauve and orange) long-range part.

Using the equations in Sec.~\ref{sec:froehlich} with the dielectric constant $\epsilon$ and Born effective charges $\vec Z^*$ as obtained from the \ac{cDFPT} calculation but neglecting the term in Eq.~\eqref{eq:lr_vector} that involves the quadrupole tensors $Q$, which cannot be calculated from perturbation theory with \textsc{Quantum ESPRESSO} at present, we obtain the orange lines.
Most of the reference points are reproduced, but there are still deviations for the two modes that peak when approaching $\Gamma$ in Fig.~\ref{fig:coupling}\,(b,\,c) as well as for the third mode with appreciable coupling for small $\vec q$ in Fig.~\ref{fig:coupling}\,(b--f).
The former discrepancy is likely due to inaccuracies in $\epsilon$ or $\vec Z^*$.
The latter however occurs for a phonon mode where the atoms move in the out-of-plane direction and can be traced back to the missing term with $Q_{\text S z}$ in Eq.~\eqref{eq:lr_vector}.

\begin{table}
    \caption{Parameters used in the long-range terms of the phonons shown in Fig.~\ref{fig:subspaces}\,(f--j) and the electron-phonon coupling in Fig.~\ref{fig:coupling}\,(b--f).
    The in-plane dielectric constants $\epsilon$ and Born effective charges $\vec Z^*$~($e$) stem from \ac{cDFPT} calculations.
    The independent elements of the quadrupole tensors $Q$~($e$\,bohr) have been optimized by fitting the interpolants to reference \ac{cDFPT} data.
    The out-of-plane elements $Q_{\text S \alpha \alpha z}$ do not contribute.
    The other elements are either zero or follow from $Q_{\kappa x x y} = Q_{\kappa x y x} = Q_{\kappa y x x} = -Q_{\kappa y y y}$ and $Q_{\kappa z x x} = Q_{\kappa z y y}$, with $Q_{\text S' \alpha} = (1 - 2 \delta_{\alpha z}) Q_{\text S \alpha}$, where $\text S'$ denotes the other S~atom.
    The range-separation parameters $L$ (bohr) have been optimized at $Q = 0$ for simplicity.
    See Appendix~\ref{app:bare} for more information about the bare values (``All''), where the last line is for different pseudopotentials with \ac{SC} states.
    We report values for neighboring Ta and S~atoms with $\bm \uptau \sub S - \bm \uptau \sub{Ta} = (0, a / \sqrt 3, -d / 2)$.
    Note that the optimized values of the bare $Q_{\text S z y y}$ are remarkably close to $Z^* \sub S d = 35.44~e\,\text{bohr}$ and $Z^* \sub{S,\ac{SC}} d = 82.69~e\,\text{bohr}$.}
    \label{tab:lr}
    \medskip
    \setlength\tabcolsep{4.2pt}
    \begin{tabular}{*8r}
        Bands
        & $\epsilon$
        & $Z^* \sub{Ta}$
        & $Z^* \sub S$
        & $Q_{\text Ta y y y}$
        & $Q_{\text S y y y}$
        & $Q_{\text S z y y}$
        & $L$
        \\[2pt]
        1   & $3.93$ & $ 2.13$ & $-0.53$ & $6.20$ & $ 1.25$ & $ 4.39$ & $6.8$ \\
        5   & $3.30$ & $ 2.84$ & $-0.43$ & $6.21$ & $ 1.82$ & $ 5.48$ & $6.4$ \\
        13  & $1.62$ & $ 3.97$ & $ 2.07$ & $3.84$ & $-2.64$ & $15.01$ & $5.4$ \\
        17  & $1.61$ & $ 6.75$ & $ 2.06$ & $6.22$ & $-2.59$ & $14.80$ & $5.3$ \\
        22  & $1.13$ & $ 8.74$ & $ 3.96$ & $9.71$ & $-1.01$ & $19.73$ & $5.5$ \\
        All & $1.00$ & $13.00$ & $ 6.00$ & $0.23$ & $ 0.38$ & $35.06$ & $3.8$ \\
\ac{SC} all & $1.00$ & $27.00$ & $14.00$ & $7.31$ & $ 0.23$ & $82.08$ & $3.8$
    \end{tabular}
\end{table}
The quadrupole tensors $Q$ could be calculated \emph{ab initio} with \textsc{Abinit}~\cite{Gonze2016, Gonze2020}, albeit only within the local-density approximation and \ac{PBE} exchange-correlation functionals and without nonlinear core correction~\cite{Royo2019}.
We instead choose an approach similar to the one in Ref.~\citenum{Ponce2021} and fit the quadrupole tensors minimizing the error in the interpolated phonons and coupling for all reference $\vec q$~points marked with black dots in Fig.~\ref{fig:coupling}\,(b--f)~%
\footnote{We simultaneously minimize the mean squared error of the (complex) dynamical matrix and of the shown coupling (absolute value), preserving the symmetries of the quadrupole tensors imposed by the point groups of the associated atoms.}.
The resulting contributing elements of $Q$ together with those of $\epsilon$ and $\vec Z^*$ from \ac{cDFPT} and the optimal $L$ are listed in Table~\ref{tab:lr}.
As shown using mauve lines in Fig.~\ref{fig:coupling}\,(b--f), the coupling to the out-of-plane mode is correctly interpolated when the quadrupole term is taken into account.

\subsection{Comparison of approaches}
\label{sec:renorm}

\begin{figure*}
    \includegraphics{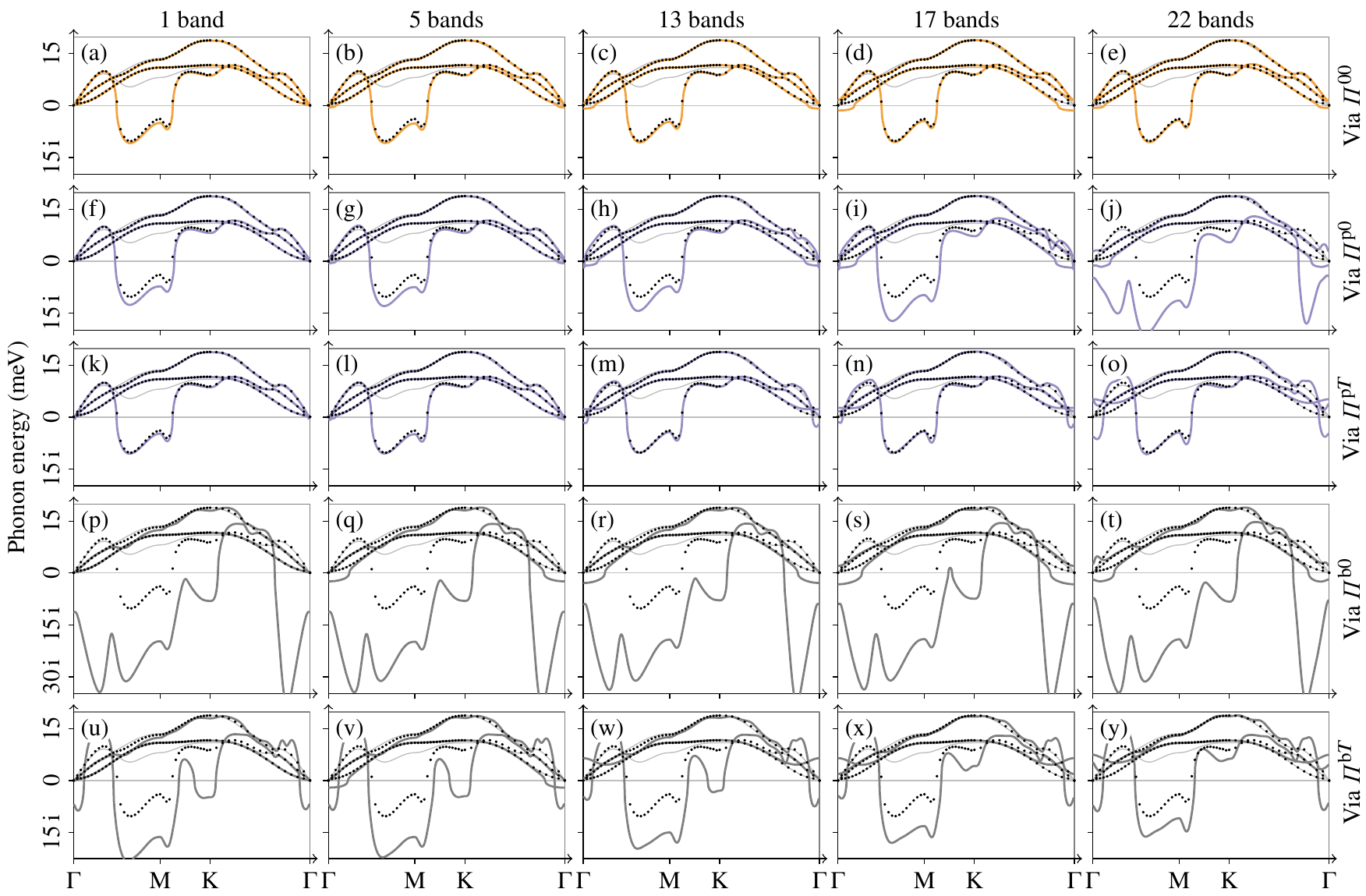}
    \caption{Renormalized acoustic phonon dispersion of monolayer TaS\s2 for a Fermi-Dirac smearing of $T = 1.9$~mRy based on \emph{ab initio} calculations performed at a Marzari-Vanderbilt smearing of $\sigma = 20$~mRy for different sizes of the active subspace according to (a--e)~Eq.~\eqref{eq:d00} (``screened-screened''), (f--j)~Eq.~\eqref{eq:dp0} and (k--o)~Eq.~\eqref{eq:dpt} (``partially screened-screened''), and (p--t)~Eq.~\eqref{eq:db0} and (u--y)~Eq.~\eqref{eq:dbt} (``bare-screened'').
    The thin gray lines and black dots are the same in all panels and indicate converged direct \ac{DFPT} results for smearings $\sigma$ (starting point) and $T$ (reference), respectively.
    Panel~(f) corresponds to the downfolding approach with optimal active subspace.}
    \label{fig:renorm}
\end{figure*}

After the analysis of the screened and partially screened phonons and interactions calculated for the high Marzari-Vanderbilt smearing of $\sigma = 20$~mRy and a coarse momentum grid, we will now use Eq.~\eqref{eq:d00} (with two screened vertices), Eqs.~\eqref{eq:dp0} and \eqref{eq:dpt} (with one partially screened vertex), as well as Eqs.~\eqref{eq:db0} and \eqref{eq:dbt} (with one bare vertex) to estimate the screened phonons for a low electronic temperature (Fermi-Dirac smearing) of $T = 1.9$~mRy and dense momentum grids.
On this basis, we will compare the different approaches with special focus on the influence of the number of active bands.
The results are shown in Fig.~\ref{fig:renorm} together with converged reference points from direct \ac{DFPT} calculations.
At this electronic temperature, the system is dynamically unstable, as indicated by imaginary frequencies, and exhibits relatively sharp Kohn anomalies.
The adiabatically renormalized phonon dispersions have not been interpolated but calculated for each $\vec q$~point along the path separately, using a converged $96 \times 96$ $\vec k$~mesh.
Only the underlying quantities, namely the electronic energies $\varepsilon$, the screened dynamical matrix $D(\sigma, 0)$, and the screened, partially screened, and bare electron-phonon coupling $g(\sigma, 0)$, $g \super p (\sigma, 0)$, and $g \super b$, have been interpolated.
Alternatively, the interpolation could be performed at the level of the unscreened $D \super u (\sigma, 0)$ and partially screened $D \super p (\sigma, 0)$ instead of the screened $D(\sigma, 0)$, but this degrades the results because errors accumulate; see Appendix~\ref{app:unscreen}.
In all self-energy calculations, the chemical potential has been adjusted to the respective smearing function $f$ and electronic temperature $\sigma$ or $T$.

Figure~\ref{fig:renorm}\,(a--e) displays the renormalized phonons from $D \super{00} (T, 0)$ according to Eq.~\eqref{eq:d00}, which is equivalent to the approach suggested in Ref.~\citenum{Calandra2010}.
Most reference points are well reproduced---even the shape of the soft mode, which is remarkable since it is completely absent in the high-smearing data used as the starting point, shown using thin gray lines.
The agreement is even better for larger active subspaces, for which the approximations leading to Eq.~\eqref{eq:approx} are less severe since the $T$~dependence of $U$ and $g \super p$ is strongly suppressed.
We note that here the active subspace merely defines the number of bands summed over since no downfolding to partially screened quantities is involved in this approach.
Only near $\Gamma$, the acoustic sum rule is slightly broken, resulting in unphysical finite energies of the acoustic modes.
This might be related to small changes in the atomic positions upon cooling the system, which are not captured by the discussed methods.

The corresponding results from $D \super{p0} (T, 0)$ according to Eq.~\eqref{eq:dp0} are shown in Fig.~\ref{fig:renorm}\,(f--j), where Fig.~\ref{fig:renorm}\,(f) represents the proposed use of an optimal active subspace and Fig.~\ref{fig:renorm}\,(j) is the closest we get to using a bare vertex (cf.\@ Appendix~\ref{app:bare}).
The \ac{cDFPT}-based approach throughout overestimates the phonon softening, more severely the larger the active subspace.
This may be surprising since the diagrammatically correct combination of partially screened and screened vertices is used.
However, opposed to the partially screened coupling $g \super p (T, 0) \approx g \super p (\sigma, 0)$, the correct screened coupling $g(T, 0) \not\approx g(\sigma, 0)$ depends significantly on $T$, a fact that is not properly accounted for.
When using the larger $g(\sigma, 0)$ in place of $g(T, 0)$ in $\Pi \super{p0} (T, 0)$, we underestimate the screening of the coupling and thus overestimate the screening of the phonons.
In the approach of Ref.~\citenum{Calandra2010} in turn, the error in the phonon self-energy correction only enters at second order in the error of the coupling, as quantified in Eq.~\eqref{eq:quadratic}.

To prove that the overscreening seen in the \ac{cDFPT}-based approach is indeed due to the failure to adjust the screened coupling to the target temperature, we also calculate the phonons from $D \super{p$T$} (T, 0)$ according to Eq.~\eqref{eq:dpt}.
The only difference to the previous approach is that we take the correctly screened coupling $g(T, 0)$ from the low-temperature reference calculation~%
\footnote{The \ac{DFPT} calculation has been done using $48 \times 48$ $\vec k$~points, but $g(T, 0)$ is only calculated on the coarser $12 \times 12$ $\vec k$- and $\vec q$-point meshes for subsequent Fourier interpolation.}.
Since the direct \ac{DFPT} calculation of $g(T, 0)$ is computationally as expensive as the direct calculation of the $D(T, 0)$ we are interested in---indeed they are calculated at the same time---this approach has no practical utility beyond this proof of concept.
As expected, in Fig.~\ref{fig:renorm}\,(k--o), the reference points are largely reproduced now, showing that it is in principle possible to obtain accurate results based on partially screened quantities.
``However\rlap,'' as already stated by Calandra \emph{et al.}, ``such a procedure requires an accurate self-consistent determination of the screened potential''~\cite{Calandra2010}.

The growing overscreening with the number of active bands in Fig.~\ref{fig:renorm}\,(f--j) presages large errors in the limit of using a bare vertex, when the screened vertex is not corrected.
This is confirmed by the phonon dispersions from $D \super{b0} (T, 0)$ according to Eq.~\eqref{eq:db0} in Fig.~\ref{fig:renorm}\,(p--t), which exhibit deviations similar to the ones already seen in Fig.~\ref{fig:renorm}\,(j), yet much more pronounced since the bare exceeds the partially screened coupling [cf.\@ Eq.~\eqref{eq:linear}].
Just like the results from $D \super{00} (T, 0)$ in Fig.~\ref{fig:renorm}\,(a--e), $D \super{b0} (T, 0)$ converges very fast with the number of bands summed over, as the vertices are fixed and the temperature dependence of the bare susceptibility stems largely from the single band at the Fermi level.
Note that the bare vertex (unlike the partially screened ones) and derived quantities depend on the pseudopotential.
The influence of \ac{SC} states is discussed in Appendix~\ref{app:bare}.

Finally, we also repeat the calculation with the bare vertex using the temperature-adjusted screened vertex.
This approach is in principle exact, at least in the limit of an infinite number of bands.
Indeed, $D \super{b$T$}$ according to Eq.~\eqref{eq:dbt} yields phonons with an overscreening error, which however decreases slowly with the number of bands summed over; see Fig.~\ref{fig:renorm}\,(u--y).
In practice, a partially screened vertex that matches the number of bands promises to be a good alternative to the bare vertex that is incompatible with the concept of a finite active subspace.

Taken together, it is clear that the method of Ref.~\citenum{Calandra2010} is the easiest to use and best performing one in this context.
However, we would like to argue in favor of using a partially screened vertex for the optimal subspace, see Fig.~\ref{fig:renorm}\,(f), for two reasons:
(i)~The Friedel long-rangedness is exactly removed, guaranteeing a smooth partially screened phonon dispersion as in Fig.~\ref{fig:subspaces}\,(f), and (ii)~the result can be systematically improved as shown in the following.

\subsection{Correction of the screened vertex}
\label{sec:corr}

To overcome the problem with the \ac{cDFPT}-based approach, we need to have precise control of the screened vertex and solve Eq.~\eqref{eq:gp2g}.
However, to our knowledge, it is currently not possible to calculate the necessary partially screened electron-electron interaction $U$ as a function of all three momenta and four electronic band indices and consistent with existing \ac{cDFPT} implementations.
Even though eventually there will be no way around this, in this section we present two alternative correction methods that approximate or circumvent the calculation of $U$ at no significant additional computational cost.

First, we can make the simplistic assumption that the dependence on electronic degrees of freedom can be neglected or averaged out.
Then we can approximately solve Eq.~\eqref{eq:gp2g} for
\begin{equation}
    U_{\vec q}
    \approx \frac{\displaystyle
        \sum_{\kappa \alpha \vec k m n}
        \abs{g_{\vec q \kappa \alpha \vec k m n} \ps (\sigma, 0)
        {-} g_{\vec q \kappa \alpha \vec k m n} \super p (\sigma, 0)}
    }{\displaystyle
        \sum_{\kappa \alpha \vec k m n}
        \abs{g_{\vec q \kappa \alpha \vec k m n} \ps (\sigma, 0)
        \chi_{\vec q \vec k m n} \super{b,A} (\sigma, 0)}
    }.
\end{equation}
The corrected screened electron-phonon coupling follows as
\begin{equation}
    \label{eq:corr1}
    g_{\vec q \kappa \alpha \vec k m n} \super{corr~I} (T, 0)
    \approx g_{\vec q \kappa \alpha \vec k m n} (\sigma, 0)
    \frac{\epsilon_{\vec q}(\sigma)} {\epsilon_{\vec q}(T)},
\end{equation}
with the $\vec q$-dependent and otherwise scalar dielectric function
\begin{equation}
    \epsilon_{\vec q}(T)
    \approx 1 - U_{\vec q}
    \sum_{\vec k m n}
    \chi_{\vec q \vec k m n} \super{b,A} (T, 0).
\end{equation}

Second, we make the ansatz that the change in the electron-phonon coupling for the electronic degrees of freedom is linear,
\begin{multline}
    \label{eq:corr2}
    g_{\vec q \kappa \alpha \vec k m n} \super{corr~II} (T, 0)
    = g_{\vec q \kappa \alpha \vec k m n} \super p (\sigma, 0)
    \\
    + x_{\vec q \kappa \alpha} \ps (T)
    [g_{\vec q \kappa \alpha \vec k m n} \ps (\sigma, 0)
    - g_{\vec q \kappa \alpha \vec k m n} \super p (\sigma, 0)],
\end{multline}
with an unknown $x$ that has to be determined for each phonon displacement separately.
Further, again assuming that the smearing dependence of $U$ and $g \super p$ is weak [cf.\@ Eq.~\eqref{eq:approx}] and can be neglected, Eq.~\eqref{eq:gp2g} for $\sigma$ and $T$ can be written as
\begin{align}
    \label{eq:gp2g_sigma}
    g(\sigma)
    &= g \super p
    + U \chi \super{b,A} (\sigma) g(\sigma),
    \\
    \label{eq:gp2g_t}
    g(T)
    &= g \super p
    + U \chi \super{b,A} (T) g(T),
\end{align}
where we have left all subscripts, summations, and prefactors understood for brevity and the only unknown is $U$.
Using the ansatz from Eq.~\eqref{eq:corr2} and inserting Eq.~\eqref{eq:gp2g_sigma} into Eq.~\eqref{eq:gp2g_t}, we obtain
\begin{multline}
    U \bigl\{
        x(T) \chi \super{b,A} (\sigma) g(\sigma)
        - x(T) \chi \super{b,A} (T) [g(\sigma) - g \super p]
        \\
        - \chi \super{b,A} (T) g \super p
    \bigr\} = 0,
\end{multline}
where $U$ and $\chi \super{b,A} g \super{(p)}$ are matrices and vectors in the electronic degrees of freedom, respectively.
We equate the expression in curly braces with zero and approximately solve for
\begin{equation}
    x_{\vec q \kappa \alpha}(T)
    = \frac{\displaystyle
        \sum_{\vec k m n}
        \abs{
            \chi \super{b,A} (T) g \super p
        }_{\vec q \kappa \alpha \vec k m n}
    }{\displaystyle
        \sum_{\vec k m n}
        \abs{
            \chi \super{b,A} (\sigma) g(\sigma)
            {-} \chi \super{b,A} (T) [g(\sigma) {-} g \super p]
        }_{\vec q \kappa \alpha \vec k m n}
    }.
\end{equation}

\begin{figure}
    \includegraphics{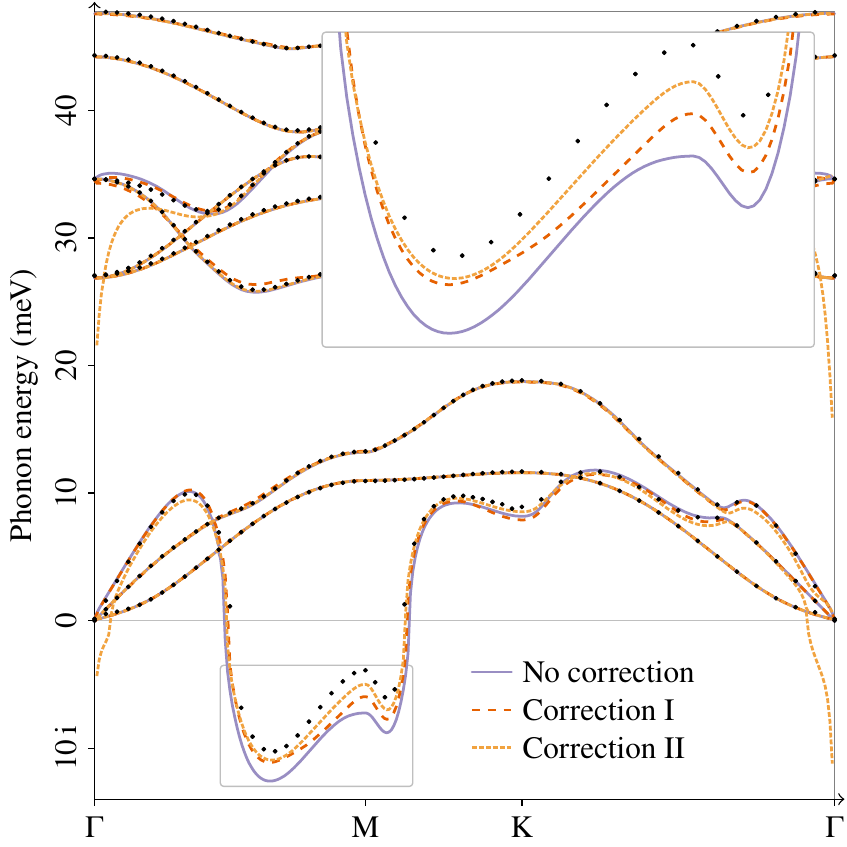}
    \caption{Comparison of correction schemes for the screened electron-phonon vertex.
    We show the same data as in Fig.~\ref{fig:renorm}\,(f), supplemented with corrected results according to Eqs.~\eqref{eq:corr1} and \eqref{eq:corr2}.
    The inset is a close-up of the framed region including the leading soft modes.}
    \label{fig:corr}
\end{figure}

In Fig.~\ref{fig:corr}, we compare the renormalized phonons according to Eq.~\eqref{eq:dp0} with and without the correction of the screened vertex from Eqs.~\eqref{eq:corr1} and \eqref{eq:corr2} for the \emph{optimal} case of a single active band.
Here we stress that the discussed corrections are not suitable for much larger active subspaces because their number of free parameters does not increase with the number of bands.
We find that the simple correction from Eq.~\eqref{eq:corr1} (red dashed lines) reduces the deviation from the reference data by about half on average.
Still, the quality of the correction depends rather strongly on $\vec q$, being more effective at the leading instability near $2/3\,\mathrm M$ than at $\mathrm M$.
In turn, the correction from Eq.~\eqref{eq:corr2} (orange dashed lines), which also takes into account changes in the $\vec k$~dependence of the coupling, yields almost the same accuracy as the method of Eq.~\eqref{eq:d00} with two screened vertices.
However, this correction fails for small momenta since the ansatz of a linear change of the coupling becomes unsuitable as soon as the long-range part of $g \super p$ dominates.

To summarize, it is to some extent possible to correct the screening of the electron-phonon coupling for changes in the electronic temperature, even without performing a full \emph{ab initio} calculation of the electron-electron interaction.
These corrections are, however, not universally applicable and limit the possibilities for systematic improvements; e.g., there is no obvious way to include an $\omega$~dependence in Eq.~\eqref{eq:corr2}.
Finally, it is important to bear in mind that the error bars from the approximations made in \ac{DFT} are likely as large as, if not larger than, the discussed deviations from the converged \ac{DFPT} calculation.

\subsection{Spectral function}
\label{sec:specfun}

Having convinced ourselves that the approach with two screened vertices and the one with one partially screened vertex yield excellent adiabatic results, we will now turn to the nonadiabatic case, $\omega \neq 0$.
According to the original Ref.~\citenum{Calandra2010}, the former approach should provide error cancellation also with respect to the frequency dependence.
With this work, we believe to have settled the debate on which method to use in the static case for practical calculation and shown that the approach with the bare vertex should not be used.
However, we emphasize that the dynamical case~\cite{Stefanucci2023, Marini2023b} is still an open question for which the community has no reference calculation to compare with, beyond experimental data and model results as discussed in Sec.~\ref{sec:models}.
In particular, it has been claimed that the classical phonon concept breaks down in the nonadiabatic case, and that theories at the level of time-dependent \ac{DFT} thus cannot have access to phonon linewidths at all~\cite{Marini2023b}.

Since the computational time scales approximately quadratically with the number of bands, and because the results in Fig.~\ref{fig:renorm}\,(a,\,f) are adequate, we will work with a single active band.
The quantity of interest is the phonon spectral function as defined in Eq.~\eqref{eq:specfun}.
In practice, the imaginary infinitesimal $0^+$ is approximated by two different finite smearing parameters~\cite{Monacelli2021}:
\begin{equation}
    \label{eq:double_smearing}
    G_{\vec q}(T, \omega + \I 0^+)
    \approx \frac
        {\mathds 1}
        {(\omega + \I \delta)^2 \mathds 1 - D_{\vec q}(T, \omega + \I \eta)}.
\end{equation}
While $\eta$ only affects phonon branches with nonzero electron-phonon coupling, $\delta$ broadens all branches equally.
The former must be large enough to ensure that the spectral function is converged with respect to the chosen $\vec k$~mesh, but small enough to avoid artificial frequency shifts.
The latter aids the graphical representation, since it prevents infinitely sharp delta peaks.

\begin{figure*}
    \includegraphics{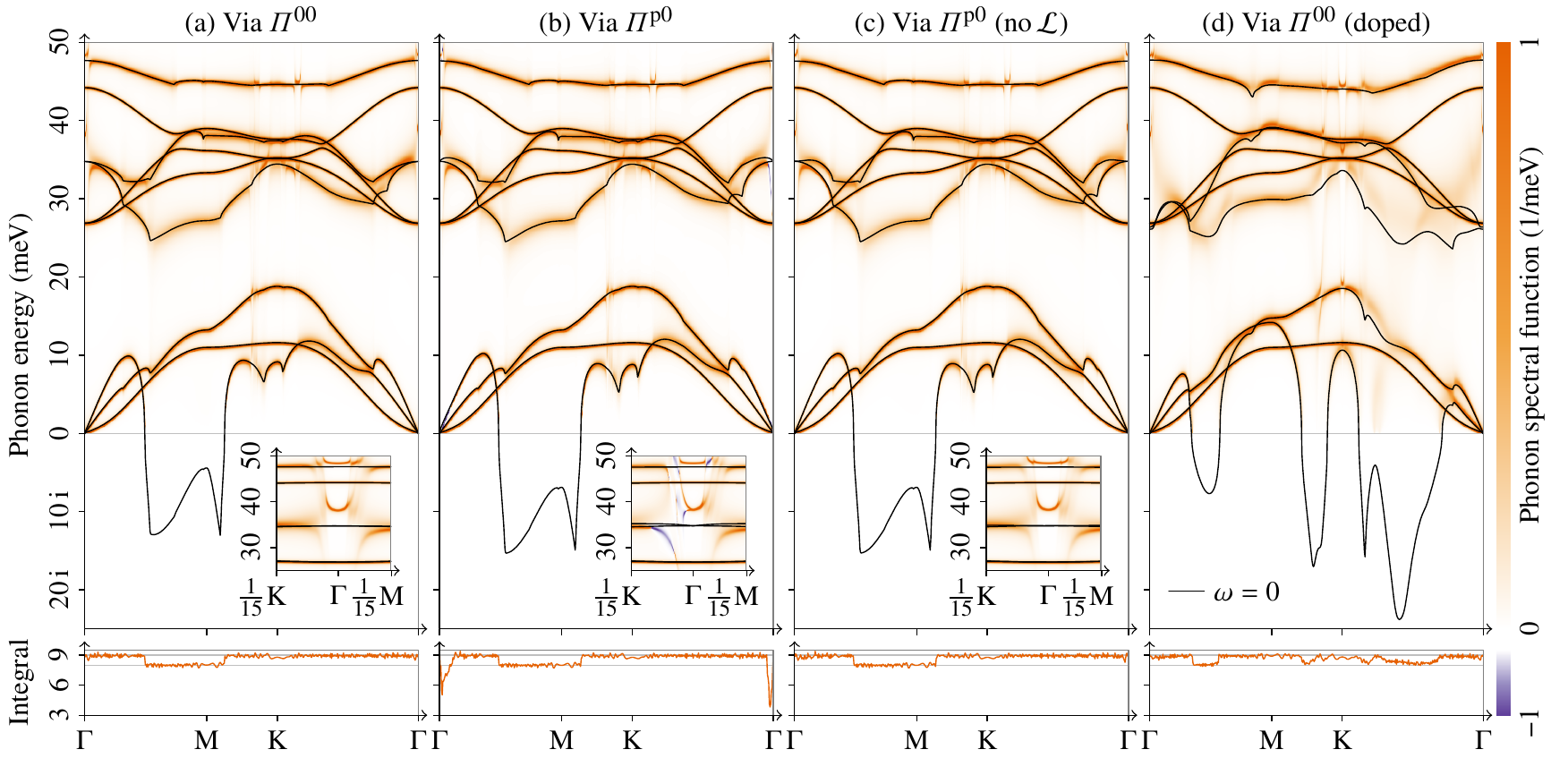}
    \caption{Phonon spectral function of monolayer TaS\s2 together with adiabatic phonon dispersion at an electronic temperature of $T = 1$~meV for (a--c)~the undoped case and (d)~the hole-doped case, where the Van Hove singularity is at the Fermi level.
    The spectral function is evaluated following Eq.~\eqref{eq:double_smearing} starting from a nonadiabatic dynamical matrix according to (a,\,d)~Eq.~\eqref{eq:d00} and (b,\,c)~Eq.~\eqref{eq:dp0}, where the long-range part is properly handled only in (b).
    The insets contain close-ups of the long-wavelength and high-frequency region.
    The bottom panels show the frequency integral of the spectral function from 0 to 50~meV; horizontal lines are drawn at values of 8 and 9 for reference.}
    \label{fig:specfun}
\end{figure*}

The results are shown in Fig.~\ref{fig:specfun}, together with the corresponding adiabatic $\omega = 0$ result for comparison.
We used a Fermi-Dirac smearing of $T = 1$~meV in combination with $\eta = 2$~meV, $\delta = 0.05$~meV, and $2000 \times 2000$ $\vec k$~points.
Note that the resulting sharp features cannot be expected in a self-consistent theory~\cite{Stefanucci2023}, where the screening of electrons and phonons is reciprocal, or in the presence of electronic correlations, due to the resulting breakdown of the single-electron picture and associated broadening.

The phonon spectral functions in Fig.~\ref{fig:specfun}\,(a--c) have been calculated for pristine TaS\s2 without electron doping.
In Fig.~\ref{fig:specfun}\,(a) the dynamical matrix $D(T, \omega)$ is approximated by $D \super{00} (T, \omega)$ according to Eq.~\eqref{eq:d00}, and in Fig.~\ref{fig:specfun}\,(b,\,c) by $D \super{p0} (T, \omega)$ according to Eq.~\eqref{eq:dp0}, with and without handling of the long-range terms as in Fig.~\ref{fig:coupling}\,(b).
Overall, the results from the three approaches are very similar.
One prominent feature is the discontinuity of long-wavelength optical modes, which separates regions of nonadiabatic phonon hardening and significant broadening on the side of smaller and larger $\vec q$, respectively.
This can be explained by which intraband electron-hole excitations---we only deal with a single band---are allowed:
The range of possible excitation energies is zero for $\vec q = 0$ and fans out into a continuum with increasing $\vec q$.
As soon as $\omega$ falls below the maximum excitation energy, the denominator of the bare susceptibility in Eq.~\eqref{eq:susc} can become arbitrarily small.
Resemblant features can also be observed in the vicinity of the $\mathrm K$~point.
Traces of these discontinuities extend vertically across the whole frequency range, similar to previous results on n-doped monolayer MoS\s2~\cite{GarciaGoiricelaya2020}.
Beyond that, we find an overall broadening of the branches that couple to the low-energy electronic band, which also have a nonzero renormalization in the adiabatic case [cf.\@ Fig.~\ref{fig:subspaces}\,(f)].

However, there is one important difference between the approaches.
The phonon spectral function from $D \super{00} (T, \omega)$ in Fig.~\ref{fig:specfun}\,(a) is strictly positive by construction and in good approximation fulfills the sum rule requiring that its frequency integral amounts to the number of phonon modes (except where the lowest mode becomes unstable and thus falls below the integration range).
In contrast, the result from $D \super{p0} (T, \omega)$ in Fig.~\ref{fig:specfun}\,(b) exhibits some negative spectral weight in the vicinity of the $\Gamma$ point, shown in mauve and more visible in the inset.
This is unphysical, breaks the sum rule, and is a direct consequence of the approximations involved.
Note that none of the affected $\vec q$~points are part of the original \emph{ab initio} mesh and interpolation errors are involved.
Interestingly, this problem is absent when the long-range part of the partially screened dynamical matrix and coupling $g \super p$ is not handled as described in Sec.~\ref{sec:froehlich} and a naive Fourier interpolation is used instead; see Fig.~\ref{fig:specfun}\,(c).
This does not imply that $g \super p$ with long-range handling is less accurate, the opposite is true, but unlike the naively interpolated one it is very different from the screened $g$ for small $\vec q$.
More precisely, there is a significant orthogonal component as a consequence of which the symmetrized outer product $\bar g \super p g$ is no longer positive-definite as anticipated in Sec.~\ref{sec:sym}.
This effect is quantified in Appendix~\ref{app:geff}.
In combination with the fact that $g \super p$ and thus the linear error in Eq.~\eqref{eq:linear} is larger when accounting for the long-range part, this leads to the unphysical features in Fig.~\ref{fig:specfun}\,(b).

An interesting question is how nonadiabatic renormalization influences the adiabatic Kohn anomalies, i.e., the stability of the system.
To investigate this, we resort to $D \super{00} (T, \omega)$ again.
In Fig.~\ref{fig:specfun}\,(a), no such effect is visible in the sense that the longitudinal-acoustic mode does not display any significant nonadiabatic renormalization close to the instability.
Indeed, processes in a large energy window of the order of 100~meV contribute to the softening~\cite{Berges2020a}, while nonadiabatic effects occur on an energy scale that is about one order of magnitude smaller.
This changes when hole doping moves the Van Hove singularity---located at the minimum of the low-energy band along $\Gamma$--$\mathrm K$ [cf.\@ Fig.~\ref{fig:subspaces}\,(a--e)], which is actually a saddle point---to the Fermi level, where it leads to a logarithmic divergence of the phonon self-energy~\cite{Berges2020a}.
If we realize this situation via a rigid shift by about 0.15~eV of the low-energy band for the calculation of $\Pi \super{00} (T, \omega)$---the unscreening via $\Pi \super{00} (\sigma, 0)$ is still done without doping---we obtain the result in Fig.~\ref{fig:specfun}\,(d).
Most effects seen in the undoped case are even more pronounced here, but we also observe that some spectral weight of the soft modes---now located at different $\vec q$~points, especially between $\Gamma$ and $\mathrm K$---remains in the positive energy range.
While this system is dynamically unstable both with and without nonadiabatic effects, there might be similar scenarios or materials where nonadiabatic damping of the \ac{CDW} occurs~\cite{Alidoosti2021, Alidoosti2022, Girotto2023}.

\section{Conclusions}
\label{sec:conclusions}

Comparing different approaches to calculate phonon dispersions at low electronic temperature with an affordable computational cost, we have explored the central findings of Ref.~\citenum{Calandra2010}.
First, correcting a static phonon self-energy with two screened electron-phonon vertices is an excellent approximation, which in particular allows us to work with a constant approximate coupling as obtained from usual \emph{ab initio} calculations.
Second, it is in principle possible to work with one bare vertex~\cite{Giustino2017, Reichardt2018, Paleari2021a, Paleari2021b, Marini2023a}, but this requires precise control of the static screened vertex and is not advantageous in practice.
We have given expressions for the associated errors, and we have shown that using one partially screened vertex is a viable alternative.

The static results suggest that the cancellation benefit of the approach with two screened vertices with respect to changes in the electronic temperature could be extended to frequency dependence, but this remains to be definitely proven, beyond our encouraging findings for simple models.
After all, changes in the adiabatic dynamical matrix are in many regards different from the complete phonon self-energy of out-of-equilibrium many-body theory~\cite{Marini2023a}.
In this context, the approach with a partially screened vertex could be useful as it allows us to incorporate the frequency dependence not only in the bare susceptibility but also in the active-subspace electron-phonon coupling in a controlled manner.
An important step in this direction would be the affordable and consistent computation of the partially screened electron-electron interaction with all relevant dependences, which occurs in the equations for the renormalization of the electron-phonon vertex [Eq.~\eqref{eq:gp2g}].

On top of this, the stationary functional of Ref.~\citenum{Calandra2010} [Eq.~\eqref{eq:stationary}], where all \ac{DFPT} vertices---both in the discussed first term and in the double-counting term---are replaced by properly screened vertices from \ac{RPA} or beyond, can still be useful to minimize the error in the phonon self-energy~\cite{Note1}.
The fact that the double-counting term is not accessible in \ac{DFPT} and can be successfully circumvented in the discussed correction method does not make it less relevant for the theory.
Without it, the phonon self-energy with two screened vertices will in general be too small, which is important when comparing absolute values rather than differences, as done here.

We have provided an easy to use implementation of consistent screened, partially screened, and bare phonons and electron-phonon interactions from \ac{DFPT} and \ac{cDFPT} in the \textsc{PHonon} and \textsc{EPW} codes of \textsc{Quantum ESPRESSO}.
Here, special care has been taken of dipolar and quadrupolar long-range terms present in the partially screened and bare quantities.

Finally, we remark that we are here limited to the harmonic approximation and that anharmonic effects can be important in materials close to a lattice instability, where the energy landscape is by definition anharmonic~\cite{Leroux2012}, or in systems such as superconducting hydrides, which have attracted a lot of attention recently.
It is likely that not only nonadiabatic but also anharmonic effects are often generated by the low-energy electronic system~\cite{Schobert2023}.
An interesting open question in this context is whether similar properties as the stationary functional can also be derived and taken advantage of for higher-order terms.

\begin{acknowledgments}
We thank Luca Binci, Matteo Calandra, Claus Falter, Jae-Mo Lihm, Andrea Marini, Francesco Mauri, Dino Novko, Fulvio Paleari, Junfeng Qiao, and Sven Reichardt for fruitful discussions, Ryotaro Arita and Yusuke Nomura for sharing their \ac{cDFPT} code, and Lucio Colombi Ciacchi for hosting the ``U Bremen Excellence Chair Program\rlap.''
J.B. and N.M. acknowledge support from the \ac{DFG} under Germany's Excellence Strategy (EXC~2077, No.~390741603, University Allowance, University of Bremen).
J.B. acknowledges computational resources of the \ac{HLRN}.
N.G. acknowledges support from the Croatian Science Foundation (UIP-2019-04-6869).
T.W. acknowledges support from the \ac{DFG} through QUAST (FOR~5249, No.~449872909) and via the Cluster of Excellence ``CUI: Advanced Imaging of Matter'' (EXC~2056, No.~390715994).
N.M. and S.P. acknowledge support from the NCCR MARVEL, a National Centre of Competence in Research, funded by the \ac{SNSF} (No.~205602).
S.P. acknowledges support from the F.R.S.-FNRS as well as from the European Union's Horizon 2020 Research and Innovation Programme, under the Marie Sk\l odowska-Curie Grant Agreement (SELPH2D, No.~839217), and computational resources awarded on the Belgian share of the EuroHPC LUMI supercomputer and by the PRACE-21 resources MareNostrum at BSC-CNS.
\end{acknowledgments}

\appendix

\section{Error of approach with two screened vertices}
\label{app:error}

Here we quantify the error associated with using the ``approximated force-constant functional'' of Ref.~\citenum{Calandra2010}, i.e., the error in the phonon self-energy resulting from an approximate electronic response $\chi \super b g$ in Eq.~\eqref{eq:stationary}.
In compact notation, Eq.~\eqref{eq:stationary} reads
\begin{align}
    \label{eq:exact}
    g \super b \chi \super b g
    &= \textcolor{mauve}{g \chi \super b g - g \chi \super b v \chi \super b g}
    \\
    \label{eq:stationary_compact}
    &= [g \super b + g \chi \super b v] \chi \super b [g \super b + v \chi \super b g] - g \chi \super b v \chi \super b g.
\end{align}
We now add an error $\delta$ to $g \chi \super b$ and $\chi \super b g$, which corresponds to an error $\Delta g = v \delta$ in the screened electron-phonon coupling $g$ [cf.\@ Eq.~\eqref{eq:gb2g}].
Equation~\eqref{eq:stationary_compact} then becomes
\begin{align}
    [g \super b &+ (g \chi \super b + \delta) v] \chi \super b [g \super b + v (\chi \super b g + \delta)]
    \notag
    \\
    &\qquad - (g \chi \super b + \delta) v (\chi \super b g + \delta)
    \notag
    \\
    &= [g + \delta v] \chi \super b [g + v \delta] - (g \chi \super b + \delta) v (\chi \super b g + \delta)
    \\
    &= \textcolor{mauve}{g \chi \super b g} \textcolor{orange}{{} + g \chi \super b v \delta + \delta v \chi \super b g} + \delta v \chi \super b v \delta
    \notag
    \\
    &\qquad\textcolor{mauve}{{} - g \chi \super b v \chi \super b g} \textcolor{orange}{{} - g \chi \super b v \delta - \delta v \chi \super b g} - \delta v \delta
    \\
    &= g \super b \chi \super b g + \delta (v \chi \super b v - v) \delta
    \\
    &= g \super b \chi \super b g + \delta v (\chi \super b - v \inv) v \delta
    \\
    &= g \super b \chi \super b g - \Delta g W \inv \Delta g,
\end{align}
where we have used Eqs.~\eqref{eq:gb2g} and \eqref{eq:exact} and defined the inverse screened electron-electron interaction $W \inv = v \inv \epsilon$ [cf.\@ Eq.~\eqref{eq:v2w}].
The error in the phonon self-energy is quadratic in $\Delta g$.

\section{Application to n-doped MoS\s2}
\label{app:mos2}

The same procedure applied previously to monolayer TaS\s2 is here utilized for another transition-metal dichalcogenide, monolayer MoS\s2.
While TaS\s2 is intrinsically metallic, in semiconducting MoS\s2 doping is needed to introduce some metallic screening and enhance electron-phonon interaction.
The unavoidable Lifshitz transition changes the phase space for electron scattering, leading to significant nonadiabatic renormalization of optical phonons~\cite{GarciaGoiricelaya2020, Novko2020a}.
Electron doping also induces a predicted~\cite{Rosner2014} and experimentally observed~\cite{BinSubhan2021} \ac{CDW}.
Moreover, enhanced electron-phonon interaction upon doping has been theoretically proposed to induce superconductivity~\cite{Ge2013}, which was later measured~\cite{Costanzo2016}.

\begin{figure}
    \includegraphics{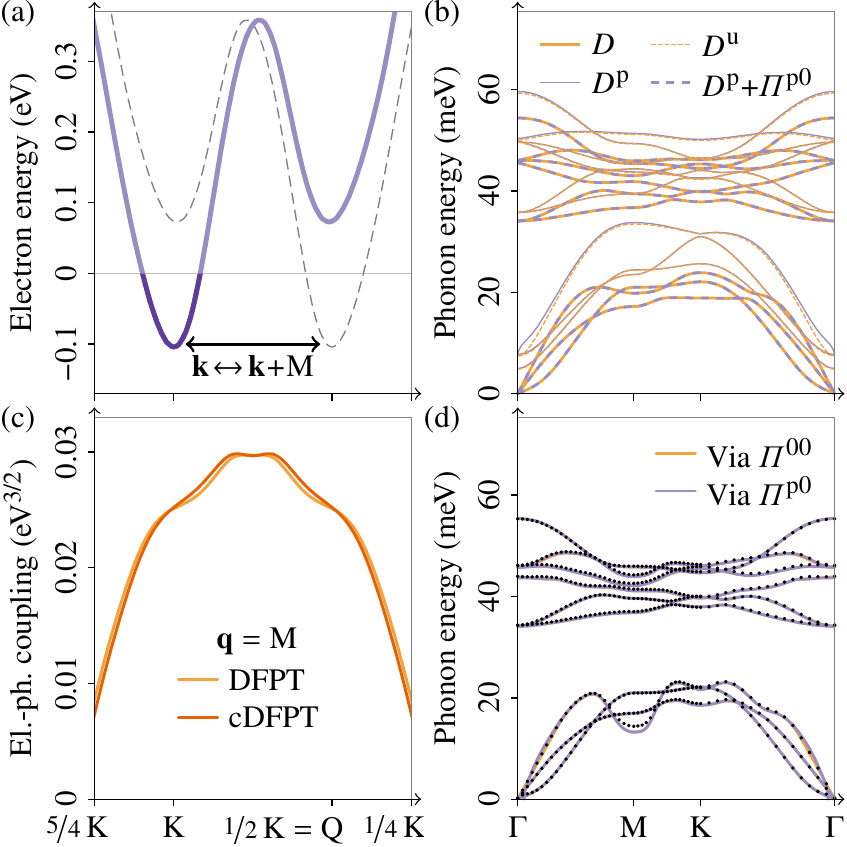}
    \caption{Electron-phonon interaction of n-doped monolayer MoS\s2.
    (a)~Electronic band structure with the conduction-band valleys at $\mathrm K$ (partially occupied) and close to $1/2\,\mathrm K = \mathrm Q$ (empty).
    Note that $\vec q = \mathrm M$ connects occupied and empty valleys.
    (b)~High-smearing phonon dispersions [cf.\@ Fig.~\ref{fig:subspaces}\,(g)].
    (c)~Electron-phonon coupling between the conduction band and the longitudinal-acoustic phonon mode at $\vec q = \mathrm M$ as a function of initial electron momentum $\vec k$.
    There is a significant coupling at $\vec k = \mathrm K$ and $\mathrm Q$, i.e., between the occupied and empty valleys.
    The maximum coupling is, however, in between these two points, where the bare electronic susceptibility is low.
    (d)~Renormalized phonon dispersion according to Eqs.~\eqref{eq:d00} and \eqref{eq:dp0} [cf.\@ Fig.~\ref{fig:renorm}\,(b,\,g)], where the black dots are the reference low-smearing \ac{DFPT} results.}
    \label{fig:mos2}
\end{figure}

Here we consider a total charge of $-0.1 e$ per unit cell, and the Fermi level is accordingly increased and crossing the $\mathrm K$~valley, as seen in Fig.~\ref{fig:mos2}\,(a).
The corresponding relaxed lattice constant is $3.19$~\AA.
For consistency, all \textit{ab initio} parameters are kept identical to the ones chosen for TaS\s2 (cf.\@ Sec.~\ref{sec:results}), and we verified that convergence was reached also in this case.
The valence band and the first four conduction bands have $d$-orbital character.
Since the valence band is energetically isolated from the lower bands, the active subspace is naturally defined by its lowest point and the top of the fourth conduction band.
This choice of the five-band active subspace is similar to the one shown in Fig.~\ref{fig:subspaces}\,(b) for TaS\s2.

The corresponding phonon dispersions for a smearing of $\sigma = 20$~mRy are shown in Fig.~\ref{fig:mos2}\,(b).
A small doping-induced phonon softening at the $\mathrm M$~point can be observed in the \ac{DFPT} phonon dispersion (solid orange lines) and is expected to increase with decreasing electronic temperature.
The partially screened dispersion (solid mauve lines) lacks this feature and also violates the acoustic sum rule; the entire dispersion is shifted upward in energy.
Unscreened phonon frequencies (dashed orange lines), obtained by subtracting the approximate self-energy with two screened vertices [Eq.~\eqref{eq:du}], are generally equal to or lower in energy than the \ac{cDFPT} phonons.
Finally, the \ac{DFPT} result can be restored (dashed mauve lines) by renormalizing the \ac{cDFPT} phonons using the phonon self-energy with one partially screened vertex [Eq.~\eqref{eq:dp0}], confirming all the conclusions from Sec.~\ref{sec:phonons}.

The partially screened and unscreened phonons in Fig.~\ref{fig:mos2}\,(b) are very similar, except for the slope near $\Gamma$, which is due to the long-range terms present in \ac{cDFPT}~%
\footnote{For n-doped monolayer MoS\s2, we use ${\epsilon = 3.84}$, ${Z^* \sub{Mo} = 1.09 e}$, ${Z^* \sub S = 0.59 e}$, ${Q_{\text{Mo} y y y} = 2.75~e\,\text{bohr}}$, ${Q_{\text S y y y} = -0.13~e\,\text{bohr}}$, ${Q_{\text S z y y} = 4.77~e\,\text{bohr}}$, and ${L = 3.8~\text{bohr}}$.}.
This similarity is attributed to the fact that the electron-phonon coupling, shown in Fig.~\ref{fig:mos2}\,(c), does not differ significantly between \ac{cDFPT} and \ac{DFPT}.
Therefore, the two approaches to estimate the phonon dispersion at low electronic temperature $T = 1.9$~mRy yield similar results, as seen in Fig.~\ref{fig:mos2}\,(d).
Using the high-smearing coupling, both schemes soften the branch at the $\mathrm M$~point in good agreement with the low-temperature \ac{DFPT} reference (black dots).
In the case of lightly electron-doped MoS\s2, the overscreening effects are smaller than in TaS\s2 due to the much smaller Fermi surface.

\begin{figure}
    \includegraphics{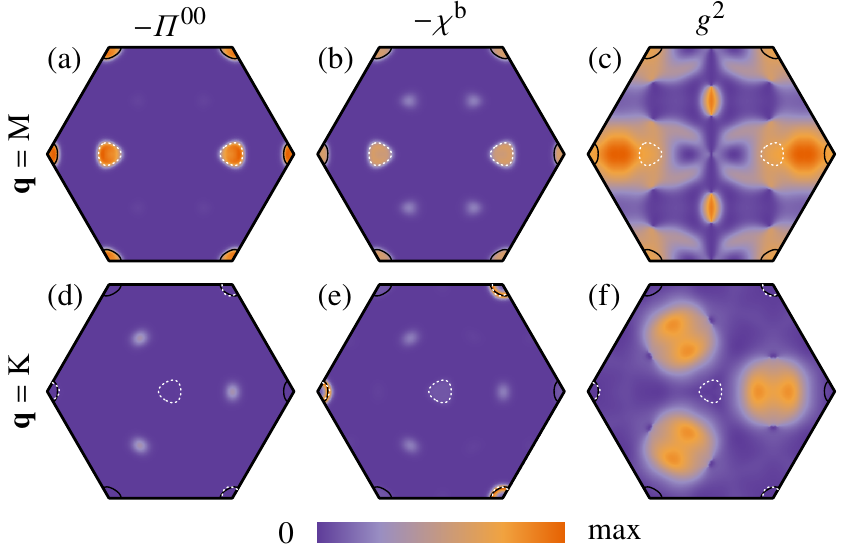}
    \caption{Fluctuation diagnostics of phonon softening in n-doped monolayer MoS\s2 at $T = 1.9$~mRy.
    We show the $\vec k$-dependent (a,\,d)~phonon self-energy, (b,\,e)~bare susceptibility, and (c,\,f)~electron-phonon coupling between the conduction band and the longitudinal-acoustic mode for (a--c)~$\vec q = \mathrm M$ and (d--f)~$\vec q = \mathrm K$.
    Solid black and dashed white lines show the Fermi contours at $\vec k$ and $\vec k + \vec q$.}
    \label{fig:fluctuations}
\end{figure}

Finally, to visualize the origin of the phonon softening in n-doped monolayer MoS\s2~\cite{Novko2020a}, we show in Fig.~\ref{fig:fluctuations} the ``fluctuation diagnostics''~\cite{Berges2020a} of the phonon self-energy due to the predominantly (84\,\%) longitudinal-acoustic mode.
The phonon momentum $\vec q = \mathrm M$ superimposes the occupied and empty conduction-band valleys at $\vec k = \mathrm K$ and $\mathrm Q$, leading to the dominant contributions to the phonon self-energy [Fig.~\ref{fig:fluctuations}\,(a)].
The $\vec k$-space structure of the phonon self-energy and the bare susceptibility [Fig.~\ref{fig:fluctuations}\,(b)] are very similar.
This is because in all regions of significant bare susceptibility there is also an appreciable electron-phonon coupling [Fig.~\ref{fig:fluctuations}\,(c)], although the maximum of the coupling is \emph{dormant} due to a vanishing bare susceptibility.
The situation is different for $\vec q = \mathrm K$, where only a weak phonon softening is observed, notwithstanding that this $\vec q$~point exactly superimposes Fermi contours at neighboring $\mathrm K$~points.
Indeed, the corresponding phonon self-energy [Fig.~\ref{fig:fluctuations}\,(d)] is overall small in spite of a significant bare susceptibility [Fig.~\ref{fig:fluctuations}\,(e)], since the latter occurs in $\vec k$~space regions where the electron-phonon coupling is small [Fig.~\ref{fig:fluctuations}\,(f)].
This analysis therefore explains why the phonon-induced softening is strongest at the M~point in lightly doped monolayer MoS\s2.

\section{Bare quantities in a pseudopotential framework}
\label{app:bare}

As noted in the main text, a \ac{cDFPT} calculation with all bands explicitly accounted for in \ac{DFT} does \emph{not} yield bare phonons and interactions since excitations between these active bands and an infinite number of empty bands still contribute to the screening.
In turn, the bare quantities can be calculated in practice by simply setting the electronic response to ionic displacements to zero.
However, the results for bare atomic nuclei will be different from that in a pseudopotential context.

\begin{figure*}
    \includegraphics{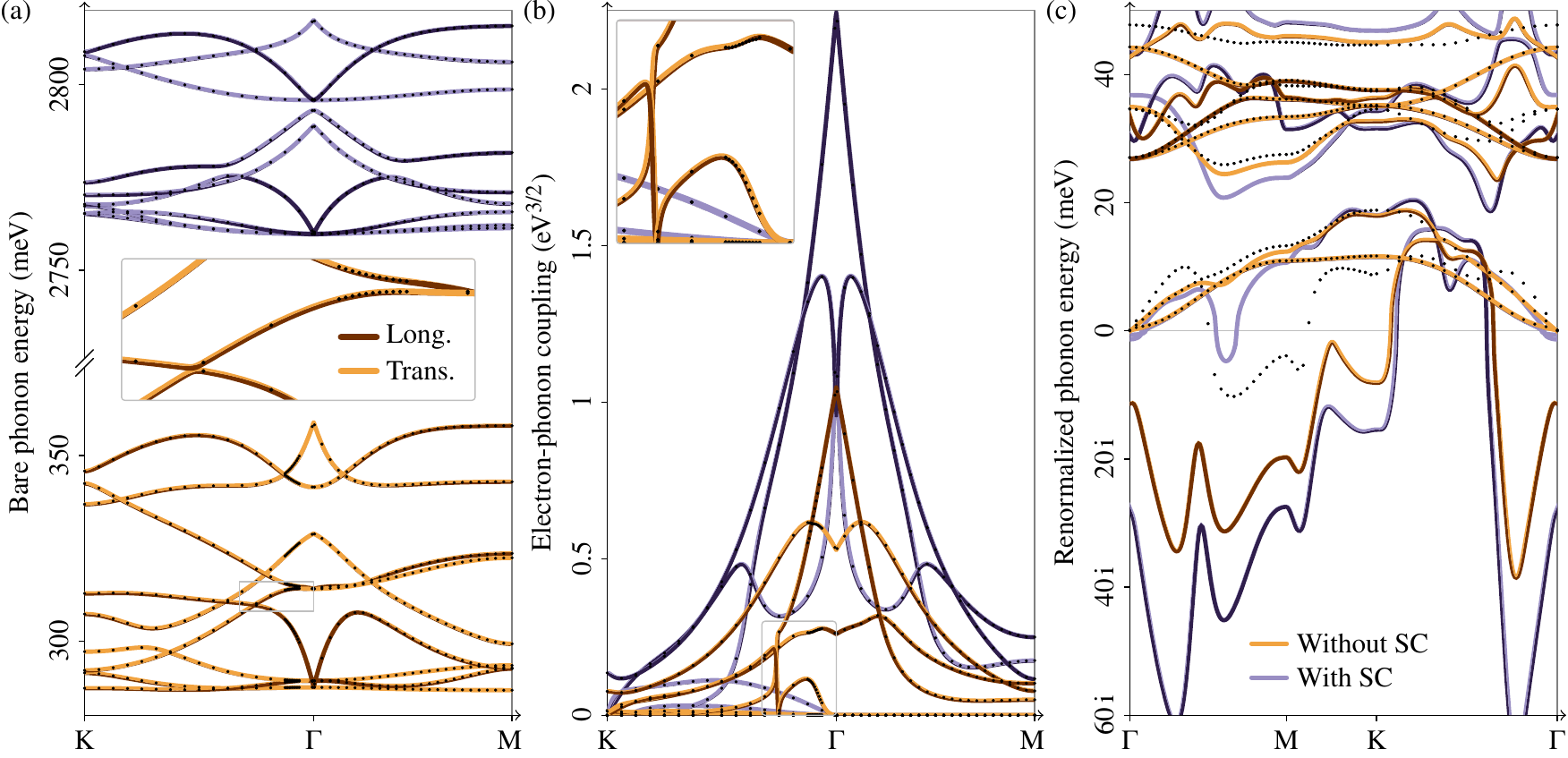}
    \caption{Bare (a)~phonon dispersion [cf.\@ Fig.~\ref{fig:subspaces}\,(f--j)], (b)~electron-phonon coupling (cf.\@ Fig.~\ref{fig:coupling}), and (c)~corresponding estimated low-temperature phonon dispersion from Eq.~\eqref{eq:db0} [cf.\@ Fig.~\ref{fig:renorm}\,(p)] of monolayer TaS\s2 for pseudopotentials with (mauve) and without (orange) \ac{SC} states in the valence shell.
    Longitudinal and transverse mode characters are represented by dark and light color shades, respectively.
    The insets in panels~(a) and (b) are close-ups of avoided crossings.}
    \label{fig:bare}
\end{figure*}

To quantify the influence of the pseudoization, in Fig.~\ref{fig:bare} we present results obtained for two different choices of pseudopotentials.
On the one hand, we use the same potentials as for the results shown in the main text, see Sec.~\ref{sec:results}, which have a combined charge of $25 e$ in the unit cell of TaS\s2.
On the other hand, we include \ac{SC} states in the valence shell, which increases the combined charge to $55 e$ per cell.

The bare phonon dispersions are depicted in Fig.~\ref{fig:bare}\,(a)~%
\footnote{It was not possible to neglect the effect of the out-of-plane polarizability (cf.\@ Sec.~\ref{sec:froehlich}) in the bare phonon dispersions.}.
While the bare phonons without \ac{SC} states reach about twice the energies of the partially screened phonons for 22 active bands [Fig.~\ref{fig:subspaces}\,(j)], they are still separated from those with \ac{SC} states by a gap of about 2.4~eV.
Again, the acoustic sum rule is broken, which has a straightforward physical interpretation in this case:
Suppressing any electronic response will prevent the ``frozen'' electron density to move along with uniformly displaced ions, thus inducing a restoring force and associated plasma oscillations~%
\footnote{Alternative definitions of the bare phonons, which fulfill the acoustic sum rule, are possible~\cite{Pickett1976}.}.
Moreover, the Born effective charges will be identical to the bare (pseudo)ionic charges (cf.\@ Table~\ref{tab:lr}), which are clearly positive and thus cannot add up to zero.

Also the bare electron-phonon coupling shown in Fig.~\ref{fig:bare}\,(b) significantly exceeds the largest calculated partially screened values [cf.\@ Fig.~\ref{fig:coupling}\,(f)] and grows with the number of states in the valence manifold.
To correctly interpolate them, it is again necessary to consider not only dipolar but also quadrupolar long-range terms.
The respective parameters are listed in Table~\ref{tab:lr}.
Note the mentioned relation between the optimized quadrupole-tensor elements and the Born effective charges.

Finally, in the same way the bare coupling depends on the pseudopotential, the renormalized phonons according to Eq.~\eqref{eq:db0} are expected to change.
Indeed, Fig.~\ref{fig:bare}\,(c) reveals that the overscreening of the longitudinal-acoustic mode, already observed in Fig.~\ref{fig:renorm}\,(p), is more pronounced with \ac{SC} states.
As anticipated in Eq.~\eqref{eq:linear}, the error in the dynamical matrix is linear in the bare coupling.
We show the one-band case here, which proved to be sufficiently converged.
We stress that when the screened vertex is correctly dealt with as in Eq.~\eqref{eq:dbt}, the result will still converge toward the correct solution with increasing number of bands, albeit slower than shown in Fig.~\ref{fig:renorm}\,(u--y).

The surprising behavior of the bare electron-phonon coupling without \ac{SC} states near $1/4\,\mathrm K$, a close-up of which is shown in the inset of Fig.~\ref{fig:bare}\,(b), can be explained in terms of a peculiar hybridization between phonon modes.
As shown in the inset of Fig.~\ref{fig:bare}\,(a), the respective modes experience two avoided crossings.
The first one near $1/4\,\mathrm K$ is hardly distinguishable from a real crossing in the full plot.
The second one near $1/12\,\mathrm K$ is very subtle but manifests in a swapping of longitudinal and transverse character.
We have performed additional reference calculations in its vicinity to rule out a real crossing.
In both cases, the ability to couple to the low-energy electrons is transferred from one continuous mode to the other, resulting in abrupt changes of the mode-resolved electron-phonon coupling.

\section{Unscreening on coarse versus fine \ac{BZ} mesh}
\label{app:unscreen}

In Sec.~\ref{sec:renorm}, we have obtained the unscreened and partially screened dynamical matrices $D \super u (\sigma, 0)$ and $D \super p (\sigma, 0)$ by subtracting phonon self-energies $\Pi \super{00} (\sigma, 0)$ and $\Pi \super{p0} (\sigma, 0)$ from the interpolated \ac{DFPT} dynamical matrix $D(\sigma, 0)$ that have been calculated for the same $\vec q$~points and dense $\vec k$~mesh as the subsequently added $\Pi \super{00} (T, 0)$ and $\Pi \super{p0} (T, 0)$ [cf.\@ Eqs.~\eqref{eq:du} and \eqref{eq:dp}].
The results have been shown in Fig.~\ref{fig:renorm}\,(a--j).

\begin{figure*}
    \includegraphics{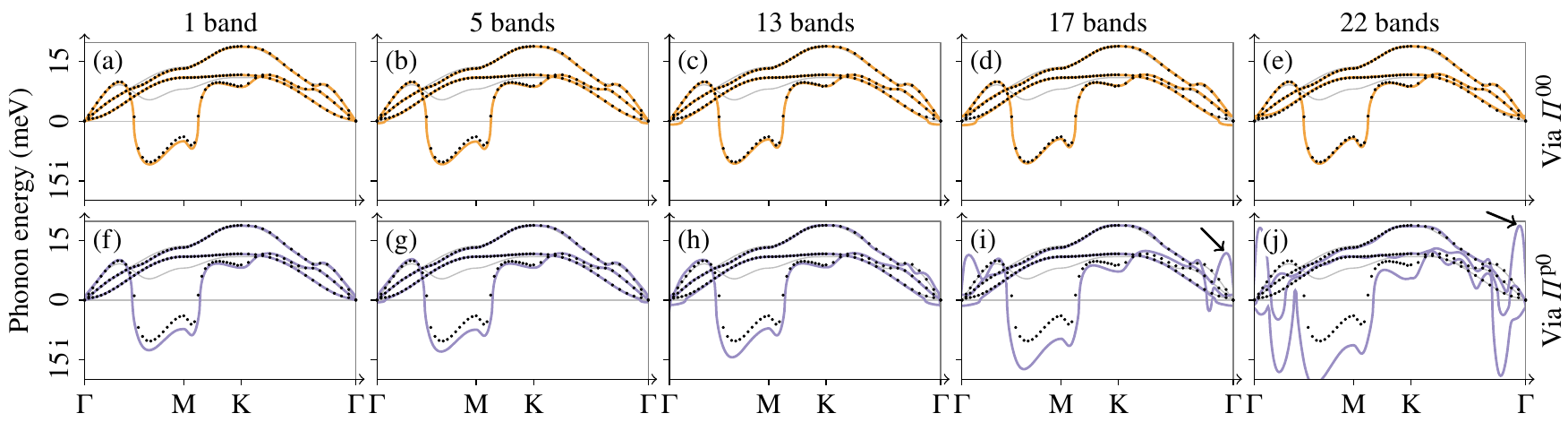}
    \caption{Renormalized phonon dispersion according to (a--e)~Eq.~\eqref{eq:d00} and (f--j)~Eq.~\eqref{eq:dp0} as in Fig.~\ref{fig:renorm}\,(a--j), except that the Fourier interpolation of the dynamical matrix has been performed at the level of $D \super u (\sigma, 0)$ and $D \super p (\sigma, 0)$ instead of $D(\sigma, 0)$.
    Arrows point to the resulting numerical artifacts.}
    \label{fig:unscreen}
\end{figure*}

Alternatively, we can obtain $D \super u (\sigma, 0)$ and $D \super p (\sigma, 0)$---the latter even self-consistently with \ac{cDFPT}---for coarse $\vec q$ and $\vec k$~meshes, and Fourier interpolate them to the $\vec q$~points on which $\Pi \super{00} (T, 0)$ and $\Pi \super{p0} (T, 0)$ have been calculated.
Corresponding results are shown in Fig.~\ref{fig:unscreen}.

Starting from the self-consistent \ac{cDFPT} phonons demonstrates the exactness of the equations in Sec.~\ref{sec:rpa}, but it suffers from independent numerical errors---mostly due to interpolation and the modeling of long-range interactions---in $D \super p (\sigma, 0)$ and the electron-phonon couplings $g \super p (\sigma, 0)$ and $g (\sigma, 0)$ entering $\Pi \super{p0} (T, 0)$, which add up in the resulting screened $D \super{p0} (T, 0)$.
Further taking into account that both $D \super p (\sigma, 0)$ and $\Pi \super{p0} (T, 0)$ are in our case up to one order of magnitude larger than their sum $D \super{p0} (\sigma, 0)$~\cite{Allen1980}, we run the risk of ``catastrophic cancellation\rlap,'' which always occurs when adding imprecise large numbers of similar magnitude and opposite sign.
As a consequence, in Fig.~\ref{fig:unscreen}\,(f--j) we find significant errors near $\Gamma$, especially for large active subspaces (arrows), which have been absent in Fig.~\ref{fig:renorm}\,(f--j) due to a cancellation of interpolation errors in the couplings.
For 22 bands, the interpolation suffers from an inferior localization of Wannier functions [cf.\@ inset of Fig.~\ref{fig:coupling}\,(f)].
Away from $\Gamma$, the results hardly differ.

In turn, the renormalized phonon dispersions in Figs.~\ref{fig:renorm}\,(a--e) and \ref{fig:unscreen}\,(a--e) are almost indistinguishable.
This is because the \ac{DFPT} electron-phonon coupling is much smaller and smooth at $\Gamma$ and can accordingly be interpolated more precisely.

\section{Indefiniteness of combined coupling matrix}
\label{app:geff}

\begin{figure}
    \includegraphics{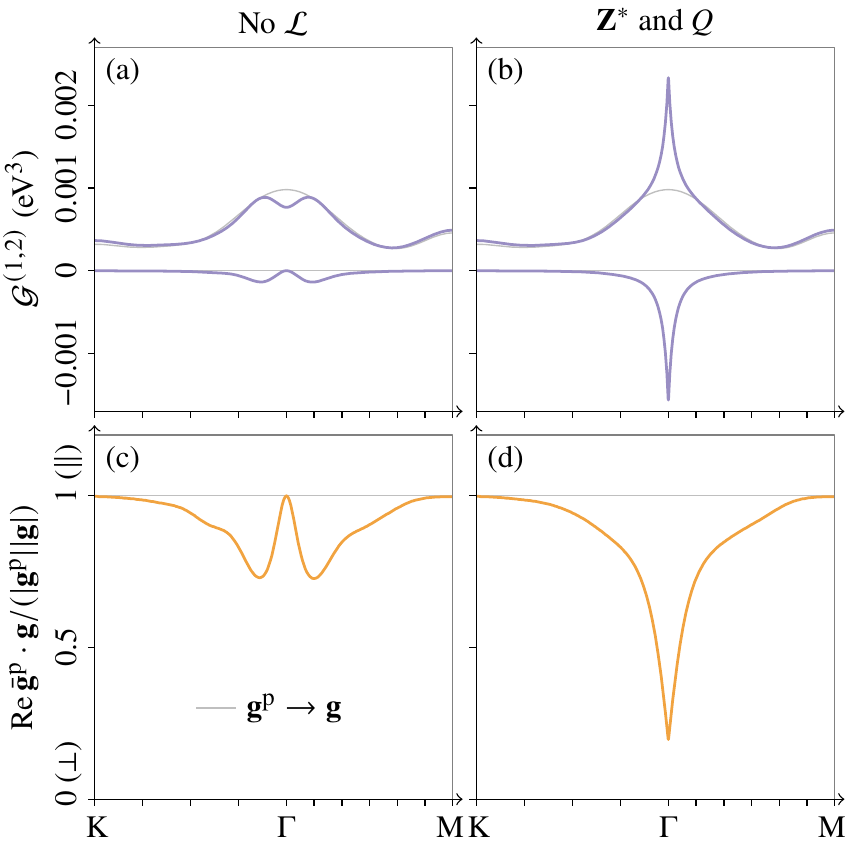}
    \caption{Indefiniteness of the symmetrized partially screened-screened coupling matrix $\mathcal G$ [Eq.~\eqref{eq:g2}].
    As in Fig.~\ref{fig:coupling}\,(a,\,b), we show data for the single active band of monolayer TaS\s2 as a function of $\vec q$ with $\vec k = 0$.
    Gray lines represent corresponding screened-screened results.
    (a,\,b)~Nonzero eigenvalues without and with handling of the long-range part.
    (c,\,d)~Relative orientation of partially screened and screened coupling.}
    \label{fig:geff}
\end{figure}

As mentioned in Sec.~\ref{sec:sym}, the matrix $\mathcal G$ [Eq.~\eqref{eq:g2}], formed as a symmetrized outer product of a partially screened vertex $\vec g \super p$ and an approximate screened vertex $\vec g$, is in general not positive-definite, except if $\vec g = \alpha \vec g \super p$ with a non-negative scalar scaling factor $\alpha$.
As a consequence, the approximations using one bare or partially screened vertex can lead to artifacts such as the negative spectral weight in Fig.~\ref{fig:specfun}\,(b), which notably has only been observed when correctly handling the long-range part of $\vec g \super p$.

To better understand this, in Fig.~\ref{fig:geff}\,(a,\,b), we display the two nonzero eigenvalues of $\mathcal G$ (mauve lines) for the single active band of monolayer TaS\s2 as a function of $\vec q$ with $\vec k = 0$ [cf.\@ Fig.~\ref{fig:coupling}\,(a,\,b)], both without and with handling of the long-range part.
For comparison, we also show the corresponding result obtained when replacing $\vec g \super p$ by $\vec g$ (gray lines), which yields the symmetric formulation with two screened vertices.
Away from $\Gamma$, the negative eigenvalue is negligibly small, which means that the phonon self-energy with one partially screened vertex and one screened vertex could be written using an effective intermediate vertex on both sides of the electron-hole bubble; see Sec.~\ref{sec:sym}.
As expected, the positive eigenvalue is larger than its screened-screened counterpart.
Approaching $\Gamma$ however, the negative value becomes significant, and the behavior is very different depending on whether the long-range part is handled or not, except on the original $\vec q$~mesh where the curves coincide by definition.
For the naive Fourier interpolation [Fig.~\ref{fig:geff}\,(a)], the negative eigenvalue vanishes again for smallest $\vec q$; with long-range handling [Fig.~\ref{fig:geff}\,(b)], it peaks instead.

This can be understood by looking at the relative orientation of $\vec g \super p$ and $\vec g$ as vectors in $3 N \sub{at}$-dimensional space, where $N \sub{at}$  is the number of basis atoms.
In Fig.~\ref{fig:geff}\,(c,\,d), we show the normalized real part of their scalar product, $\Re \bar{\vec g} \super p \cdot \vec g / (\abs{\vec g \super p} \abs{\vec g})$, the inverse cosine of which can be seen as the angle between $\vec g \super p$ and $\vec g$.
Note that $\Re \bar{\vec g} \super p \cdot \vec g = \Tr \mathcal G$.
Values of $1$, $0$, and $-1$ one indicate that $\vec g \super p$ and $\vec g$ are parallel, orthogonal, and antiparallel.
Toward $\Gamma$, the physically correct vertices become increasingly orthogonal [Fig.~\ref{fig:geff}\,(d)], as $\vec g \super p$ can be large for modes for which $\vec g$ is fully suppressed by metallic screening (cf.\@ Fig.~\ref{fig:coupling}).
However, the erroneous naive interpolation makes $\vec g \super p$ behave similarly to $\vec g$ for smallest $\vec q$, restoring parallelism [Fig.~\ref{fig:geff}\,(c)].
Coincidentally, this turns out to be beneficial in the context of calculating the approximate $D \super{p0} (T, \omega)$, as seen in Fig.~\ref{fig:specfun}\,(c).
Note that we have only discussed selected matrix elements here.
For instance, for some initial electron momenta $\vec k \neq 0$ we have observed that the negative eigenvalue can even dominate, indicating that the vertices are partially antiparallel.
Discarding the negative eigenvalue would also guarantee a positive spectral weight but lead to other problems such as a broken acoustic sum rule.

The presence of the negative eigenvalue implies that it is in general impossible to evenly spread the dielectric screening over two electron-phonon vertices attached to the \emph{bare} electronic susceptibility.
Hence, symmetric formulations must move some or all screening to the susceptibility; e.g., $\Pi = g \super b \chi g \super b$ with $\chi = \epsilon \inv \chi \super b$ or $\Pi = g (\chi \super b - v \chi \super b v) g$ as used in Ref.~\citenum{Calandra2010}.

\bibliography{ms}

\begin{thebibliography}{141}%
\makeatletter
\providecommand \@ifxundefined [1]{%
 \@ifx{#1\undefined}
}%
\providecommand \@ifnum [1]{%
 \ifnum #1\expandafter \@firstoftwo
 \else \expandafter \@secondoftwo
 \fi
}%
\providecommand \@ifx [1]{%
 \ifx #1\expandafter \@firstoftwo
 \else \expandafter \@secondoftwo
 \fi
}%
\providecommand \natexlab [1]{#1}%
\providecommand \enquote  [1]{``#1''}%
\providecommand \bibnamefont  [1]{#1}%
\providecommand \bibfnamefont [1]{#1}%
\providecommand \citenamefont [1]{#1}%
\providecommand \href@noop [0]{\@secondoftwo}%
\providecommand \href [0]{\begingroup \@sanitize@url \@href}%
\providecommand \@href[1]{\@@startlink{#1}\@@href}%
\providecommand \@@href[1]{\endgroup#1\@@endlink}%
\providecommand \@sanitize@url [0]{\catcode `\\12\catcode `\$12\catcode
  `\&12\catcode `\#12\catcode `\^12\catcode `\_12\catcode `\%12\relax}%
\providecommand \@@startlink[1]{}%
\providecommand \@@endlink[0]{}%
\providecommand \url  [0]{\begingroup\@sanitize@url \@url }%
\providecommand \@url [1]{\endgroup\@href {#1}{\urlprefix }}%
\providecommand \urlprefix  [0]{URL }%
\providecommand \Eprint [0]{\href }%
\providecommand \doibase [0]{https://doi.org/}%
\providecommand \selectlanguage [0]{\@gobble}%
\providecommand \bibinfo  [0]{\@secondoftwo}%
\providecommand \bibfield  [0]{\@secondoftwo}%
\providecommand \translation [1]{[#1]}%
\providecommand \BibitemOpen [0]{}%
\providecommand \bibitemStop [0]{}%
\providecommand \bibitemNoStop [0]{.\EOS\space}%
\providecommand \EOS [0]{\spacefactor3000\relax}%
\providecommand \BibitemShut  [1]{\csname bibitem#1\endcsname}%
\let\auto@bib@innerbib\@empty
\bibitem [{\citenamefont {Giustino}(2017)}]{Giustino2017}%
  \BibitemOpen
  \bibfield  {author} {\bibinfo {author} {\bibfnamefont {F.}~\bibnamefont
  {Giustino}},\ }\bibfield  {title} {\bibinfo {title} {\emph {Electron-phonon
  interactions from first principles}},\ }\href
  {https://doi.org/10.1103/RevModPhys.89.015003} {\bibfield  {journal}
  {\bibinfo  {journal} {Rev. Mod. Phys.}\ }\textbf {\bibinfo {volume} {89}},\
  \bibinfo {pages} {015003} (\bibinfo {year} {2017})},\ \Eprint
  {https://arxiv.org/abs/1603.06965} {arXiv:1603.06965}\BibitemShut {NoStop}%
\bibitem [{\citenamefont {Calandra}\ \emph {et~al.}(2010)\citenamefont
  {Calandra}, \citenamefont {Profeta},\ and\ \citenamefont
  {Mauri}}]{Calandra2010}%
  \BibitemOpen
  \bibfield  {author} {\bibinfo {author} {\bibfnamefont {M.}~\bibnamefont
  {Calandra}}, \bibinfo {author} {\bibfnamefont {G.}~\bibnamefont {Profeta}},\
  and\ \bibinfo {author} {\bibfnamefont {F.}~\bibnamefont {Mauri}},\ }\bibfield
   {title} {\bibinfo {title} {\emph {Adiabatic and nonadiabatic phonon
  dispersion in a {Wannier} function approach}},\ }\href
  {https://doi.org/10.1103/PhysRevB.82.165111} {\bibfield  {journal} {\bibinfo
  {journal} {Phys. Rev. B}\ }\textbf {\bibinfo {volume} {82}},\ \bibinfo
  {pages} {165111} (\bibinfo {year} {2010})},\ \Eprint
  {https://arxiv.org/abs/1007.2098} {arXiv:1007.2098}\BibitemShut {NoStop}%
\bibitem [{\citenamefont {Nomura}\ and\ \citenamefont
  {Arita}(2015)}]{Nomura2015}%
  \BibitemOpen
  \bibfield  {author} {\bibinfo {author} {\bibfnamefont {Y.}~\bibnamefont
  {Nomura}}\ and\ \bibinfo {author} {\bibfnamefont {R.}~\bibnamefont {Arita}},\
  }\bibfield  {title} {\bibinfo {title} {\emph {Ab initio downfolding for
  electron-phonon-coupled systems: Constrained density-functional perturbation
  theory}},\ }\href {https://doi.org/10.1103/PhysRevB.92.245108} {\bibfield
  {journal} {\bibinfo  {journal} {Phys. Rev. B}\ }\textbf {\bibinfo {volume}
  {92}},\ \bibinfo {pages} {245108} (\bibinfo {year} {2015})},\ \Eprint
  {https://arxiv.org/abs/1509.01138} {arXiv:1509.01138}\BibitemShut {NoStop}%
\bibitem [{\citenamefont {Ponc\'e}\ \emph {et~al.}(2020)\citenamefont
  {Ponc\'e}, \citenamefont {Li}, \citenamefont {Reichardt},\ and\ \citenamefont
  {Giustino}}]{Ponce2020}%
  \BibitemOpen
  \bibfield  {author} {\bibinfo {author} {\bibfnamefont {S.}~\bibnamefont
  {Ponc\'e}}, \bibinfo {author} {\bibfnamefont {W.}~\bibnamefont {Li}},
  \bibinfo {author} {\bibfnamefont {S.}~\bibnamefont {Reichardt}},\ and\
  \bibinfo {author} {\bibfnamefont {F.}~\bibnamefont {Giustino}},\ }\bibfield
  {title} {\bibinfo {title} {\emph {First-principles calculations of charge
  carrier mobility and conductivity in bulk semiconductors and two-dimensional
  materials}},\ }\href {https://doi.org/10.1088/1361-6633/ab6a43} {\bibfield
  {journal} {\bibinfo  {journal} {Rep. Prog. Phys.}\ }\textbf {\bibinfo
  {volume} {83}},\ \bibinfo {pages} {036501} (\bibinfo {year} {2020})},\
  \Eprint {https://arxiv.org/abs/1908.01733} {arXiv:1908.01733}\BibitemShut
  {NoStop}%
\bibitem [{\citenamefont {Knoop}\ \emph {et~al.}(2020)\citenamefont {Knoop},
  \citenamefont {Purcell}, \citenamefont {Scheffler},\ and\ \citenamefont
  {Carbogno}}]{Knoop2020}%
  \BibitemOpen
  \bibfield  {author} {\bibinfo {author} {\bibfnamefont {F.}~\bibnamefont
  {Knoop}}, \bibinfo {author} {\bibfnamefont {T.~A.~R.}\ \bibnamefont
  {Purcell}}, \bibinfo {author} {\bibfnamefont {M.}~\bibnamefont {Scheffler}},\
  and\ \bibinfo {author} {\bibfnamefont {C.}~\bibnamefont {Carbogno}},\
  }\bibfield  {title} {\bibinfo {title} {\emph {Anharmonicity measure for
  materials}},\ }\href {https://doi.org/10.1103/PhysRevMaterials.4.083809}
  {\bibfield  {journal} {\bibinfo  {journal} {Phys. Rev. Materials}\ }\textbf
  {\bibinfo {volume} {4}},\ \bibinfo {pages} {083809} (\bibinfo {year}
  {2020})},\ \Eprint {https://arxiv.org/abs/2006.14672}
  {arXiv:2006.14672}\BibitemShut {NoStop}%
\bibitem [{\citenamefont {Wilson}\ \emph {et~al.}(1974)\citenamefont {Wilson},
  \citenamefont {Di~Salvo},\ and\ \citenamefont {Mahajan}}]{Wilson1974}%
  \BibitemOpen
  \bibfield  {author} {\bibinfo {author} {\bibfnamefont {J.~A.}\ \bibnamefont
  {Wilson}}, \bibinfo {author} {\bibfnamefont {F.~J.}\ \bibnamefont
  {Di~Salvo}},\ and\ \bibinfo {author} {\bibfnamefont {S.}~\bibnamefont
  {Mahajan}},\ }\bibfield  {title} {\bibinfo {title} {\emph {Charge-density
  waves in metallic, layered, transition-metal dichalcogenides}},\ }\href
  {https://doi.org/10.1103/PhysRevLett.32.882} {\bibfield  {journal} {\bibinfo
  {journal} {Phys. Rev. Lett.}\ }\textbf {\bibinfo {volume} {32}},\ \bibinfo
  {pages} {882} (\bibinfo {year} {1974})}\BibitemShut {NoStop}%
\bibitem [{\citenamefont {Revolinsky}\ \emph {et~al.}(1963)\citenamefont
  {Revolinsky}, \citenamefont {Lautenschlager},\ and\ \citenamefont
  {Armitage}}]{Revolinsky1963}%
  \BibitemOpen
  \bibfield  {author} {\bibinfo {author} {\bibfnamefont {E.}~\bibnamefont
  {Revolinsky}}, \bibinfo {author} {\bibfnamefont {E.~P.}\ \bibnamefont
  {Lautenschlager}},\ and\ \bibinfo {author} {\bibfnamefont {C.~H.}\
  \bibnamefont {Armitage}},\ }\bibfield  {title} {\bibinfo {title} {\emph
  {Layer structure superconductor}},\ }\href
  {https://doi.org/10.1016/0038-1098(63)90358-2} {\bibfield  {journal}
  {\bibinfo  {journal} {Solid State Commun.}\ }\textbf {\bibinfo {volume}
  {1}},\ \bibinfo {pages} {59} (\bibinfo {year} {1963})}\BibitemShut {NoStop}%
\bibitem [{\citenamefont {Baroni}\ \emph {et~al.}(2001)\citenamefont {Baroni},
  \citenamefont {de~Gironcoli}, \citenamefont {Dal~Corso},\ and\ \citenamefont
  {Giannozzi}}]{Baroni2001}%
  \BibitemOpen
  \bibfield  {author} {\bibinfo {author} {\bibfnamefont {S.}~\bibnamefont
  {Baroni}}, \bibinfo {author} {\bibfnamefont {S.}~\bibnamefont
  {de~Gironcoli}}, \bibinfo {author} {\bibfnamefont {A.}~\bibnamefont
  {Dal~Corso}},\ and\ \bibinfo {author} {\bibfnamefont {P.}~\bibnamefont
  {Giannozzi}},\ }\bibfield  {title} {\bibinfo {title} {\emph {Phonons and
  related crystal properties from density-functional perturbation theory}},\
  }\href {https://doi.org/10.1103/RevModPhys.73.515} {\bibfield  {journal}
  {\bibinfo  {journal} {Rev. Mod. Phys.}\ }\textbf {\bibinfo {volume} {73}},\
  \bibinfo {pages} {515} (\bibinfo {year} {2001})},\ \Eprint
  {https://arxiv.org/abs/cond-mat/0012092} {arXiv:cond-mat/0012092}\BibitemShut
  {NoStop}%
\bibitem [{\citenamefont {Hohenberg}\ and\ \citenamefont
  {Kohn}(1964)}]{Hohenberg1964}%
  \BibitemOpen
  \bibfield  {author} {\bibinfo {author} {\bibfnamefont {P.}~\bibnamefont
  {Hohenberg}}\ and\ \bibinfo {author} {\bibfnamefont {W.}~\bibnamefont
  {Kohn}},\ }\bibfield  {title} {\bibinfo {title} {\emph {Inhomogeneous
  electron gas}},\ }\href {https://doi.org/10.1103/PhysRev.136.B864} {\bibfield
   {journal} {\bibinfo  {journal} {Phys. Rev.}\ }\textbf {\bibinfo {volume}
  {136}},\ \bibinfo {pages} {B864} (\bibinfo {year} {1964})}\BibitemShut
  {NoStop}%
\bibitem [{\citenamefont {Kohn}\ and\ \citenamefont {Sham}(1965)}]{Kohn1965}%
  \BibitemOpen
  \bibfield  {author} {\bibinfo {author} {\bibfnamefont {W.}~\bibnamefont
  {Kohn}}\ and\ \bibinfo {author} {\bibfnamefont {L.~J.}\ \bibnamefont
  {Sham}},\ }\bibfield  {title} {\bibinfo {title} {\emph {Self-consistent
  equations including exchange and correlation effects}},\ }\href
  {https://doi.org/10.1103/PhysRev.140.A1133} {\bibfield  {journal} {\bibinfo
  {journal} {Phys. Rev.}\ }\textbf {\bibinfo {volume} {140}},\ \bibinfo {pages}
  {A1133} (\bibinfo {year} {1965})}\BibitemShut {NoStop}%
\bibitem [{\citenamefont {Zein}(1984)}]{Zein1984}%
  \BibitemOpen
  \bibfield  {author} {\bibinfo {author} {\bibfnamefont {N.~E.}\ \bibnamefont
  {Zein}},\ }\bibfield  {title} {\bibinfo {title} {\emph {Density functional
  calculations of elastic moduli and phonon spectra of crystals}},\ }\href@noop
  {} {\bibfield  {journal} {\bibinfo  {journal} {Sov. Phys. Solid State}\
  }\textbf {\bibinfo {volume} {26}},\ \bibinfo {pages} {1825} (\bibinfo {year}
  {1984})}\BibitemShut {NoStop}%
\bibitem [{\citenamefont {Gonze}\ and\ \citenamefont {Lee}(1997)}]{Gonze1997}%
  \BibitemOpen
  \bibfield  {author} {\bibinfo {author} {\bibfnamefont {X.}~\bibnamefont
  {Gonze}}\ and\ \bibinfo {author} {\bibfnamefont {C.}~\bibnamefont {Lee}},\
  }\bibfield  {title} {\bibinfo {title} {\emph {Dynamical matrices, {Born}
  effective charges, dielectric permittivity tensors, and interatomic force
  constants from density-functional perturbation theory}},\ }\href
  {https://doi.org/10.1103/PhysRevB.55.10355} {\bibfield  {journal} {\bibinfo
  {journal} {Phys. Rev. B}\ }\textbf {\bibinfo {volume} {55}},\ \bibinfo
  {pages} {10355} (\bibinfo {year} {1997})}\BibitemShut {NoStop}%
\bibitem [{\citenamefont {Petretto}\ \emph {et~al.}(2018)\citenamefont
  {Petretto}, \citenamefont {Dwaraknath}, \citenamefont {P.~C.~Miranda},
  \citenamefont {Winston}, \citenamefont {Giantomassi}, \citenamefont {van
  Setten}, \citenamefont {Gonze}, \citenamefont {Persson}, \citenamefont
  {Hautier},\ and\ \citenamefont {Rignanese}}]{Petretto2018}%
  \BibitemOpen
  \bibfield  {author} {\bibinfo {author} {\bibfnamefont {G.}~\bibnamefont
  {Petretto}}, \bibinfo {author} {\bibfnamefont {S.}~\bibnamefont
  {Dwaraknath}}, \bibinfo {author} {\bibfnamefont {H.}~\bibnamefont
  {P.~C.~Miranda}}, \bibinfo {author} {\bibfnamefont {D.}~\bibnamefont
  {Winston}}, \bibinfo {author} {\bibfnamefont {M.}~\bibnamefont
  {Giantomassi}}, \bibinfo {author} {\bibfnamefont {M.~J.}\ \bibnamefont {van
  Setten}}, \bibinfo {author} {\bibfnamefont {X.}~\bibnamefont {Gonze}},
  \bibinfo {author} {\bibfnamefont {K.~A.}\ \bibnamefont {Persson}}, \bibinfo
  {author} {\bibfnamefont {G.}~\bibnamefont {Hautier}},\ and\ \bibinfo {author}
  {\bibfnamefont {G.-M.}\ \bibnamefont {Rignanese}},\ }\bibfield  {title}
  {\bibinfo {title} {\emph {High-throughput density-functional perturbation
  theory phonons for inorganic materials}},\ }\href
  {https://doi.org/10.1038/sdata.2018.65} {\bibfield  {journal} {\bibinfo
  {journal} {Sci. Data}\ }\textbf {\bibinfo {volume} {5}},\ \bibinfo {pages}
  {180065} (\bibinfo {year} {2018})}\BibitemShut {NoStop}%
\bibitem [{\citenamefont {Mounet}\ \emph {et~al.}(2018)\citenamefont {Mounet},
  \citenamefont {Gibertini}, \citenamefont {Schwaller}, \citenamefont {Campi},
  \citenamefont {Merkys}, \citenamefont {Marrazzo}, \citenamefont {Sohier},
  \citenamefont {Castelli}, \citenamefont {Cepellotti}, \citenamefont {Pizzi},\
  and\ \citenamefont {Marzari}}]{Mounet2018}%
  \BibitemOpen
  \bibfield  {author} {\bibinfo {author} {\bibfnamefont {N.}~\bibnamefont
  {Mounet}}, \bibinfo {author} {\bibfnamefont {M.}~\bibnamefont {Gibertini}},
  \bibinfo {author} {\bibfnamefont {P.}~\bibnamefont {Schwaller}}, \bibinfo
  {author} {\bibfnamefont {D.}~\bibnamefont {Campi}}, \bibinfo {author}
  {\bibfnamefont {A.}~\bibnamefont {Merkys}}, \bibinfo {author} {\bibfnamefont
  {A.}~\bibnamefont {Marrazzo}}, \bibinfo {author} {\bibfnamefont
  {T.}~\bibnamefont {Sohier}}, \bibinfo {author} {\bibfnamefont {I.~E.}\
  \bibnamefont {Castelli}}, \bibinfo {author} {\bibfnamefont {A.}~\bibnamefont
  {Cepellotti}}, \bibinfo {author} {\bibfnamefont {G.}~\bibnamefont {Pizzi}},\
  and\ \bibinfo {author} {\bibfnamefont {N.}~\bibnamefont {Marzari}},\
  }\bibfield  {title} {\bibinfo {title} {\emph {Two-dimensional materials from
  high-throughput computational exfoliation of experimentally known
  compounds}},\ }\href {https://doi.org/10.1038/s41565-017-0035-5} {\bibfield
  {journal} {\bibinfo  {journal} {Nat. Nanotechnol.}\ }\textbf {\bibinfo
  {volume} {13}},\ \bibinfo {pages} {246} (\bibinfo {year} {2018})},\ \Eprint
  {https://arxiv.org/abs/1611.05234} {arXiv:1611.05234}\BibitemShut {NoStop}%
\bibitem [{\citenamefont {Baroni}\ \emph {et~al.}(1987)\citenamefont {Baroni},
  \citenamefont {Giannozzi},\ and\ \citenamefont {Testa}}]{Baroni1987}%
  \BibitemOpen
  \bibfield  {author} {\bibinfo {author} {\bibfnamefont {S.}~\bibnamefont
  {Baroni}}, \bibinfo {author} {\bibfnamefont {P.}~\bibnamefont {Giannozzi}},\
  and\ \bibinfo {author} {\bibfnamefont {A.}~\bibnamefont {Testa}},\ }\bibfield
   {title} {\bibinfo {title} {\emph {Green's-function approach to linear
  response in solids}},\ }\href {https://doi.org/10.1103/PhysRevLett.58.1861}
  {\bibfield  {journal} {\bibinfo  {journal} {Phys. Rev. Lett.}\ }\textbf
  {\bibinfo {volume} {58}},\ \bibinfo {pages} {1861} (\bibinfo {year}
  {1987})}\BibitemShut {NoStop}%
\bibitem [{\citenamefont {Giannozzi}\ \emph {et~al.}(1991)\citenamefont
  {Giannozzi}, \citenamefont {de~Gironcoli}, \citenamefont {Pavone},\ and\
  \citenamefont {Baroni}}]{Giannozzi1991}%
  \BibitemOpen
  \bibfield  {author} {\bibinfo {author} {\bibfnamefont {P.}~\bibnamefont
  {Giannozzi}}, \bibinfo {author} {\bibfnamefont {S.}~\bibnamefont
  {de~Gironcoli}}, \bibinfo {author} {\bibfnamefont {P.}~\bibnamefont
  {Pavone}},\ and\ \bibinfo {author} {\bibfnamefont {S.}~\bibnamefont
  {Baroni}},\ }\bibfield  {title} {\bibinfo {title} {\emph {Ab initio
  calculation of phonon dispersions in semiconductors}},\ }\href
  {https://doi.org/10.1103/PhysRevB.43.7231} {\bibfield  {journal} {\bibinfo
  {journal} {Phys. Rev. B}\ }\textbf {\bibinfo {volume} {43}},\ \bibinfo
  {pages} {7231} (\bibinfo {year} {1991})}\BibitemShut {NoStop}%
\bibitem [{\citenamefont {de~Gironcoli}(1995)}]{deGironcoli1995}%
  \BibitemOpen
  \bibfield  {author} {\bibinfo {author} {\bibfnamefont {S.}~\bibnamefont
  {de~Gironcoli}},\ }\bibfield  {title} {\bibinfo {title} {\emph {Lattice
  dynamics of metals from density-functional perturbation theory}},\ }\href
  {https://doi.org/10.1103/PhysRevB.51.6773} {\bibfield  {journal} {\bibinfo
  {journal} {Phys. Rev. B}\ }\textbf {\bibinfo {volume} {51}},\ \bibinfo
  {pages} {6773} (\bibinfo {year} {1995})}\BibitemShut {NoStop}%
\bibitem [{\citenamefont {Born}\ and\ \citenamefont
  {Oppenheimer}(1927)}]{Born1927}%
  \BibitemOpen
  \bibfield  {author} {\bibinfo {author} {\bibfnamefont {M.}~\bibnamefont
  {Born}}\ and\ \bibinfo {author} {\bibfnamefont {R.}~\bibnamefont
  {Oppenheimer}},\ }\bibfield  {title} {\bibinfo {title} {\emph {Zur
  {Quantentheorie} der {Molekeln}}},\ }\href
  {https://doi.org/10.1002/andp.19273892002} {\bibfield  {journal} {\bibinfo
  {journal} {Ann. Phys.}\ }\textbf {\bibinfo {volume} {389}},\ \bibinfo {pages}
  {457} (\bibinfo {year} {1927})}\BibitemShut {NoStop}%
\bibitem [{\citenamefont {Maksimov}\ and\ \citenamefont
  {Shulga}(1996)}]{Maksimov1996}%
  \BibitemOpen
  \bibfield  {author} {\bibinfo {author} {\bibfnamefont {E.~G.}\ \bibnamefont
  {Maksimov}}\ and\ \bibinfo {author} {\bibfnamefont {S.~V.}\ \bibnamefont
  {Shulga}},\ }\bibfield  {title} {\bibinfo {title} {\emph {Nonadiabatic
  effects in optical phonon self-energy}},\ }\href
  {https://doi.org/10.1016/0038-1098(95)00745-8} {\bibfield  {journal}
  {\bibinfo  {journal} {Solid State Commun.}\ }\textbf {\bibinfo {volume}
  {97}},\ \bibinfo {pages} {553} (\bibinfo {year} {1996})}\BibitemShut
  {NoStop}%
\bibitem [{\citenamefont {Lazzeri}\ and\ \citenamefont
  {Mauri}(2006)}]{Lazzeri2006}%
  \BibitemOpen
  \bibfield  {author} {\bibinfo {author} {\bibfnamefont {M.}~\bibnamefont
  {Lazzeri}}\ and\ \bibinfo {author} {\bibfnamefont {F.}~\bibnamefont
  {Mauri}},\ }\bibfield  {title} {\bibinfo {title} {\emph {Nonadiabatic {Kohn}
  anomaly in a doped graphene monolayer}},\ }\href
  {https://doi.org/10.1103/PhysRevLett.97.266407} {\bibfield  {journal}
  {\bibinfo  {journal} {Phys. Rev. Lett.}\ }\textbf {\bibinfo {volume} {97}},\
  \bibinfo {pages} {266407} (\bibinfo {year} {2006})},\ \Eprint
  {https://arxiv.org/abs/cond-mat/0611708} {arXiv:cond-mat/0611708}\BibitemShut
  {NoStop}%
\bibitem [{\citenamefont {Pisana}\ \emph {et~al.}(2007)\citenamefont {Pisana},
  \citenamefont {Lazzeri}, \citenamefont {Casiraghi}, \citenamefont
  {Novoselov}, \citenamefont {Geim}, \citenamefont {Ferrari},\ and\
  \citenamefont {Mauri}}]{Pisana2007}%
  \BibitemOpen
  \bibfield  {author} {\bibinfo {author} {\bibfnamefont {S.}~\bibnamefont
  {Pisana}}, \bibinfo {author} {\bibfnamefont {M.}~\bibnamefont {Lazzeri}},
  \bibinfo {author} {\bibfnamefont {C.}~\bibnamefont {Casiraghi}}, \bibinfo
  {author} {\bibfnamefont {K.~S.}\ \bibnamefont {Novoselov}}, \bibinfo {author}
  {\bibfnamefont {A.~K.}\ \bibnamefont {Geim}}, \bibinfo {author}
  {\bibfnamefont {A.~C.}\ \bibnamefont {Ferrari}},\ and\ \bibinfo {author}
  {\bibfnamefont {F.}~\bibnamefont {Mauri}},\ }\bibfield  {title} {\bibinfo
  {title} {\emph {Breakdown of the adiabatic {Born}-{Oppenheimer} approximation
  in graphene}},\ }\href {https://doi.org/10.1038/nmat1846} {\bibfield
  {journal} {\bibinfo  {journal} {Nat. Mater.}\ }\textbf {\bibinfo {volume}
  {6}},\ \bibinfo {pages} {198} (\bibinfo {year} {2007})},\ \Eprint
  {https://arxiv.org/abs/cond-mat/0611714} {arXiv:cond-mat/0611714}\BibitemShut
  {NoStop}%
\bibitem [{\citenamefont {Bauer}\ and\ \citenamefont
  {Falter}(2009)}]{Bauer2009}%
  \BibitemOpen
  \bibfield  {author} {\bibinfo {author} {\bibfnamefont {T.}~\bibnamefont
  {Bauer}}\ and\ \bibinfo {author} {\bibfnamefont {C.}~\bibnamefont {Falter}},\
  }\bibfield  {title} {\bibinfo {title} {\emph {Impact of dynamical screening
  on the phonon dynamics of metallic {La\s2CuO\s4}}},\ }\href
  {https://doi.org/10.1103/PhysRevB.80.094525} {\bibfield  {journal} {\bibinfo
  {journal} {Phys. Rev. B}\ }\textbf {\bibinfo {volume} {80}},\ \bibinfo
  {pages} {094525} (\bibinfo {year} {2009})},\ \Eprint
  {https://arxiv.org/abs/0808.2765} {arXiv:0808.2765}\BibitemShut {NoStop}%
\bibitem [{\citenamefont {Ponc\'e}\ \emph {et~al.}(2015)\citenamefont
  {Ponc\'e}, \citenamefont {Gillet}, \citenamefont {Laflamme~Janssen},
  \citenamefont {Marini}, \citenamefont {Verstraete},\ and\ \citenamefont
  {Gonze}}]{Ponce2015}%
  \BibitemOpen
  \bibfield  {author} {\bibinfo {author} {\bibfnamefont {S.}~\bibnamefont
  {Ponc\'e}}, \bibinfo {author} {\bibfnamefont {Y.}~\bibnamefont {Gillet}},
  \bibinfo {author} {\bibfnamefont {J.}~\bibnamefont {Laflamme~Janssen}},
  \bibinfo {author} {\bibfnamefont {A.}~\bibnamefont {Marini}}, \bibinfo
  {author} {\bibfnamefont {M.}~\bibnamefont {Verstraete}},\ and\ \bibinfo
  {author} {\bibfnamefont {X.}~\bibnamefont {Gonze}},\ }\bibfield  {title}
  {\bibinfo {title} {\emph {Temperature dependence of the electronic structure
  of semiconductors and insulators}},\ }\href
  {https://doi.org/10.1063/1.4927081} {\bibfield  {journal} {\bibinfo
  {journal} {J. Chem. Phys.}\ }\textbf {\bibinfo {volume} {143}},\ \bibinfo
  {pages} {102813} (\bibinfo {year} {2015})},\ \Eprint
  {https://arxiv.org/abs/1504.05992} {arXiv:1504.05992}\BibitemShut {NoStop}%
\bibitem [{\citenamefont {Caruso}\ \emph {et~al.}(2017)\citenamefont {Caruso},
  \citenamefont {Hoesch}, \citenamefont {Achatz}, \citenamefont {Serrano},
  \citenamefont {Krisch}, \citenamefont {Bustarret},\ and\ \citenamefont
  {Giustino}}]{Caruso2017}%
  \BibitemOpen
  \bibfield  {author} {\bibinfo {author} {\bibfnamefont {F.}~\bibnamefont
  {Caruso}}, \bibinfo {author} {\bibfnamefont {M.}~\bibnamefont {Hoesch}},
  \bibinfo {author} {\bibfnamefont {P.}~\bibnamefont {Achatz}}, \bibinfo
  {author} {\bibfnamefont {J.}~\bibnamefont {Serrano}}, \bibinfo {author}
  {\bibfnamefont {M.}~\bibnamefont {Krisch}}, \bibinfo {author} {\bibfnamefont
  {E.}~\bibnamefont {Bustarret}},\ and\ \bibinfo {author} {\bibfnamefont
  {F.}~\bibnamefont {Giustino}},\ }\bibfield  {title} {\bibinfo {title} {\emph
  {Nonadiabatic {Kohn} anomaly in heavily boron-doped diamond}},\ }\href
  {https://doi.org/10.1103/PhysRevLett.119.017001} {\bibfield  {journal}
  {\bibinfo  {journal} {Phys. Rev. Lett.}\ }\textbf {\bibinfo {volume} {119}},\
  \bibinfo {pages} {017001} (\bibinfo {year} {2017})},\ \Eprint
  {https://arxiv.org/abs/1706.02151} {arXiv:1706.02151}\BibitemShut {NoStop}%
\bibitem [{\citenamefont {Verdi}\ \emph {et~al.}(2017)\citenamefont {Verdi},
  \citenamefont {Caruso},\ and\ \citenamefont {Giustino}}]{Verdi2017}%
  \BibitemOpen
  \bibfield  {author} {\bibinfo {author} {\bibfnamefont {C.}~\bibnamefont
  {Verdi}}, \bibinfo {author} {\bibfnamefont {F.}~\bibnamefont {Caruso}},\ and\
  \bibinfo {author} {\bibfnamefont {F.}~\bibnamefont {Giustino}},\ }\bibfield
  {title} {\bibinfo {title} {\emph {Origin of the crossover from polarons to
  {Fermi} liquids in transition metal oxides}},\ }\href
  {https://doi.org/10.1038/ncomms15769} {\bibfield  {journal} {\bibinfo
  {journal} {Nat. Commun.}\ }\textbf {\bibinfo {volume} {8}},\ \bibinfo {pages}
  {15769} (\bibinfo {year} {2017})},\ \Eprint
  {https://arxiv.org/abs/1705.02967} {arXiv:1705.02967}\BibitemShut {NoStop}%
\bibitem [{\citenamefont {Miglio}\ \emph {et~al.}(2020)\citenamefont {Miglio},
  \citenamefont {Brousseau-Couture}, \citenamefont {Godbout}, \citenamefont
  {Antonius}, \citenamefont {Chan}, \citenamefont {Louie}, \citenamefont
  {C\^ot\'e}, \citenamefont {Giantomassi},\ and\ \citenamefont
  {Gonze}}]{Miglio2020}%
  \BibitemOpen
  \bibfield  {author} {\bibinfo {author} {\bibfnamefont {A.}~\bibnamefont
  {Miglio}}, \bibinfo {author} {\bibfnamefont {V.}~\bibnamefont
  {Brousseau-Couture}}, \bibinfo {author} {\bibfnamefont {E.}~\bibnamefont
  {Godbout}}, \bibinfo {author} {\bibfnamefont {G.}~\bibnamefont {Antonius}},
  \bibinfo {author} {\bibfnamefont {Y.-H.}\ \bibnamefont {Chan}}, \bibinfo
  {author} {\bibfnamefont {S.~G.}\ \bibnamefont {Louie}}, \bibinfo {author}
  {\bibfnamefont {M.}~\bibnamefont {C\^ot\'e}}, \bibinfo {author}
  {\bibfnamefont {M.}~\bibnamefont {Giantomassi}},\ and\ \bibinfo {author}
  {\bibfnamefont {X.}~\bibnamefont {Gonze}},\ }\bibfield  {title} {\bibinfo
  {title} {\emph {Predominance of non-adiabatic effects in zero-point
  renormalization of the electronic band gap}},\ }\href
  {https://doi.org/10.1038/s41524-020-00434-z} {\bibfield  {journal} {\bibinfo
  {journal} {npj Comput. Mater.}\ }\textbf {\bibinfo {volume} {6}},\ \bibinfo
  {pages} {1} (\bibinfo {year} {2020})},\ \Eprint
  {https://arxiv.org/abs/2011.12765} {arXiv:2011.12765}\BibitemShut {NoStop}%
\bibitem [{\citenamefont {Girotto}\ and\ \citenamefont
  {Novko}(2023)}]{Girotto2023}%
  \BibitemOpen
  \bibfield  {author} {\bibinfo {author} {\bibfnamefont {N.}~\bibnamefont
  {Girotto}}\ and\ \bibinfo {author} {\bibfnamefont {D.}~\bibnamefont
  {Novko}},\ }\bibfield  {title} {\bibinfo {title} {\emph {Dynamical
  renormalization of electron-phonon coupling in conventional
  superconductors}},\ }\href {https://doi.org/10.1103/PhysRevB.107.064310}
  {\bibfield  {journal} {\bibinfo  {journal} {Phys. Rev. B}\ }\textbf {\bibinfo
  {volume} {107}},\ \bibinfo {pages} {064310} (\bibinfo {year} {2023})},\
  \Eprint {https://arxiv.org/abs/2212.08117} {arXiv:2212.08117}\BibitemShut
  {NoStop}%
\bibitem [{\citenamefont {Kohn}(1959)}]{Kohn1959}%
  \BibitemOpen
  \bibfield  {author} {\bibinfo {author} {\bibfnamefont {W.}~\bibnamefont
  {Kohn}},\ }\bibfield  {title} {\bibinfo {title} {\emph {Image of the {Fermi}
  surface in the vibration spectrum of a metal}},\ }\href
  {https://doi.org/10.1103/PhysRevLett.2.393} {\bibfield  {journal} {\bibinfo
  {journal} {Phys. Rev. Lett.}\ }\textbf {\bibinfo {volume} {2}},\ \bibinfo
  {pages} {393} (\bibinfo {year} {1959})}\BibitemShut {NoStop}%
\bibitem [{\citenamefont {Hubbard}\ and\ \citenamefont
  {Flowers}(1963)}]{Hubbard1963}%
  \BibitemOpen
  \bibfield  {author} {\bibinfo {author} {\bibfnamefont {J.}~\bibnamefont
  {Hubbard}}\ and\ \bibinfo {author} {\bibfnamefont {B.~H.}\ \bibnamefont
  {Flowers}},\ }\bibfield  {title} {\bibinfo {title} {\emph {Electron
  correlations in narrow energy bands}},\ }\href
  {https://doi.org/10.1098/rspa.1963.0204} {\bibfield  {journal} {\bibinfo
  {journal} {Proc. R. Soc. Lond. A}\ }\textbf {\bibinfo {volume} {276}},\
  \bibinfo {pages} {238} (\bibinfo {year} {1963})}\BibitemShut {NoStop}%
\bibitem [{\citenamefont {Ge{\u\i}likman}(1971)}]{Geilikman1971}%
  \BibitemOpen
  \bibfield  {author} {\bibinfo {author} {\bibfnamefont {B.~T.}\ \bibnamefont
  {Ge{\u\i}likman}},\ }\bibfield  {title} {\bibinfo {title} {\emph {The
  adiabatic approximation and {Fr\"ohlich} model in the theory of metals}},\
  }\href {https://doi.org/10.1007/BF00628391} {\bibfield  {journal} {\bibinfo
  {journal} {J. Low Temp. Phys.}\ }\textbf {\bibinfo {volume} {4}},\ \bibinfo
  {pages} {189} (\bibinfo {year} {1971})}\BibitemShut {NoStop}%
\bibitem [{\citenamefont {Ko\c{c}er}\ \emph {et~al.}(2020)\citenamefont
  {Ko\c{c}er}, \citenamefont {Haule}, \citenamefont {Pascut},\ and\
  \citenamefont {Monserrat}}]{Kocer2020}%
  \BibitemOpen
  \bibfield  {author} {\bibinfo {author} {\bibfnamefont {C.~P.}\ \bibnamefont
  {Ko\c{c}er}}, \bibinfo {author} {\bibfnamefont {K.}~\bibnamefont {Haule}},
  \bibinfo {author} {\bibfnamefont {G.~L.}\ \bibnamefont {Pascut}},\ and\
  \bibinfo {author} {\bibfnamefont {B.}~\bibnamefont {Monserrat}},\ }\bibfield
  {title} {\bibinfo {title} {\emph {Efficient lattice dynamics calculations for
  correlated materials with {DFT+DMFT}}},\ }\href
  {https://doi.org/10.1103/PhysRevB.102.245104} {\bibfield  {journal} {\bibinfo
   {journal} {Phys. Rev. B}\ }\textbf {\bibinfo {volume} {102}},\ \bibinfo
  {pages} {245104} (\bibinfo {year} {2020})},\ \Eprint
  {https://arxiv.org/abs/2008.06058} {arXiv:2008.06058}\BibitemShut {NoStop}%
\bibitem [{\citenamefont {Brovman}\ and\ \citenamefont
  {Kagan}(1967)}]{Brovman1967}%
  \BibitemOpen
  \bibfield  {author} {\bibinfo {author} {\bibfnamefont {E.~G.}\ \bibnamefont
  {Brovman}}\ and\ \bibinfo {author} {\bibfnamefont {Y.}~\bibnamefont
  {Kagan}},\ }\bibfield  {title} {\bibinfo {title} {\emph {The phonon spectrum
  of metals}},\ }\href
  {http://www.jetp.ras.ru/cgi-bin/e/index/e/25/2/p365?a=list} {\bibfield
  {journal} {\bibinfo  {journal} {Sov. Phys. JETP}\ }\textbf {\bibinfo {volume}
  {25}},\ \bibinfo {pages} {365} (\bibinfo {year} {1967})}\BibitemShut
  {NoStop}%
\bibitem [{\citenamefont {Ipatova}\ and\ \citenamefont
  {Subashiev}(1974)}]{Ipatova1974}%
  \BibitemOpen
  \bibfield  {author} {\bibinfo {author} {\bibfnamefont {I.~P.}\ \bibnamefont
  {Ipatova}}\ and\ \bibinfo {author} {\bibfnamefont {A.~V.}\ \bibnamefont
  {Subashiev}},\ }\bibfield  {title} {\bibinfo {title} {\emph {Long-wave
  optical-phonon spectrum in metals and heavily doped semiconductors}},\ }\href
  {http://www.jetp.ras.ru/cgi-bin/e/index/r/66/2/p722?a=list} {\bibfield
  {journal} {\bibinfo  {journal} {Sov. Phys. JETP}\ }\textbf {\bibinfo {volume}
  {39}},\ \bibinfo {pages} {349} (\bibinfo {year} {1974})}\BibitemShut
  {NoStop}%
\bibitem [{\citenamefont {Falter}\ and\ \citenamefont
  {Selmke}(1981)}]{Falter1981}%
  \BibitemOpen
  \bibfield  {author} {\bibinfo {author} {\bibfnamefont {C.}~\bibnamefont
  {Falter}}\ and\ \bibinfo {author} {\bibfnamefont {M.}~\bibnamefont
  {Selmke}},\ }\bibfield  {title} {\bibinfo {title} {\emph {Renormalization of
  the dielectric response with applications to effective ion interactions and
  phonons}},\ }\href {https://doi.org/10.1103/PhysRevB.24.586} {\bibfield
  {journal} {\bibinfo  {journal} {Phys. Rev. B}\ }\textbf {\bibinfo {volume}
  {24}},\ \bibinfo {pages} {586} (\bibinfo {year} {1981})}\BibitemShut
  {NoStop}%
\bibitem [{\citenamefont {Falter}(1988)}]{Falter1988}%
  \BibitemOpen
  \bibfield  {author} {\bibinfo {author} {\bibfnamefont {C.}~\bibnamefont
  {Falter}},\ }\bibfield  {title} {\bibinfo {title} {\emph {A unifying approach
  to lattice dynamical and electronic properties of solids}},\ }\href
  {https://doi.org/10.1016/0370-1573(88)90058-0} {\bibfield  {journal}
  {\bibinfo  {journal} {Phys. Rep.}\ }\textbf {\bibinfo {volume} {164}},\
  \bibinfo {pages} {1} (\bibinfo {year} {1988})}\BibitemShut {NoStop}%
\bibitem [{\citenamefont {Paleari}\ and\ \citenamefont
  {Marini}(2021{\natexlab{a}})}]{Paleari2021a}%
  \BibitemOpen
  \bibfield  {author} {\bibinfo {author} {\bibfnamefont {F.}~\bibnamefont
  {Paleari}}\ and\ \bibinfo {author} {\bibfnamefont {A.}~\bibnamefont
  {Marini}},\ }\bibfield  {title} {\bibinfo {title} {\emph {Electron-phonon
  interaction without overscreening: A strategy for first-principles
  modelling}},\ }\Eprint {https://arxiv.org/abs/2105.09828} {arXiv:2105.09828}
  (\bibinfo {year} {2021}{\natexlab{a}})\BibitemShut {NoStop}%
\bibitem [{\citenamefont {Paleari}\ and\ \citenamefont
  {Marini}(2021{\natexlab{b}})}]{Paleari2021b}%
  \BibitemOpen
  \bibfield  {author} {\bibinfo {author} {\bibfnamefont {F.}~\bibnamefont
  {Paleari}}\ and\ \bibinfo {author} {\bibfnamefont {A.}~\bibnamefont
  {Marini}},\ }\bibfield  {title} {\bibinfo {title} {\emph {Overscreening-free
  electron-phonon interaction in realistic materials}},\ }\Eprint
  {https://arxiv.org/abs/2105.09823} {arXiv:2105.09823} (\bibinfo {year}
  {2021}{\natexlab{b}})\BibitemShut {NoStop}%
\bibitem [{\citenamefont {Marini}(2023{\natexlab{a}})}]{Marini2023a}%
  \BibitemOpen
  \bibfield  {author} {\bibinfo {author} {\bibfnamefont {A.}~\bibnamefont
  {Marini}},\ }\bibfield  {title} {\bibinfo {title} {\emph {Equilibrium and
  out-of-equilibrium realistic phonon self-energy free from overscreening}},\
  }\href {https://doi.org/10.1103/PhysRevB.107.024305} {\bibfield  {journal}
  {\bibinfo  {journal} {Phys. Rev. B}\ }\textbf {\bibinfo {volume} {107}},\
  \bibinfo {pages} {024305} (\bibinfo {year} {2023}{\natexlab{a}})},\ \Eprint
  {https://arxiv.org/abs/2211.02573} {arXiv:2211.02573}\BibitemShut {NoStop}%
\bibitem [{\citenamefont {Aryasetiawan}\ \emph {et~al.}(2004)\citenamefont
  {Aryasetiawan}, \citenamefont {Imada}, \citenamefont {Georges}, \citenamefont
  {Kotliar}, \citenamefont {Biermann},\ and\ \citenamefont
  {Lichtenstein}}]{Aryasetiawan2004}%
  \BibitemOpen
  \bibfield  {author} {\bibinfo {author} {\bibfnamefont {F.}~\bibnamefont
  {Aryasetiawan}}, \bibinfo {author} {\bibfnamefont {M.}~\bibnamefont {Imada}},
  \bibinfo {author} {\bibfnamefont {A.}~\bibnamefont {Georges}}, \bibinfo
  {author} {\bibfnamefont {G.}~\bibnamefont {Kotliar}}, \bibinfo {author}
  {\bibfnamefont {S.}~\bibnamefont {Biermann}},\ and\ \bibinfo {author}
  {\bibfnamefont {A.~I.}\ \bibnamefont {Lichtenstein}},\ }\bibfield  {title}
  {\bibinfo {title} {\emph {Frequency-dependent local interactions and
  low-energy effective models from electronic structure calculations}},\ }\href
  {https://doi.org/10.1103/PhysRevB.70.195104} {\bibfield  {journal} {\bibinfo
  {journal} {Phys. Rev. B}\ }\textbf {\bibinfo {volume} {70}},\ \bibinfo
  {pages} {195104} (\bibinfo {year} {2004})},\ \Eprint
  {https://arxiv.org/abs/cond-mat/0401620} {arXiv:cond-mat/0401620}\BibitemShut
  {NoStop}%
\bibitem [{\citenamefont {van Loon}\ \emph
  {et~al.}(2021{\natexlab{a}})\citenamefont {van Loon}, \citenamefont
  {Berges},\ and\ \citenamefont {Wehling}}]{vanLoon2021a}%
  \BibitemOpen
  \bibfield  {author} {\bibinfo {author} {\bibfnamefont {E.~G. C.~P.}\
  \bibnamefont {van Loon}}, \bibinfo {author} {\bibfnamefont {J.}~\bibnamefont
  {Berges}},\ and\ \bibinfo {author} {\bibfnamefont {T.~O.}\ \bibnamefont
  {Wehling}},\ }\bibfield  {title} {\bibinfo {title} {\emph {Downfolding
  approaches to electron-ion coupling: Constrained density-functional
  perturbation theory for molecules}},\ }\href
  {https://doi.org/10.1103/PhysRevB.103.205103} {\bibfield  {journal} {\bibinfo
   {journal} {Phys. Rev. B}\ }\textbf {\bibinfo {volume} {103}},\ \bibinfo
  {pages} {205103} (\bibinfo {year} {2021}{\natexlab{a}})},\ \Eprint
  {https://arxiv.org/abs/2102.10072} {arXiv:2102.10072}\BibitemShut {NoStop}%
\bibitem [{\citenamefont {van Loon}\ \emph
  {et~al.}(2021{\natexlab{b}})\citenamefont {van Loon}, \citenamefont
  {R\"osner}, \citenamefont {Katsnelson},\ and\ \citenamefont
  {Wehling}}]{vanLoon2021b}%
  \BibitemOpen
  \bibfield  {author} {\bibinfo {author} {\bibfnamefont {E.~G. C.~P.}\
  \bibnamefont {van Loon}}, \bibinfo {author} {\bibfnamefont {M.}~\bibnamefont
  {R\"osner}}, \bibinfo {author} {\bibfnamefont {M.~I.}\ \bibnamefont
  {Katsnelson}},\ and\ \bibinfo {author} {\bibfnamefont {T.~O.}\ \bibnamefont
  {Wehling}},\ }\bibfield  {title} {\bibinfo {title} {\emph {Random phase
  approximation for gapped systems: Role of vertex corrections and
  applicability of the constrained random phase approximation}},\ }\href
  {https://doi.org/10.1103/PhysRevB.104.045134} {\bibfield  {journal} {\bibinfo
   {journal} {Phys. Rev. B}\ }\textbf {\bibinfo {volume} {104}},\ \bibinfo
  {pages} {045134} (\bibinfo {year} {2021}{\natexlab{b}})},\ \Eprint
  {https://arxiv.org/abs/2103.04419} {arXiv:2103.04419}\BibitemShut {NoStop}%
\bibitem [{\citenamefont {Imada}\ and\ \citenamefont
  {Miyake}(2010)}]{Imada2010}%
  \BibitemOpen
  \bibfield  {author} {\bibinfo {author} {\bibfnamefont {M.}~\bibnamefont
  {Imada}}\ and\ \bibinfo {author} {\bibfnamefont {T.}~\bibnamefont {Miyake}},\
  }\bibfield  {title} {\bibinfo {title} {\emph {Electronic structure
  calculation by first principles for strongly correlated electron systems}},\
  }\href {https://doi.org/10.1143/JPSJ.79.112001} {\bibfield  {journal}
  {\bibinfo  {journal} {J. Phys. Soc. Jpn.}\ }\textbf {\bibinfo {volume}
  {79}},\ \bibinfo {pages} {112001} (\bibinfo {year} {2010})},\ \Eprint
  {https://arxiv.org/abs/1009.3851} {arXiv:1009.3851}\BibitemShut {NoStop}%
\bibitem [{\citenamefont {Giovannetti}\ \emph {et~al.}(2014)\citenamefont
  {Giovannetti}, \citenamefont {Casula}, \citenamefont {Werner}, \citenamefont
  {Mauri},\ and\ \citenamefont {Capone}}]{Giovannetti2014}%
  \BibitemOpen
  \bibfield  {author} {\bibinfo {author} {\bibfnamefont {G.}~\bibnamefont
  {Giovannetti}}, \bibinfo {author} {\bibfnamefont {M.}~\bibnamefont {Casula}},
  \bibinfo {author} {\bibfnamefont {P.}~\bibnamefont {Werner}}, \bibinfo
  {author} {\bibfnamefont {F.}~\bibnamefont {Mauri}},\ and\ \bibinfo {author}
  {\bibfnamefont {M.}~\bibnamefont {Capone}},\ }\bibfield  {title} {\bibinfo
  {title} {\emph {Downfolding electron-phonon {Hamiltonians} from ab initio
  calculations: Application to {K\s3} picene}},\ }\href
  {https://doi.org/10.1103/PhysRevB.90.115435} {\bibfield  {journal} {\bibinfo
  {journal} {Phys. Rev. B}\ }\textbf {\bibinfo {volume} {90}},\ \bibinfo
  {pages} {115435} (\bibinfo {year} {2014})},\ \Eprint
  {https://arxiv.org/abs/1406.4108} {arXiv:1406.4108}\BibitemShut {NoStop}%
\bibitem [{\citenamefont {Aryasetiawan}\ \emph {et~al.}(2006)\citenamefont
  {Aryasetiawan}, \citenamefont {Karlsson}, \citenamefont {Jepsen},\ and\
  \citenamefont {Sch\"onberger}}]{Aryasetiawan2006}%
  \BibitemOpen
  \bibfield  {author} {\bibinfo {author} {\bibfnamefont {F.}~\bibnamefont
  {Aryasetiawan}}, \bibinfo {author} {\bibfnamefont {K.}~\bibnamefont
  {Karlsson}}, \bibinfo {author} {\bibfnamefont {O.}~\bibnamefont {Jepsen}},\
  and\ \bibinfo {author} {\bibfnamefont {U.}~\bibnamefont {Sch\"onberger}},\
  }\bibfield  {title} {\bibinfo {title} {\emph {Calculations of {Hubbard} {$U$}
  from first-principles}},\ }\href {https://doi.org/10.1103/PhysRevB.74.125106}
  {\bibfield  {journal} {\bibinfo  {journal} {Phys. Rev. B}\ }\textbf {\bibinfo
  {volume} {74}},\ \bibinfo {pages} {125106} (\bibinfo {year} {2006})},\
  \Eprint {https://arxiv.org/abs/cond-mat/0603138}
  {arXiv:cond-mat/0603138}\BibitemShut {NoStop}%
\bibitem [{\citenamefont {Friedrich}\ \emph {et~al.}(2010)\citenamefont
  {Friedrich}, \citenamefont {Bl\"ugel},\ and\ \citenamefont
  {Schindlmayr}}]{Friedrich2010}%
  \BibitemOpen
  \bibfield  {author} {\bibinfo {author} {\bibfnamefont {C.}~\bibnamefont
  {Friedrich}}, \bibinfo {author} {\bibfnamefont {S.}~\bibnamefont
  {Bl\"ugel}},\ and\ \bibinfo {author} {\bibfnamefont {A.}~\bibnamefont
  {Schindlmayr}},\ }\bibfield  {title} {\bibinfo {title} {\emph {Efficient
  implementation of the {$GW$} approximation within the all-electron {FLAPW}
  method}},\ }\href {https://doi.org/10.1103/PhysRevB.81.125102} {\bibfield
  {journal} {\bibinfo  {journal} {Phys. Rev. B}\ }\textbf {\bibinfo {volume}
  {81}},\ \bibinfo {pages} {125102} (\bibinfo {year} {2010})},\ \Eprint
  {https://arxiv.org/abs/1003.0316} {arXiv:1003.0316}\BibitemShut {NoStop}%
\bibitem [{\citenamefont {\c{S}a\c{s}{\i}o\u{g}lu}\ \emph
  {et~al.}(2013)\citenamefont {\c{S}a\c{s}{\i}o\u{g}lu}, \citenamefont
  {Galanakis}, \citenamefont {Friedrich},\ and\ \citenamefont
  {Bl\"ugel}}]{Sasioglu2013}%
  \BibitemOpen
  \bibfield  {author} {\bibinfo {author} {\bibfnamefont {E.}~\bibnamefont
  {\c{S}a\c{s}{\i}o\u{g}lu}}, \bibinfo {author} {\bibfnamefont
  {I.}~\bibnamefont {Galanakis}}, \bibinfo {author} {\bibfnamefont
  {C.}~\bibnamefont {Friedrich}},\ and\ \bibinfo {author} {\bibfnamefont
  {S.}~\bibnamefont {Bl\"ugel}},\ }\bibfield  {title} {\bibinfo {title} {\emph
  {Ab initio calculation of the effective on-site {Coulomb} interaction
  parameters for half-metallic magnets}},\ }\href
  {https://doi.org/10.1103/PhysRevB.88.134402} {\bibfield  {journal} {\bibinfo
  {journal} {Phys. Rev. B}\ }\textbf {\bibinfo {volume} {88}},\ \bibinfo
  {pages} {134402} (\bibinfo {year} {2013})},\ \Eprint
  {https://arxiv.org/abs/1309.6657} {arXiv:1309.6657}\BibitemShut {NoStop}%
\bibitem [{\citenamefont {Amadon}\ \emph {et~al.}(2014)\citenamefont {Amadon},
  \citenamefont {Applencourt},\ and\ \citenamefont {Bruneval}}]{Amadon2014}%
  \BibitemOpen
  \bibfield  {author} {\bibinfo {author} {\bibfnamefont {B.}~\bibnamefont
  {Amadon}}, \bibinfo {author} {\bibfnamefont {T.}~\bibnamefont
  {Applencourt}},\ and\ \bibinfo {author} {\bibfnamefont {F.}~\bibnamefont
  {Bruneval}},\ }\bibfield  {title} {\bibinfo {title} {\emph {Screened
  {Coulomb} interaction calculations: {CRPA} implementation and applications to
  dynamical screening and self-consistency in uranium dioxide and cerium}},\
  }\href {https://doi.org/10.1103/PhysRevB.89.125110} {\bibfield  {journal}
  {\bibinfo  {journal} {Phys. Rev. B}\ }\textbf {\bibinfo {volume} {89}},\
  \bibinfo {pages} {125110} (\bibinfo {year} {2014})},\ \Eprint
  {https://arxiv.org/abs/1403.5386} {arXiv:1403.5386}\BibitemShut {NoStop}%
\bibitem [{\citenamefont {Kaltak}(2015)}]{Kaltak2015}%
  \BibitemOpen
  \bibfield  {author} {\bibinfo {author} {\bibfnamefont {M.}~\bibnamefont
  {Kaltak}},\ }\emph {\bibinfo {title} {Merging {GW} with {DMFT}}},\ \href
  {https://doi.org/10.25365/thesis.38099} {\bibinfo {type} {Dissertation}},\
  \bibinfo  {school} {Universit\"at Wien} (\bibinfo {year} {2015})\BibitemShut
  {NoStop}%
\bibitem [{\citenamefont {Nakamura}\ \emph {et~al.}(2021)\citenamefont
  {Nakamura}, \citenamefont {Yoshimoto}, \citenamefont {Nomura}, \citenamefont
  {Tadano}, \citenamefont {Kawamura}, \citenamefont {Kosugi}, \citenamefont
  {Yoshimi}, \citenamefont {Misawa},\ and\ \citenamefont
  {Motoyama}}]{Nakamura2021}%
  \BibitemOpen
  \bibfield  {author} {\bibinfo {author} {\bibfnamefont {K.}~\bibnamefont
  {Nakamura}}, \bibinfo {author} {\bibfnamefont {Y.}~\bibnamefont {Yoshimoto}},
  \bibinfo {author} {\bibfnamefont {Y.}~\bibnamefont {Nomura}}, \bibinfo
  {author} {\bibfnamefont {T.}~\bibnamefont {Tadano}}, \bibinfo {author}
  {\bibfnamefont {M.}~\bibnamefont {Kawamura}}, \bibinfo {author}
  {\bibfnamefont {T.}~\bibnamefont {Kosugi}}, \bibinfo {author} {\bibfnamefont
  {K.}~\bibnamefont {Yoshimi}}, \bibinfo {author} {\bibfnamefont
  {T.}~\bibnamefont {Misawa}},\ and\ \bibinfo {author} {\bibfnamefont
  {Y.}~\bibnamefont {Motoyama}},\ }\bibfield  {title} {\bibinfo {title} {\emph
  {{RESPACK}: An ab initio tool for derivation of effective low-energy model of
  material}},\ }\href {https://doi.org/10.1016/j.cpc.2020.107781} {\bibfield
  {journal} {\bibinfo  {journal} {Comput. Phys. Commun.}\ }\textbf {\bibinfo
  {volume} {261}},\ \bibinfo {pages} {107781} (\bibinfo {year} {2021})},\
  \Eprint {https://arxiv.org/abs/2001.02351} {arXiv:2001.02351}\BibitemShut
  {NoStop}%
\bibitem [{\citenamefont {Berges}\ \emph {et~al.}(2020)\citenamefont {Berges},
  \citenamefont {van Loon}, \citenamefont {Schobert}, \citenamefont
  {R\"osner},\ and\ \citenamefont {Wehling}}]{Berges2020a}%
  \BibitemOpen
  \bibfield  {author} {\bibinfo {author} {\bibfnamefont {J.}~\bibnamefont
  {Berges}}, \bibinfo {author} {\bibfnamefont {E.~G. C.~P.}\ \bibnamefont {van
  Loon}}, \bibinfo {author} {\bibfnamefont {A.}~\bibnamefont {Schobert}},
  \bibinfo {author} {\bibfnamefont {M.}~\bibnamefont {R\"osner}},\ and\
  \bibinfo {author} {\bibfnamefont {T.~O.}\ \bibnamefont {Wehling}},\
  }\bibfield  {title} {\bibinfo {title} {\emph {Ab initio phonon self-energies
  and fluctuation diagnostics of phonon anomalies: Lattice instabilities from
  {Dirac} pseudospin physics in transition metal dichalcogenides}},\ }\href
  {https://doi.org/10.1103/PhysRevB.101.155107} {\bibfield  {journal} {\bibinfo
   {journal} {Phys. Rev. B}\ }\textbf {\bibinfo {volume} {101}},\ \bibinfo
  {pages} {155107} (\bibinfo {year} {2020})},\ \Eprint
  {https://arxiv.org/abs/1911.02450} {arXiv:1911.02450}\BibitemShut {NoStop}%
\bibitem [{\citenamefont {Novko}(2020)}]{Novko2020a}%
  \BibitemOpen
  \bibfield  {author} {\bibinfo {author} {\bibfnamefont {D.}~\bibnamefont
  {Novko}},\ }\bibfield  {title} {\bibinfo {title} {\emph {Broken adiabaticity
  induced by {Lifshitz} transition in {MoS\s2} and {WS\s2} single layers}},\
  }\href {https://doi.org/10.1038/s42005-020-0299-1} {\bibfield  {journal}
  {\bibinfo  {journal} {Commun. Phys.}\ }\textbf {\bibinfo {volume} {3}},\
  \bibinfo {pages} {30} (\bibinfo {year} {2020})},\ \Eprint
  {https://arxiv.org/abs/1907.04766} {arXiv:1907.04766}\BibitemShut {NoStop}%
\bibitem [{\citenamefont {Reichardt}(2018)}]{Reichardt2018}%
  \BibitemOpen
  \bibfield  {author} {\bibinfo {author} {\bibfnamefont {S.}~\bibnamefont
  {Reichardt}},\ }\emph {\bibinfo {title} {Many-Body Perturbation Theory
  Approach to {Raman} Spectroscopy and Its Application to {2D} Materials}},\
  \href {https://orbilu.uni.lu/handle/10993/35610} {\bibinfo {type}
  {Dissertation}},\ \bibinfo  {school} {University of Luxembourg} (\bibinfo
  {year} {2018})\BibitemShut {NoStop}%
\bibitem [{\citenamefont {Allen}(1972)}]{Allen1972}%
  \BibitemOpen
  \bibfield  {author} {\bibinfo {author} {\bibfnamefont {P.~B.}\ \bibnamefont
  {Allen}},\ }\bibfield  {title} {\bibinfo {title} {\emph {Neutron spectroscopy
  of superconductors}},\ }\href {https://doi.org/10.1103/PhysRevB.6.2577}
  {\bibfield  {journal} {\bibinfo  {journal} {Phys. Rev. B}\ }\textbf {\bibinfo
  {volume} {6}},\ \bibinfo {pages} {2577} (\bibinfo {year} {1972})}\BibitemShut
  {NoStop}%
\bibitem [{\citenamefont {Giustino}\ \emph
  {et~al.}(2007{\natexlab{a}})\citenamefont {Giustino}, \citenamefont {Yates},
  \citenamefont {Souza}, \citenamefont {Cohen},\ and\ \citenamefont
  {Louie}}]{Giustino2007a}%
  \BibitemOpen
  \bibfield  {author} {\bibinfo {author} {\bibfnamefont {F.}~\bibnamefont
  {Giustino}}, \bibinfo {author} {\bibfnamefont {J.~R.}\ \bibnamefont {Yates}},
  \bibinfo {author} {\bibfnamefont {I.}~\bibnamefont {Souza}}, \bibinfo
  {author} {\bibfnamefont {M.~L.}\ \bibnamefont {Cohen}},\ and\ \bibinfo
  {author} {\bibfnamefont {S.~G.}\ \bibnamefont {Louie}},\ }\bibfield  {title}
  {\bibinfo {title} {\emph {Electron-phonon interaction via electronic and
  lattice {Wannier} functions: Superconductivity in boron-doped diamond
  reexamined}},\ }\href {https://doi.org/10.1103/PhysRevLett.98.047005}
  {\bibfield  {journal} {\bibinfo  {journal} {Phys. Rev. Lett.}\ }\textbf
  {\bibinfo {volume} {98}},\ \bibinfo {pages} {047005} (\bibinfo {year}
  {2007}{\natexlab{a}})}\BibitemShut {NoStop}%
\bibitem [{\citenamefont {Novko}(2018)}]{Novko2018}%
  \BibitemOpen
  \bibfield  {author} {\bibinfo {author} {\bibfnamefont {D.}~\bibnamefont
  {Novko}},\ }\bibfield  {title} {\bibinfo {title} {\emph {Nonadiabatic
  coupling effects in {MgB\s2} reexamined}},\ }\href
  {https://doi.org/10.1103/PhysRevB.98.041112} {\bibfield  {journal} {\bibinfo
  {journal} {Phys. Rev. B}\ }\textbf {\bibinfo {volume} {98}},\ \bibinfo
  {pages} {041112(R)} (\bibinfo {year} {2018})},\ \Eprint
  {https://arxiv.org/abs/1806.03212} {arXiv:1806.03212}\BibitemShut {NoStop}%
\bibitem [{\citenamefont {Novko}\ \emph {et~al.}(2020)\citenamefont {Novko},
  \citenamefont {Caruso}, \citenamefont {Draxl},\ and\ \citenamefont
  {Cappelluti}}]{Novko2020b}%
  \BibitemOpen
  \bibfield  {author} {\bibinfo {author} {\bibfnamefont {D.}~\bibnamefont
  {Novko}}, \bibinfo {author} {\bibfnamefont {F.}~\bibnamefont {Caruso}},
  \bibinfo {author} {\bibfnamefont {C.}~\bibnamefont {Draxl}},\ and\ \bibinfo
  {author} {\bibfnamefont {E.}~\bibnamefont {Cappelluti}},\ }\bibfield  {title}
  {\bibinfo {title} {\emph {Ultrafast hot phonon dynamics in {MgB\s2} driven by
  anisotropic electron-phonon coupling}},\ }\href
  {https://doi.org/10.1103/PhysRevLett.124.077001} {\bibfield  {journal}
  {\bibinfo  {journal} {Phys. Rev. Lett.}\ }\textbf {\bibinfo {volume} {124}},\
  \bibinfo {pages} {077001} (\bibinfo {year} {2020})},\ \Eprint
  {https://arxiv.org/abs/1904.03062} {arXiv:1904.03062}\BibitemShut {NoStop}%
\bibitem [{\citenamefont {Lihm}\ and\ \citenamefont {Park}(2021)}]{Lihm2021}%
  \BibitemOpen
  \bibfield  {author} {\bibinfo {author} {\bibfnamefont {J.-M.}\ \bibnamefont
  {Lihm}}\ and\ \bibinfo {author} {\bibfnamefont {C.-H.}\ \bibnamefont
  {Park}},\ }\bibfield  {title} {\bibinfo {title} {\emph {Wannier function
  perturbation theory: Localized representation and interpolation of wave
  function perturbation}},\ }\href {https://doi.org/10.1103/PhysRevX.11.041053}
  {\bibfield  {journal} {\bibinfo  {journal} {Phys. Rev. X}\ }\textbf {\bibinfo
  {volume} {11}},\ \bibinfo {pages} {041053} (\bibinfo {year} {2021})},\
  \Eprint {https://arxiv.org/abs/2112.10778} {arXiv:2112.10778}\BibitemShut
  {NoStop}%
\bibitem [{\citenamefont {Prasad~Kafle}\ \emph {et~al.}(2020)\citenamefont
  {Prasad~Kafle}, \citenamefont {Heil}, \citenamefont {Paudyal},\ and\
  \citenamefont {R.~Margine}}]{PrasadKafle2020}%
  \BibitemOpen
  \bibfield  {author} {\bibinfo {author} {\bibfnamefont {G.}~\bibnamefont
  {Prasad~Kafle}}, \bibinfo {author} {\bibfnamefont {C.}~\bibnamefont {Heil}},
  \bibinfo {author} {\bibfnamefont {H.}~\bibnamefont {Paudyal}},\ and\ \bibinfo
  {author} {\bibfnamefont {E.}~\bibnamefont {R.~Margine}},\ }\bibfield  {title}
  {\bibinfo {title} {\emph {Electronic, vibrational, and electron-phonon
  coupling properties in {SnSe\s2} and {SnS\s2} under pressure}},\ }\href
  {https://doi.org/10.1039/D0TC04356G} {\bibfield  {journal} {\bibinfo
  {journal} {J. Mater. Chem. C}\ }\textbf {\bibinfo {volume} {8}},\ \bibinfo
  {pages} {16404} (\bibinfo {year} {2020})},\ \Eprint
  {https://arxiv.org/abs/2010.11493} {arXiv:2010.11493}\BibitemShut {NoStop}%
\bibitem [{\citenamefont {Savrasov}\ and\ \citenamefont
  {Kotliar}(2003)}]{Savrasov2003}%
  \BibitemOpen
  \bibfield  {author} {\bibinfo {author} {\bibfnamefont {S.~Y.}\ \bibnamefont
  {Savrasov}}\ and\ \bibinfo {author} {\bibfnamefont {G.}~\bibnamefont
  {Kotliar}},\ }\bibfield  {title} {\bibinfo {title} {\emph {Linear response
  calculations of lattice dynamics in strongly correlated systems}},\ }\href
  {https://doi.org/10.1103/PhysRevLett.90.056401} {\bibfield  {journal}
  {\bibinfo  {journal} {Phys. Rev. Lett.}\ }\textbf {\bibinfo {volume} {90}},\
  \bibinfo {pages} {056401} (\bibinfo {year} {2003})},\ \Eprint
  {https://arxiv.org/abs/cond-mat/0206136} {arXiv:cond-mat/0206136}\BibitemShut
  {NoStop}%
\bibitem [{Note1()}]{Note1}%
  \BibitemOpen
  \bibinfo {note} {Francesco Mauri (private communication).}\BibitemShut
  {Stop}%
\bibitem [{\citenamefont {Engelsberg}\ and\ \citenamefont
  {Schrieffer}(1963)}]{Engelsberg1963}%
  \BibitemOpen
  \bibfield  {author} {\bibinfo {author} {\bibfnamefont {S.}~\bibnamefont
  {Engelsberg}}\ and\ \bibinfo {author} {\bibfnamefont {J.~R.}\ \bibnamefont
  {Schrieffer}},\ }\bibfield  {title} {\bibinfo {title} {\emph {Coupled
  electron-phonon system}},\ }\href {https://doi.org/10.1103/PhysRev.131.993}
  {\bibfield  {journal} {\bibinfo  {journal} {Phys. Rev.}\ }\textbf {\bibinfo
  {volume} {131}},\ \bibinfo {pages} {993} (\bibinfo {year}
  {1963})}\BibitemShut {NoStop}%
\bibitem [{\citenamefont {Saitta}\ \emph {et~al.}(2008)\citenamefont {Saitta},
  \citenamefont {Lazzeri}, \citenamefont {Calandra},\ and\ \citenamefont
  {Mauri}}]{Saitta2008}%
  \BibitemOpen
  \bibfield  {author} {\bibinfo {author} {\bibfnamefont {A.~M.}\ \bibnamefont
  {Saitta}}, \bibinfo {author} {\bibfnamefont {M.}~\bibnamefont {Lazzeri}},
  \bibinfo {author} {\bibfnamefont {M.}~\bibnamefont {Calandra}},\ and\
  \bibinfo {author} {\bibfnamefont {F.}~\bibnamefont {Mauri}},\ }\bibfield
  {title} {\bibinfo {title} {\emph {Giant nonadiabatic effects in layer metals:
  Raman spectra of intercalated graphite explained}},\ }\href
  {https://doi.org/10.1103/PhysRevLett.100.226401} {\bibfield  {journal}
  {\bibinfo  {journal} {Phys. Rev. Lett.}\ }\textbf {\bibinfo {volume} {100}},\
  \bibinfo {pages} {226401} (\bibinfo {year} {2008})},\ \Eprint
  {https://arxiv.org/abs/0802.4426} {arXiv:0802.4426}\BibitemShut {NoStop}%
\bibitem [{\citenamefont {Giannozzi}\ \emph {et~al.}(2009)\citenamefont
  {Giannozzi} \emph {et~al.}}]{Giannozzi2009}%
  \BibitemOpen
  \bibfield  {author} {\bibinfo {author} {\bibfnamefont {P.}~\bibnamefont
  {Giannozzi}} \emph {et~al.},\ }\bibfield  {title} {\bibinfo {title} {\emph
  {\textsc{Quantum ESPRESSO}: A modular and open-source software project for
  quantum simulations of materials}},\ }\href
  {https://doi.org/10.1088/0953-8984/21/39/395502} {\bibfield  {journal}
  {\bibinfo  {journal} {J. Phys. Condens. Matter}\ }\textbf {\bibinfo {volume}
  {21}},\ \bibinfo {pages} {395502} (\bibinfo {year} {2009})},\ \Eprint
  {https://arxiv.org/abs/0906.2569} {arXiv:0906.2569}\BibitemShut {NoStop}%
\bibitem [{\citenamefont {Giannozzi}\ \emph {et~al.}(2017)\citenamefont
  {Giannozzi} \emph {et~al.}}]{Giannozzi2017}%
  \BibitemOpen
  \bibfield  {author} {\bibinfo {author} {\bibfnamefont {P.}~\bibnamefont
  {Giannozzi}} \emph {et~al.},\ }\bibfield  {title} {\bibinfo {title} {\emph
  {Advanced capabilities for materials modelling with \textsc{Quantum
  ESPRESSO}}},\ }\href {https://doi.org/10.1088/1361-648X/aa8f79} {\bibfield
  {journal} {\bibinfo  {journal} {J. Phys. Condens. Matter}\ }\textbf {\bibinfo
  {volume} {29}},\ \bibinfo {pages} {465901} (\bibinfo {year} {2017})},\
  \Eprint {https://arxiv.org/abs/1709.10010} {arXiv:1709.10010}\BibitemShut
  {NoStop}%
\bibitem [{\citenamefont {Giannozzi}\ \emph {et~al.}(2020)\citenamefont
  {Giannozzi} \emph {et~al.}}]{Giannozzi2020}%
  \BibitemOpen
  \bibfield  {author} {\bibinfo {author} {\bibfnamefont {P.}~\bibnamefont
  {Giannozzi}} \emph {et~al.},\ }\bibfield  {title} {\bibinfo {title} {\emph
  {\textsc{Quantum ESPRESSO} toward the exascale}},\ }\href
  {https://doi.org/10.1063/5.0005082} {\bibfield  {journal} {\bibinfo
  {journal} {J. Chem. Phys.}\ }\textbf {\bibinfo {volume} {152}},\ \bibinfo
  {pages} {154105} (\bibinfo {year} {2020})},\ \Eprint
  {https://arxiv.org/abs/2104.10502} {arXiv:2104.10502}\BibitemShut {NoStop}%
\bibitem [{\citenamefont {Giustino}\ \emph
  {et~al.}(2007{\natexlab{b}})\citenamefont {Giustino}, \citenamefont {Cohen},\
  and\ \citenamefont {Louie}}]{Giustino2007b}%
  \BibitemOpen
  \bibfield  {author} {\bibinfo {author} {\bibfnamefont {F.}~\bibnamefont
  {Giustino}}, \bibinfo {author} {\bibfnamefont {M.~L.}\ \bibnamefont
  {Cohen}},\ and\ \bibinfo {author} {\bibfnamefont {S.~G.}\ \bibnamefont
  {Louie}},\ }\bibfield  {title} {\bibinfo {title} {\emph {Electron-phonon
  interaction using {Wannier} functions}},\ }\href
  {https://doi.org/10.1103/PhysRevB.76.165108} {\bibfield  {journal} {\bibinfo
  {journal} {Phys. Rev. B}\ }\textbf {\bibinfo {volume} {76}},\ \bibinfo
  {pages} {165108} (\bibinfo {year} {2007}{\natexlab{b}})}\BibitemShut
  {NoStop}%
\bibitem [{\citenamefont {Noffsinger}\ \emph {et~al.}(2010)\citenamefont
  {Noffsinger}, \citenamefont {Giustino}, \citenamefont {Malone}, \citenamefont
  {Park}, \citenamefont {Louie},\ and\ \citenamefont {Cohen}}]{Noffsinger2010}%
  \BibitemOpen
  \bibfield  {author} {\bibinfo {author} {\bibfnamefont {J.}~\bibnamefont
  {Noffsinger}}, \bibinfo {author} {\bibfnamefont {F.}~\bibnamefont
  {Giustino}}, \bibinfo {author} {\bibfnamefont {B.~D.}\ \bibnamefont
  {Malone}}, \bibinfo {author} {\bibfnamefont {C.-H.}\ \bibnamefont {Park}},
  \bibinfo {author} {\bibfnamefont {S.~G.}\ \bibnamefont {Louie}},\ and\
  \bibinfo {author} {\bibfnamefont {M.~L.}\ \bibnamefont {Cohen}},\ }\bibfield
  {title} {\bibinfo {title} {\emph {{EPW}: A program for calculating the
  electron-phonon coupling using maximally localized {Wannier} functions}},\
  }\href {https://doi.org/10.1016/j.cpc.2010.08.027} {\bibfield  {journal}
  {\bibinfo  {journal} {Comput. Phys. Commun.}\ }\textbf {\bibinfo {volume}
  {181}},\ \bibinfo {pages} {2140} (\bibinfo {year} {2010})},\ \Eprint
  {https://arxiv.org/abs/1005.4418} {arXiv:1005.4418}\BibitemShut {NoStop}%
\bibitem [{\citenamefont {Ponc\'e}\ \emph {et~al.}(2016)\citenamefont
  {Ponc\'e}, \citenamefont {Margine}, \citenamefont {Verdi},\ and\
  \citenamefont {Giustino}}]{Ponce2016}%
  \BibitemOpen
  \bibfield  {author} {\bibinfo {author} {\bibfnamefont {S.}~\bibnamefont
  {Ponc\'e}}, \bibinfo {author} {\bibfnamefont {E.}~\bibnamefont {Margine}},
  \bibinfo {author} {\bibfnamefont {C.}~\bibnamefont {Verdi}},\ and\ \bibinfo
  {author} {\bibfnamefont {F.}~\bibnamefont {Giustino}},\ }\bibfield  {title}
  {\bibinfo {title} {\emph {{EPW}: Electron-phonon coupling, transport and
  superconducting properties using maximally localized {Wannier} functions}},\
  }\href {https://doi.org/10.1016/j.cpc.2016.07.028} {\bibfield  {journal}
  {\bibinfo  {journal} {Comput. Phys. Commun.}\ }\textbf {\bibinfo {volume}
  {209}},\ \bibinfo {pages} {116} (\bibinfo {year} {2016})},\ \Eprint
  {https://arxiv.org/abs/1604.03525} {arXiv:1604.03525}\BibitemShut {NoStop}%
\bibitem [{Note2()}]{Note2}%
  \BibitemOpen
  \bibinfo {note} {See Supplemental Material at \protect \url
  {https://arxiv.org/src/2212.11806/anc} for the source code and data
  associated with this work. It is also available in the Materials Cloud
  Archive at \protect \url
  {https://doi.org/10.24435/materialscloud:he-pv}.}\BibitemShut {Stop}%
\bibitem [{\citenamefont {Marini}\ \emph {et~al.}(2023)\citenamefont {Marini},
  \citenamefont {Marchese}, \citenamefont {Profeta}, \citenamefont {Sjakste},
  \citenamefont {Macheda}, \citenamefont {Vast}, \citenamefont {Mauri},\ and\
  \citenamefont {Calandra}}]{Marini2023c}%
  \BibitemOpen
  \bibfield  {author} {\bibinfo {author} {\bibfnamefont {G.}~\bibnamefont
  {Marini}}, \bibinfo {author} {\bibfnamefont {G.}~\bibnamefont {Marchese}},
  \bibinfo {author} {\bibfnamefont {G.}~\bibnamefont {Profeta}}, \bibinfo
  {author} {\bibfnamefont {J.}~\bibnamefont {Sjakste}}, \bibinfo {author}
  {\bibfnamefont {F.}~\bibnamefont {Macheda}}, \bibinfo {author} {\bibfnamefont
  {N.}~\bibnamefont {Vast}}, \bibinfo {author} {\bibfnamefont {F.}~\bibnamefont
  {Mauri}},\ and\ \bibinfo {author} {\bibfnamefont {M.}~\bibnamefont
  {Calandra}},\ }\bibfield  {title} {\bibinfo {title} {\emph {{EPIq}: An
  open-source software for the calculation of electron-phonon interaction
  related properties}},\ }\Eprint {https://arxiv.org/abs/2306.15462}
  {arXiv:2306.15462} (\bibinfo {year} {2023})\BibitemShut {NoStop}%
\bibitem [{\citenamefont {He}\ \emph {et~al.}(2020)\citenamefont {He},
  \citenamefont {Liu}, \citenamefont {Li}, \citenamefont {Rignanese},\ and\
  \citenamefont {Zhou}}]{He2020}%
  \BibitemOpen
  \bibfield  {author} {\bibinfo {author} {\bibfnamefont {L.}~\bibnamefont
  {He}}, \bibinfo {author} {\bibfnamefont {F.}~\bibnamefont {Liu}}, \bibinfo
  {author} {\bibfnamefont {J.}~\bibnamefont {Li}}, \bibinfo {author}
  {\bibfnamefont {G.-M.}\ \bibnamefont {Rignanese}},\ and\ \bibinfo {author}
  {\bibfnamefont {A.}~\bibnamefont {Zhou}},\ }\bibfield  {title} {\bibinfo
  {title} {\emph {First-principles investigation of monatomic gold wires under
  tension}},\ }\href {https://doi.org/10.1016/j.commatsci.2019.109226}
  {\bibfield  {journal} {\bibinfo  {journal} {Comput. Mater. Sci.}\ }\textbf
  {\bibinfo {volume} {171}},\ \bibinfo {pages} {109226} (\bibinfo {year}
  {2020})}\BibitemShut {NoStop}%
\bibitem [{Note3()}]{Note3}%
  \BibitemOpen
  \bibinfo {note} {Strictly speaking, the phonon Green's function is the
  correlation function of phonon ladder operators instead of displacement
  operators~\cite {Giustino2017} and thus differs from Eq.~\protect \textup
  {\hbox {\mathsurround \z@ \protect \normalfont (\ignorespaces \ref
  {eq:green_t}\unskip \@@italiccorr )}} by a factor of $2 \omega $, which thus
  appears in Eq.~\protect \textup {\hbox {\mathsurround \z@ \protect
  \normalfont (\ignorespaces \ref {eq:specfun}\unskip \@@italiccorr )}} for the
  phonon spectral function.}\BibitemShut {Stop}%
\bibitem [{\citenamefont {Abrikosov}\ \emph {et~al.}(1963)\citenamefont
  {Abrikosov}, \citenamefont {Gor'kov},\ and\ \citenamefont
  {Dzyaloshinski}}]{Abrikosov1963}%
  \BibitemOpen
  \bibfield  {author} {\bibinfo {author} {\bibfnamefont {A.~A.}\ \bibnamefont
  {Abrikosov}}, \bibinfo {author} {\bibfnamefont {L.~P.}\ \bibnamefont
  {Gor'kov}},\ and\ \bibinfo {author} {\bibfnamefont {I.~E.}\ \bibnamefont
  {Dzyaloshinski}},\ }\href@noop {} {\emph {\bibinfo {title} {Methods of
  Quantum Field Theory in Statistical Physics}}}\ (\bibinfo  {publisher} {Dover
  Publications},\ \bibinfo {address} {New York},\ \bibinfo {year}
  {1963})\BibitemShut {NoStop}%
\bibitem [{\citenamefont {Falter}\ \emph {et~al.}(2006)\citenamefont {Falter},
  \citenamefont {Bauer},\ and\ \citenamefont {Schnetg\"oke}}]{Falter2006}%
  \BibitemOpen
  \bibfield  {author} {\bibinfo {author} {\bibfnamefont {C.}~\bibnamefont
  {Falter}}, \bibinfo {author} {\bibfnamefont {T.}~\bibnamefont {Bauer}},\ and\
  \bibinfo {author} {\bibfnamefont {F.}~\bibnamefont {Schnetg\"oke}},\
  }\bibfield  {title} {\bibinfo {title} {\emph {Modeling the electronic state
  of the high-{$T_c$} superconductor {LaCuO}: Phonon dynamics and charge
  response}},\ }\href {https://doi.org/10.1103/PhysRevB.73.224502} {\bibfield
  {journal} {\bibinfo  {journal} {Phys. Rev. B}\ }\textbf {\bibinfo {volume}
  {73}},\ \bibinfo {pages} {224502} (\bibinfo {year} {2006})},\ \Eprint
  {https://arxiv.org/abs/cond-mat/0512577} {arXiv:cond-mat/0512577}\BibitemShut
  {NoStop}%
\bibitem [{\citenamefont {Bohm}\ and\ \citenamefont {Pines}(1951)}]{Bohm1951}%
  \BibitemOpen
  \bibfield  {author} {\bibinfo {author} {\bibfnamefont {D.}~\bibnamefont
  {Bohm}}\ and\ \bibinfo {author} {\bibfnamefont {D.}~\bibnamefont {Pines}},\
  }\bibfield  {title} {\bibinfo {title} {\emph {A collective description of
  electron interactions. {I}. {Magnetic} interactions}},\ }\href
  {https://doi.org/10.1103/PhysRev.82.625} {\bibfield  {journal} {\bibinfo
  {journal} {Phys. Rev.}\ }\textbf {\bibinfo {volume} {82}},\ \bibinfo {pages}
  {625} (\bibinfo {year} {1951})}\BibitemShut {NoStop}%
\bibitem [{\citenamefont {Gell-Mann}\ and\ \citenamefont
  {Brueckner}(1957)}]{GellMann1957}%
  \BibitemOpen
  \bibfield  {author} {\bibinfo {author} {\bibfnamefont {M.}~\bibnamefont
  {Gell-Mann}}\ and\ \bibinfo {author} {\bibfnamefont {K.~A.}\ \bibnamefont
  {Brueckner}},\ }\bibfield  {title} {\bibinfo {title} {\emph {Correlation
  energy of an electron gas at high density}},\ }\href
  {https://doi.org/10.1103/PhysRev.106.364} {\bibfield  {journal} {\bibinfo
  {journal} {Phys. Rev.}\ }\textbf {\bibinfo {volume} {106}},\ \bibinfo {pages}
  {364} (\bibinfo {year} {1957})}\BibitemShut {NoStop}%
\bibitem [{\citenamefont {Ren}\ \emph {et~al.}(2012)\citenamefont {Ren},
  \citenamefont {Rinke}, \citenamefont {Joas},\ and\ \citenamefont
  {Scheffler}}]{Ren2012}%
  \BibitemOpen
  \bibfield  {author} {\bibinfo {author} {\bibfnamefont {X.}~\bibnamefont
  {Ren}}, \bibinfo {author} {\bibfnamefont {P.}~\bibnamefont {Rinke}}, \bibinfo
  {author} {\bibfnamefont {C.}~\bibnamefont {Joas}},\ and\ \bibinfo {author}
  {\bibfnamefont {M.}~\bibnamefont {Scheffler}},\ }\bibfield  {title} {\bibinfo
  {title} {\emph {Random-phase approximation and its applications in
  computational chemistry and materials science}},\ }\href
  {https://doi.org/10.1007/s10853-012-6570-4} {\bibfield  {journal} {\bibinfo
  {journal} {J. Mater. Sci.}\ }\textbf {\bibinfo {volume} {47}},\ \bibinfo
  {pages} {7447} (\bibinfo {year} {2012})},\ \Eprint
  {https://arxiv.org/abs/1203.5536} {arXiv:1203.5536}\BibitemShut {NoStop}%
\bibitem [{\citenamefont {Hedin}(1965)}]{Hedin1965}%
  \BibitemOpen
  \bibfield  {author} {\bibinfo {author} {\bibfnamefont {L.}~\bibnamefont
  {Hedin}},\ }\bibfield  {title} {\bibinfo {title} {\emph {New method for
  calculating the one-particle {Green's} function with application to the
  electron-gas problem}},\ }\href {https://doi.org/10.1103/PhysRev.139.A796}
  {\bibfield  {journal} {\bibinfo  {journal} {Phys. Rev.}\ }\textbf {\bibinfo
  {volume} {139}},\ \bibinfo {pages} {A796} (\bibinfo {year}
  {1965})}\BibitemShut {NoStop}%
\bibitem [{\citenamefont {Golze}\ \emph {et~al.}(2019)\citenamefont {Golze},
  \citenamefont {Dvorak},\ and\ \citenamefont {Rinke}}]{Golze2019}%
  \BibitemOpen
  \bibfield  {author} {\bibinfo {author} {\bibfnamefont {D.}~\bibnamefont
  {Golze}}, \bibinfo {author} {\bibfnamefont {M.}~\bibnamefont {Dvorak}},\ and\
  \bibinfo {author} {\bibfnamefont {P.}~\bibnamefont {Rinke}},\ }\bibfield
  {title} {\bibinfo {title} {\emph {The {GW} compendium: A practical guide to
  theoretical photoemission spectroscopy}},\ }\href
  {https://doi.org/10.3389/fchem.2019.00377} {\bibfield  {journal} {\bibinfo
  {journal} {Front. Chem.}\ }\textbf {\bibinfo {volume} {7}},\ \bibinfo {pages}
  {377} (\bibinfo {year} {2019})},\ \Eprint {https://arxiv.org/abs/1912.04893}
  {arXiv:1912.04893}\BibitemShut {NoStop}%
\bibitem [{\citenamefont {Li}\ \emph {et~al.}(2019)\citenamefont {Li},
  \citenamefont {Antonius}, \citenamefont {Wu}, \citenamefont {da~Jornada},\
  and\ \citenamefont {Louie}}]{Li2019}%
  \BibitemOpen
  \bibfield  {author} {\bibinfo {author} {\bibfnamefont {Z.}~\bibnamefont
  {Li}}, \bibinfo {author} {\bibfnamefont {G.}~\bibnamefont {Antonius}},
  \bibinfo {author} {\bibfnamefont {M.}~\bibnamefont {Wu}}, \bibinfo {author}
  {\bibfnamefont {F.~H.}\ \bibnamefont {da~Jornada}},\ and\ \bibinfo {author}
  {\bibfnamefont {S.~G.}\ \bibnamefont {Louie}},\ }\bibfield  {title} {\bibinfo
  {title} {\emph {Electron-phonon coupling from ab initio linear-response
  theory within the {$GW$} method: Correlation-enhanced interactions and
  superconductivity in {Ba\s{1\ensuremath-x}K\s{x}BiO\s3}}},\ }\href
  {https://doi.org/10.1103/PhysRevLett.122.186402} {\bibfield  {journal}
  {\bibinfo  {journal} {Phys. Rev. Lett.}\ }\textbf {\bibinfo {volume} {122}},\
  \bibinfo {pages} {186402} (\bibinfo {year} {2019})},\ \Eprint
  {https://arxiv.org/abs/1902.06212} {arXiv:1902.06212}\BibitemShut {NoStop}%
\bibitem [{\citenamefont {Sternheimer}(1954)}]{Sternheimer1954}%
  \BibitemOpen
  \bibfield  {author} {\bibinfo {author} {\bibfnamefont {R.~M.}\ \bibnamefont
  {Sternheimer}},\ }\bibfield  {title} {\bibinfo {title} {\emph {Electronic
  polarizabilities of ions from the {Hartree}-{Fock} wave functions}},\ }\href
  {https://doi.org/10.1103/PhysRev.96.951} {\bibfield  {journal} {\bibinfo
  {journal} {Phys. Rev.}\ }\textbf {\bibinfo {volume} {96}},\ \bibinfo {pages}
  {951} (\bibinfo {year} {1954})}\BibitemShut {NoStop}%
\bibitem [{\citenamefont {Giustino}\ \emph {et~al.}(2010)\citenamefont
  {Giustino}, \citenamefont {Cohen},\ and\ \citenamefont
  {Louie}}]{Giustino2010}%
  \BibitemOpen
  \bibfield  {author} {\bibinfo {author} {\bibfnamefont {F.}~\bibnamefont
  {Giustino}}, \bibinfo {author} {\bibfnamefont {M.~L.}\ \bibnamefont
  {Cohen}},\ and\ \bibinfo {author} {\bibfnamefont {S.~G.}\ \bibnamefont
  {Louie}},\ }\bibfield  {title} {\bibinfo {title} {\emph {{GW} method with the
  self-consistent {Sternheimer} equation}},\ }\href
  {https://doi.org/10.1103/PhysRevB.81.115105} {\bibfield  {journal} {\bibinfo
  {journal} {Phys. Rev. B}\ }\textbf {\bibinfo {volume} {81}},\ \bibinfo
  {pages} {115105} (\bibinfo {year} {2010})},\ \Eprint
  {https://arxiv.org/abs/0912.3087} {arXiv:0912.3087}\BibitemShut {NoStop}%
\bibitem [{\citenamefont {Schlipf}\ \emph {et~al.}(2020)\citenamefont
  {Schlipf}, \citenamefont {Lambert}, \citenamefont {Zibouche},\ and\
  \citenamefont {Giustino}}]{Schlipf2020}%
  \BibitemOpen
  \bibfield  {author} {\bibinfo {author} {\bibfnamefont {M.}~\bibnamefont
  {Schlipf}}, \bibinfo {author} {\bibfnamefont {H.}~\bibnamefont {Lambert}},
  \bibinfo {author} {\bibfnamefont {N.}~\bibnamefont {Zibouche}},\ and\
  \bibinfo {author} {\bibfnamefont {F.}~\bibnamefont {Giustino}},\ }\bibfield
  {title} {\bibinfo {title} {\emph {\textsc{SternheimerGW}: A program for
  calculating {GW} quasiparticle band structures and spectral functions without
  unoccupied states}},\ }\href {https://doi.org/10.1016/j.cpc.2019.07.019}
  {\bibfield  {journal} {\bibinfo  {journal} {Comput. Phys. Commun.}\ }\textbf
  {\bibinfo {volume} {247}},\ \bibinfo {pages} {106856} (\bibinfo {year}
  {2020})},\ \Eprint {https://arxiv.org/abs/1812.03717}
  {arXiv:1812.03717}\BibitemShut {NoStop}%
\bibitem [{\citenamefont {Miyake}\ \emph {et~al.}(2009)\citenamefont {Miyake},
  \citenamefont {Aryasetiawan},\ and\ \citenamefont {Imada}}]{Miyake2009}%
  \BibitemOpen
  \bibfield  {author} {\bibinfo {author} {\bibfnamefont {T.}~\bibnamefont
  {Miyake}}, \bibinfo {author} {\bibfnamefont {F.}~\bibnamefont
  {Aryasetiawan}},\ and\ \bibinfo {author} {\bibfnamefont {M.}~\bibnamefont
  {Imada}},\ }\bibfield  {title} {\bibinfo {title} {\emph {Ab initio procedure
  for constructing effective models of correlated materials with entangled band
  structure}},\ }\href {https://doi.org/10.1103/PhysRevB.80.155134} {\bibfield
  {journal} {\bibinfo  {journal} {Phys. Rev. B}\ }\textbf {\bibinfo {volume}
  {80}},\ \bibinfo {pages} {155134} (\bibinfo {year} {2009})},\ \Eprint
  {https://arxiv.org/abs/0906.1344} {arXiv:0906.1344}\BibitemShut {NoStop}%
\bibitem [{\citenamefont {Ge{\u\i}likman}(1975)}]{Geilikman1975}%
  \BibitemOpen
  \bibfield  {author} {\bibinfo {author} {\bibfnamefont {B.~T.}\ \bibnamefont
  {Ge{\u\i}likman}},\ }\bibfield  {title} {\bibinfo {title} {\emph {Adiabatic
  perturbation theory for metals and the problem of lattice stability}},\
  }\href {https://doi.org/10.1070/PU1975v018n03ABEH001953} {\bibfield
  {journal} {\bibinfo  {journal} {Sov. Phys. Usp.}\ }\textbf {\bibinfo {volume}
  {18}},\ \bibinfo {pages} {190} (\bibinfo {year} {1975})}\BibitemShut
  {NoStop}%
\bibitem [{\citenamefont {Berges}(2020)}]{Berges2020b}%
  \BibitemOpen
  \bibfield  {author} {\bibinfo {author} {\bibfnamefont {J.}~\bibnamefont
  {Berges}},\ }\emph {\bibinfo {title} {Many-body instabilities in
  two-dimensional materials}},\ \href {https://doi.org/10.26092/elib/250}
  {\bibinfo {type} {Dissertation}},\ \bibinfo  {school} {Universit\"at Bremen}
  (\bibinfo {year} {2020})\BibitemShut {NoStop}%
\bibitem [{\citenamefont {Allen}(1980)}]{Allen1980}%
  \BibitemOpen
  \bibfield  {author} {\bibinfo {author} {\bibfnamefont {P.~B.}\ \bibnamefont
  {Allen}},\ }\bibfield  {title} {\bibinfo {title} {\emph {Phonons and the
  superconducting transition temperature}},\ }in\ \href@noop {} {\emph
  {\bibinfo {booktitle} {Dynamical Properties of Solids}}},\ Vol.~\bibinfo
  {volume} {3},\ \bibinfo {editor} {edited by\ \bibinfo {editor} {\bibfnamefont
  {G.~K.}\ \bibnamefont {Horton}}\ and\ \bibinfo {editor} {\bibfnamefont
  {A.~A.}\ \bibnamefont {Maradudin}}}\ (\bibinfo  {publisher} {North-Holland},\
  \bibinfo {address} {Amsterdam},\ \bibinfo {year} {1980})\BibitemShut
  {NoStop}%
\bibitem [{\citenamefont {Maksimov}\ and\ \citenamefont
  {Karakozov}(2008)}]{Maksimov2008}%
  \BibitemOpen
  \bibfield  {author} {\bibinfo {author} {\bibfnamefont {E.~G.}\ \bibnamefont
  {Maksimov}}\ and\ \bibinfo {author} {\bibfnamefont {A.~E.}\ \bibnamefont
  {Karakozov}},\ }\bibfield  {title} {\bibinfo {title} {\emph {On nonadiabatic
  effects in phonon spectra of metals}},\ }\href
  {https://doi.org/10.1070/PU2008v051n06ABEH006462} {\bibfield  {journal}
  {\bibinfo  {journal} {Phys. Usp.}\ }\textbf {\bibinfo {volume} {51}},\
  \bibinfo {pages} {535} (\bibinfo {year} {2008})}\BibitemShut {NoStop}%
\bibitem [{\citenamefont {Nomura}\ \emph {et~al.}(2012)\citenamefont {Nomura},
  \citenamefont {Kaltak}, \citenamefont {Nakamura}, \citenamefont {Taranto},
  \citenamefont {Sakai}, \citenamefont {Toschi}, \citenamefont {Arita},
  \citenamefont {Held}, \citenamefont {Kresse},\ and\ \citenamefont
  {Imada}}]{Nomura2012}%
  \BibitemOpen
  \bibfield  {author} {\bibinfo {author} {\bibfnamefont {Y.}~\bibnamefont
  {Nomura}}, \bibinfo {author} {\bibfnamefont {M.}~\bibnamefont {Kaltak}},
  \bibinfo {author} {\bibfnamefont {K.}~\bibnamefont {Nakamura}}, \bibinfo
  {author} {\bibfnamefont {C.}~\bibnamefont {Taranto}}, \bibinfo {author}
  {\bibfnamefont {S.}~\bibnamefont {Sakai}}, \bibinfo {author} {\bibfnamefont
  {A.}~\bibnamefont {Toschi}}, \bibinfo {author} {\bibfnamefont
  {R.}~\bibnamefont {Arita}}, \bibinfo {author} {\bibfnamefont
  {K.}~\bibnamefont {Held}}, \bibinfo {author} {\bibfnamefont {G.}~\bibnamefont
  {Kresse}},\ and\ \bibinfo {author} {\bibfnamefont {M.}~\bibnamefont
  {Imada}},\ }\bibfield  {title} {\bibinfo {title} {\emph {Effective on-site
  interaction for dynamical mean-field theory}},\ }\href
  {https://doi.org/10.1103/PhysRevB.86.085117} {\bibfield  {journal} {\bibinfo
  {journal} {Phys. Rev. B}\ }\textbf {\bibinfo {volume} {86}},\ \bibinfo
  {pages} {085117} (\bibinfo {year} {2012})},\ \Eprint
  {https://arxiv.org/abs/1205.2836} {arXiv:1205.2836}\BibitemShut {NoStop}%
\bibitem [{\citenamefont {Migdal}(1958)}]{Migdal1958}%
  \BibitemOpen
  \bibfield  {author} {\bibinfo {author} {\bibfnamefont {A.~B.}\ \bibnamefont
  {Migdal}},\ }\bibfield  {title} {\bibinfo {title} {\emph {Interaction between
  electrons and lattice vibrations in a normal metal}},\ }\href
  {http://www.jetp.ras.ru/cgi-bin/e/index/e/7/6/p996?a=list} {\bibfield
  {journal} {\bibinfo  {journal} {Sov. Phys. JETP}\ }\textbf {\bibinfo {volume}
  {7}},\ \bibinfo {pages} {996} (\bibinfo {year} {1958})}\BibitemShut {NoStop}%
\bibitem [{\citenamefont {Friedel}(1958)}]{Friedel1958}%
  \BibitemOpen
  \bibfield  {author} {\bibinfo {author} {\bibfnamefont {J.}~\bibnamefont
  {Friedel}},\ }\bibfield  {title} {\bibinfo {title} {\emph {Metallic
  alloys}},\ }\href {https://doi.org/10.1007/BF02751483} {\bibfield  {journal}
  {\bibinfo  {journal} {Nuovo Cim.}\ }\textbf {\bibinfo {volume} {7}},\
  \bibinfo {pages} {287} (\bibinfo {year} {1958})}\BibitemShut {NoStop}%
\bibitem [{\citenamefont {Peierls}(1955)}]{Peierls1955}%
  \BibitemOpen
  \bibfield  {author} {\bibinfo {author} {\bibfnamefont {R.~E.}\ \bibnamefont
  {Peierls}},\ }\href
  {https://doi.org/10.1093/acprof:oso/9780198507819.001.0001} {\emph {\bibinfo
  {title} {Quantum Theory of Solids}}}\ (\bibinfo  {publisher} {Oxford
  University Press},\ \bibinfo {address} {London},\ \bibinfo {year}
  {1955})\BibitemShut {NoStop}%
\bibitem [{\citenamefont {Gonze}\ \emph {et~al.}(1994)\citenamefont {Gonze},
  \citenamefont {Charlier}, \citenamefont {Allan},\ and\ \citenamefont
  {Teter}}]{Gonze1994}%
  \BibitemOpen
  \bibfield  {author} {\bibinfo {author} {\bibfnamefont {X.}~\bibnamefont
  {Gonze}}, \bibinfo {author} {\bibfnamefont {J.-C.}\ \bibnamefont {Charlier}},
  \bibinfo {author} {\bibfnamefont {D.}~\bibnamefont {Allan}},\ and\ \bibinfo
  {author} {\bibfnamefont {M.}~\bibnamefont {Teter}},\ }\bibfield  {title}
  {\bibinfo {title} {\emph {Interatomic force constants from first principles:
  The case of $\alpha$-quartz}},\ }\href
  {https://doi.org/10.1103/PhysRevB.50.13035} {\bibfield  {journal} {\bibinfo
  {journal} {Phys. Rev. B}\ }\textbf {\bibinfo {volume} {50}},\ \bibinfo
  {pages} {13035} (\bibinfo {year} {1994})}\BibitemShut {NoStop}%
\bibitem [{\citenamefont {Verdi}\ and\ \citenamefont
  {Giustino}(2015)}]{Verdi2015}%
  \BibitemOpen
  \bibfield  {author} {\bibinfo {author} {\bibfnamefont {C.}~\bibnamefont
  {Verdi}}\ and\ \bibinfo {author} {\bibfnamefont {F.}~\bibnamefont
  {Giustino}},\ }\bibfield  {title} {\bibinfo {title} {\emph {Fr\"ohlich
  electron-phonon vertex from first principles}},\ }\href
  {https://doi.org/10.1103/PhysRevLett.115.176401} {\bibfield  {journal}
  {\bibinfo  {journal} {Phys. Rev. Lett.}\ }\textbf {\bibinfo {volume} {115}},\
  \bibinfo {pages} {176401} (\bibinfo {year} {2015})},\ \Eprint
  {https://arxiv.org/abs/1510.06373} {arXiv:1510.06373}\BibitemShut {NoStop}%
\bibitem [{\citenamefont {Sjakste}\ \emph {et~al.}(2015)\citenamefont
  {Sjakste}, \citenamefont {Vast}, \citenamefont {Calandra},\ and\
  \citenamefont {Mauri}}]{Sjakste2015}%
  \BibitemOpen
  \bibfield  {author} {\bibinfo {author} {\bibfnamefont {J.}~\bibnamefont
  {Sjakste}}, \bibinfo {author} {\bibfnamefont {N.}~\bibnamefont {Vast}},
  \bibinfo {author} {\bibfnamefont {M.}~\bibnamefont {Calandra}},\ and\
  \bibinfo {author} {\bibfnamefont {F.}~\bibnamefont {Mauri}},\ }\bibfield
  {title} {\bibinfo {title} {\emph {Wannier interpolation of the
  electron-phonon matrix elements in polar semiconductors: Polar-optical
  coupling in {GaAs}}},\ }\href {https://doi.org/10.1103/PhysRevB.92.054307}
  {\bibfield  {journal} {\bibinfo  {journal} {Phys. Rev. B}\ }\textbf {\bibinfo
  {volume} {92}},\ \bibinfo {pages} {054307} (\bibinfo {year} {2015})},\
  \Eprint {https://arxiv.org/abs/1508.06172} {arXiv:1508.06172}\BibitemShut
  {NoStop}%
\bibitem [{\citenamefont {Sohier}\ \emph {et~al.}(2016)\citenamefont {Sohier},
  \citenamefont {Calandra},\ and\ \citenamefont {Mauri}}]{Sohier2016}%
  \BibitemOpen
  \bibfield  {author} {\bibinfo {author} {\bibfnamefont {T.}~\bibnamefont
  {Sohier}}, \bibinfo {author} {\bibfnamefont {M.}~\bibnamefont {Calandra}},\
  and\ \bibinfo {author} {\bibfnamefont {F.}~\bibnamefont {Mauri}},\ }\bibfield
   {title} {\bibinfo {title} {\emph {Two-dimensional {Fr\"ohlich} interaction
  in transition-metal dichalcogenide monolayers: Theoretical modeling and
  first-principles calculations}},\ }\href
  {https://doi.org/10.1103/PhysRevB.94.085415} {\bibfield  {journal} {\bibinfo
  {journal} {Phys. Rev. B}\ }\textbf {\bibinfo {volume} {94}},\ \bibinfo
  {pages} {085415} (\bibinfo {year} {2016})},\ \Eprint
  {https://arxiv.org/abs/1605.08207} {arXiv:1605.08207}\BibitemShut {NoStop}%
\bibitem [{\citenamefont {Royo}\ \emph {et~al.}(2020)\citenamefont {Royo},
  \citenamefont {Hahn},\ and\ \citenamefont {Stengel}}]{Royo2020}%
  \BibitemOpen
  \bibfield  {author} {\bibinfo {author} {\bibfnamefont {M.}~\bibnamefont
  {Royo}}, \bibinfo {author} {\bibfnamefont {K.~R.}\ \bibnamefont {Hahn}},\
  and\ \bibinfo {author} {\bibfnamefont {M.}~\bibnamefont {Stengel}},\
  }\bibfield  {title} {\bibinfo {title} {\emph {Using high multipolar orders to
  reconstruct the sound velocity in piezoelectrics from lattice dynamics}},\
  }\href {https://doi.org/10.1103/PhysRevLett.125.217602} {\bibfield  {journal}
  {\bibinfo  {journal} {Phys. Rev. Lett.}\ }\textbf {\bibinfo {volume} {125}},\
  \bibinfo {pages} {217602} (\bibinfo {year} {2020})},\ \Eprint
  {https://arxiv.org/abs/2004.08875} {arXiv:2004.08875}\BibitemShut {NoStop}%
\bibitem [{\citenamefont {Brunin}\ \emph {et~al.}(2020)\citenamefont {Brunin},
  \citenamefont {Miranda}, \citenamefont {Giantomassi}, \citenamefont {Royo},
  \citenamefont {Stengel}, \citenamefont {Verstraete}, \citenamefont {Gonze},
  \citenamefont {Rignanese},\ and\ \citenamefont {Hautier}}]{Brunin2020}%
  \BibitemOpen
  \bibfield  {author} {\bibinfo {author} {\bibfnamefont {G.}~\bibnamefont
  {Brunin}}, \bibinfo {author} {\bibfnamefont {H.~P.~C.}\ \bibnamefont
  {Miranda}}, \bibinfo {author} {\bibfnamefont {M.}~\bibnamefont
  {Giantomassi}}, \bibinfo {author} {\bibfnamefont {M.}~\bibnamefont {Royo}},
  \bibinfo {author} {\bibfnamefont {M.}~\bibnamefont {Stengel}}, \bibinfo
  {author} {\bibfnamefont {M.~J.}\ \bibnamefont {Verstraete}}, \bibinfo
  {author} {\bibfnamefont {X.}~\bibnamefont {Gonze}}, \bibinfo {author}
  {\bibfnamefont {G.-M.}\ \bibnamefont {Rignanese}},\ and\ \bibinfo {author}
  {\bibfnamefont {G.}~\bibnamefont {Hautier}},\ }\bibfield  {title} {\bibinfo
  {title} {\emph {Electron-phonon beyond {Fr\"ohlich}: Dynamical quadrupoles in
  polar and covalent solids}},\ }\href
  {https://doi.org/10.1103/PhysRevLett.125.136601} {\bibfield  {journal}
  {\bibinfo  {journal} {Phys. Rev. Lett.}\ }\textbf {\bibinfo {volume} {125}},\
  \bibinfo {pages} {136601} (\bibinfo {year} {2020})},\ \Eprint
  {https://arxiv.org/abs/2002.00628} {arXiv:2002.00628}\BibitemShut {NoStop}%
\bibitem [{\citenamefont {Jhalani}\ \emph {et~al.}(2020)\citenamefont
  {Jhalani}, \citenamefont {Zhou}, \citenamefont {Park}, \citenamefont
  {Dreyer},\ and\ \citenamefont {Bernardi}}]{Jhalani2020}%
  \BibitemOpen
  \bibfield  {author} {\bibinfo {author} {\bibfnamefont {V.~A.}\ \bibnamefont
  {Jhalani}}, \bibinfo {author} {\bibfnamefont {J.-J.}\ \bibnamefont {Zhou}},
  \bibinfo {author} {\bibfnamefont {J.}~\bibnamefont {Park}}, \bibinfo {author}
  {\bibfnamefont {C.~E.}\ \bibnamefont {Dreyer}},\ and\ \bibinfo {author}
  {\bibfnamefont {M.}~\bibnamefont {Bernardi}},\ }\bibfield  {title} {\bibinfo
  {title} {\emph {Piezoelectric electron-phonon interaction from ab initio
  dynamical quadrupoles: Impact on charge transport in wurtzite {GaN}}},\
  }\href {https://doi.org/10.1103/PhysRevLett.125.136602} {\bibfield  {journal}
  {\bibinfo  {journal} {Phys. Rev. Lett.}\ }\textbf {\bibinfo {volume} {125}},\
  \bibinfo {pages} {136602} (\bibinfo {year} {2020})},\ \Eprint
  {https://arxiv.org/abs/2002.08351} {arXiv:2002.08351}\BibitemShut {NoStop}%
\bibitem [{\citenamefont {Ponc\'e}\ \emph {et~al.}(2021)\citenamefont
  {Ponc\'e}, \citenamefont {Macheda}, \citenamefont {Margine}, \citenamefont
  {Marzari}, \citenamefont {Bonini},\ and\ \citenamefont
  {Giustino}}]{Ponce2021}%
  \BibitemOpen
  \bibfield  {author} {\bibinfo {author} {\bibfnamefont {S.}~\bibnamefont
  {Ponc\'e}}, \bibinfo {author} {\bibfnamefont {F.}~\bibnamefont {Macheda}},
  \bibinfo {author} {\bibfnamefont {E.~R.}\ \bibnamefont {Margine}}, \bibinfo
  {author} {\bibfnamefont {N.}~\bibnamefont {Marzari}}, \bibinfo {author}
  {\bibfnamefont {N.}~\bibnamefont {Bonini}},\ and\ \bibinfo {author}
  {\bibfnamefont {F.}~\bibnamefont {Giustino}},\ }\bibfield  {title} {\bibinfo
  {title} {\emph {First-principles predictions of {Hall} and drift mobilities
  in semiconductors}},\ }\href
  {https://doi.org/10.1103/PhysRevResearch.3.043022} {\bibfield  {journal}
  {\bibinfo  {journal} {Phys. Rev. Research}\ }\textbf {\bibinfo {volume}
  {3}},\ \bibinfo {pages} {043022} (\bibinfo {year} {2021})},\ \Eprint
  {https://arxiv.org/abs/2105.04192} {arXiv:2105.04192}\BibitemShut {NoStop}%
\bibitem [{\citenamefont {Royo}\ and\ \citenamefont
  {Stengel}(2021)}]{Royo2021}%
  \BibitemOpen
  \bibfield  {author} {\bibinfo {author} {\bibfnamefont {M.}~\bibnamefont
  {Royo}}\ and\ \bibinfo {author} {\bibfnamefont {M.}~\bibnamefont {Stengel}},\
  }\bibfield  {title} {\bibinfo {title} {\emph {Exact long-range dielectric
  screening and interatomic force constants in quasi-two-dimensional
  crystals}},\ }\href {https://doi.org/10.1103/PhysRevX.11.041027} {\bibfield
  {journal} {\bibinfo  {journal} {Phys. Rev. X}\ }\textbf {\bibinfo {volume}
  {11}},\ \bibinfo {pages} {041027} (\bibinfo {year} {2021})},\ \Eprint
  {https://arxiv.org/abs/2012.07961} {arXiv:2012.07961}\BibitemShut {NoStop}%
\bibitem [{\citenamefont {Macheda}\ \emph {et~al.}(2022)\citenamefont
  {Macheda}, \citenamefont {Barone},\ and\ \citenamefont
  {Mauri}}]{Macheda2022}%
  \BibitemOpen
  \bibfield  {author} {\bibinfo {author} {\bibfnamefont {F.}~\bibnamefont
  {Macheda}}, \bibinfo {author} {\bibfnamefont {P.}~\bibnamefont {Barone}},\
  and\ \bibinfo {author} {\bibfnamefont {F.}~\bibnamefont {Mauri}},\ }\bibfield
   {title} {\bibinfo {title} {\emph {Electron-phonon interaction and
  longitudinal-transverse phonon splitting in doped semiconductors}},\ }\href
  {https://doi.org/10.1103/PhysRevLett.129.185902} {\bibfield  {journal}
  {\bibinfo  {journal} {Phys. Rev. Lett.}\ }\textbf {\bibinfo {volume} {129}},\
  \bibinfo {pages} {185902} (\bibinfo {year} {2022})},\ \Eprint
  {https://arxiv.org/abs/2202.02835} {arXiv:2202.02835}\BibitemShut {NoStop}%
\bibitem [{\citenamefont {Macheda}\ \emph {et~al.}(2023)\citenamefont
  {Macheda}, \citenamefont {Sohier}, \citenamefont {Barone},\ and\
  \citenamefont {Mauri}}]{Macheda2023}%
  \BibitemOpen
  \bibfield  {author} {\bibinfo {author} {\bibfnamefont {F.}~\bibnamefont
  {Macheda}}, \bibinfo {author} {\bibfnamefont {T.}~\bibnamefont {Sohier}},
  \bibinfo {author} {\bibfnamefont {P.}~\bibnamefont {Barone}},\ and\ \bibinfo
  {author} {\bibfnamefont {F.}~\bibnamefont {Mauri}},\ }\bibfield  {title}
  {\bibinfo {title} {\emph {Electron-phonon interaction and phonon frequencies
  in two-dimensional doped semiconductors}},\ }\href
  {https://doi.org/10.1103/PhysRevB.107.094308} {\bibfield  {journal} {\bibinfo
   {journal} {Phys. Rev. B}\ }\textbf {\bibinfo {volume} {107}},\ \bibinfo
  {pages} {094308} (\bibinfo {year} {2023})},\ \Eprint
  {https://arxiv.org/abs/2212.12237} {arXiv:2212.12237}\BibitemShut {NoStop}%
\bibitem [{\citenamefont {Sio}\ and\ \citenamefont {Giustino}(2022)}]{Sio2022}%
  \BibitemOpen
  \bibfield  {author} {\bibinfo {author} {\bibfnamefont {W.~H.}\ \bibnamefont
  {Sio}}\ and\ \bibinfo {author} {\bibfnamefont {F.}~\bibnamefont {Giustino}},\
  }\bibfield  {title} {\bibinfo {title} {\emph {Unified ab initio description
  of {Fr\"ohlich} electron-phonon interactions in two-dimensional and
  three-dimensional materials}},\ }\href
  {https://doi.org/10.1103/PhysRevB.105.115414} {\bibfield  {journal} {\bibinfo
   {journal} {Phys. Rev. B}\ }\textbf {\bibinfo {volume} {105}},\ \bibinfo
  {pages} {115414} (\bibinfo {year} {2022})},\ \Eprint
  {https://arxiv.org/abs/2203.00568} {arXiv:2203.00568}\BibitemShut {NoStop}%
\bibitem [{\citenamefont {Ponc\'e}\ \emph
  {et~al.}(2023{\natexlab{a}})\citenamefont {Ponc\'e}, \citenamefont {Royo},
  \citenamefont {Stengel}, \citenamefont {Marzari},\ and\ \citenamefont
  {Gibertini}}]{Ponce2023a}%
  \BibitemOpen
  \bibfield  {author} {\bibinfo {author} {\bibfnamefont {S.}~\bibnamefont
  {Ponc\'e}}, \bibinfo {author} {\bibfnamefont {M.}~\bibnamefont {Royo}},
  \bibinfo {author} {\bibfnamefont {M.}~\bibnamefont {Stengel}}, \bibinfo
  {author} {\bibfnamefont {N.}~\bibnamefont {Marzari}},\ and\ \bibinfo {author}
  {\bibfnamefont {M.}~\bibnamefont {Gibertini}},\ }\bibfield  {title} {\bibinfo
  {title} {\emph {Long-range electrostatic contribution to electron-phonon
  couplings and mobilities of two-dimensional and bulk materials}},\ }\href
  {https://doi.org/10.1103/PhysRevB.107.155424} {\bibfield  {journal} {\bibinfo
   {journal} {Phys. Rev. B}\ }\textbf {\bibinfo {volume} {107}},\ \bibinfo
  {pages} {155424} (\bibinfo {year} {2023}{\natexlab{a}})},\ \Eprint
  {https://arxiv.org/abs/2207.10190} {arXiv:2207.10190}\BibitemShut {NoStop}%
\bibitem [{\citenamefont {Ponc\'e}\ \emph
  {et~al.}(2023{\natexlab{b}})\citenamefont {Ponc\'e}, \citenamefont {Royo},
  \citenamefont {Gibertini}, \citenamefont {Marzari},\ and\ \citenamefont
  {Stengel}}]{Ponce2023b}%
  \BibitemOpen
  \bibfield  {author} {\bibinfo {author} {\bibfnamefont {S.}~\bibnamefont
  {Ponc\'e}}, \bibinfo {author} {\bibfnamefont {M.}~\bibnamefont {Royo}},
  \bibinfo {author} {\bibfnamefont {M.}~\bibnamefont {Gibertini}}, \bibinfo
  {author} {\bibfnamefont {N.}~\bibnamefont {Marzari}},\ and\ \bibinfo {author}
  {\bibfnamefont {M.}~\bibnamefont {Stengel}},\ }\bibfield  {title} {\bibinfo
  {title} {\emph {Accurate prediction of {Hall} mobilities in two-dimensional
  materials through gauge-covariant quadrupolar contributions}},\ }\href
  {https://doi.org/10.1103/PhysRevLett.130.166301} {\bibfield  {journal}
  {\bibinfo  {journal} {Phys. Rev. Lett.}\ }\textbf {\bibinfo {volume} {130}},\
  \bibinfo {pages} {166301} (\bibinfo {year} {2023}{\natexlab{b}})},\ \Eprint
  {https://arxiv.org/abs/2207.10187} {arXiv:2207.10187}\BibitemShut {NoStop}%
\bibitem [{\citenamefont {Marzari}\ \emph {et~al.}(2012)\citenamefont
  {Marzari}, \citenamefont {Mostofi}, \citenamefont {Yates}, \citenamefont
  {Souza},\ and\ \citenamefont {Vanderbilt}}]{Marzari2012}%
  \BibitemOpen
  \bibfield  {author} {\bibinfo {author} {\bibfnamefont {N.}~\bibnamefont
  {Marzari}}, \bibinfo {author} {\bibfnamefont {A.~A.}\ \bibnamefont
  {Mostofi}}, \bibinfo {author} {\bibfnamefont {J.~R.}\ \bibnamefont {Yates}},
  \bibinfo {author} {\bibfnamefont {I.}~\bibnamefont {Souza}},\ and\ \bibinfo
  {author} {\bibfnamefont {D.}~\bibnamefont {Vanderbilt}},\ }\bibfield  {title}
  {\bibinfo {title} {\emph {Maximally localized {Wannier} functions: Theory and
  applications}},\ }\href {https://doi.org/10.1103/RevModPhys.84.1419}
  {\bibfield  {journal} {\bibinfo  {journal} {Rev. Mod. Phys.}\ }\textbf
  {\bibinfo {volume} {84}},\ \bibinfo {pages} {1419} (\bibinfo {year}
  {2012})},\ \Eprint {https://arxiv.org/abs/1112.5411}
  {arXiv:1112.5411}\BibitemShut {NoStop}%
\bibitem [{\citenamefont {Su}\ \emph {et~al.}(1979)\citenamefont {Su},
  \citenamefont {Schrieffer},\ and\ \citenamefont {Heeger}}]{Su1979}%
  \BibitemOpen
  \bibfield  {author} {\bibinfo {author} {\bibfnamefont {W.~P.}\ \bibnamefont
  {Su}}, \bibinfo {author} {\bibfnamefont {J.~R.}\ \bibnamefont {Schrieffer}},\
  and\ \bibinfo {author} {\bibfnamefont {A.~J.}\ \bibnamefont {Heeger}},\
  }\bibfield  {title} {\bibinfo {title} {\emph {Solitons in polyacetylene}},\
  }\href {https://doi.org/10.1103/PhysRevLett.42.1698} {\bibfield  {journal}
  {\bibinfo  {journal} {Phys. Rev. Lett.}\ }\textbf {\bibinfo {volume} {42}},\
  \bibinfo {pages} {1698} (\bibinfo {year} {1979})}\BibitemShut {NoStop}%
\bibitem [{\citenamefont {Berges}\ \emph {et~al.}(2017)\citenamefont {Berges},
  \citenamefont {Schobert}, \citenamefont {van Loon}, \citenamefont
  {R\"osner},\ and\ \citenamefont {Wehling}}]{Berges2017}%
  \BibitemOpen
  \bibfield  {author} {\bibinfo {author} {\bibfnamefont {J.}~\bibnamefont
  {Berges}}, \bibinfo {author} {\bibfnamefont {A.}~\bibnamefont {Schobert}},
  \bibinfo {author} {\bibfnamefont {E.~G. C.~P.}\ \bibnamefont {van Loon}},
  \bibinfo {author} {\bibfnamefont {M.}~\bibnamefont {R\"osner}},\ and\
  \bibinfo {author} {\bibfnamefont {T.~O.}\ \bibnamefont {Wehling}},\ }\bibinfo
  {title} {\emph {elphmod: Python modules for electron-phonon models}},\ \href
  {https://doi.org/10.5281/zenodo.5919991} {\bibinfo {howpublished}
  {doi.org/\allowbreak 10.5281/\allowbreak zenodo.5919991}} (\bibinfo {year}
  {2017})\BibitemShut {NoStop}%
\bibitem [{\citenamefont {Pizzi}\ \emph {et~al.}(2020)\citenamefont {Pizzi}
  \emph {et~al.}}]{Pizzi2020}%
  \BibitemOpen
  \bibfield  {author} {\bibinfo {author} {\bibfnamefont {G.}~\bibnamefont
  {Pizzi}} \emph {et~al.},\ }\bibfield  {title} {\bibinfo {title} {\emph
  {{Wannier90} as a community code: New features and applications}},\ }\href
  {https://doi.org/10.1088/1361-648X/ab51ff} {\bibfield  {journal} {\bibinfo
  {journal} {J. Phys. Condens. Matter}\ }\textbf {\bibinfo {volume} {32}},\
  \bibinfo {pages} {165902} (\bibinfo {year} {2020})},\ \Eprint
  {https://arxiv.org/abs/1907.09788} {arXiv:1907.09788}\BibitemShut {NoStop}%
\bibitem [{\citenamefont {Tidman}\ \emph {et~al.}(1974)\citenamefont {Tidman},
  \citenamefont {Singh}, \citenamefont {Curzon},\ and\ \citenamefont
  {Frindt}}]{Tidman1974}%
  \BibitemOpen
  \bibfield  {author} {\bibinfo {author} {\bibfnamefont {J.~P.}\ \bibnamefont
  {Tidman}}, \bibinfo {author} {\bibfnamefont {O.}~\bibnamefont {Singh}},
  \bibinfo {author} {\bibfnamefont {A.~E.}\ \bibnamefont {Curzon}},\ and\
  \bibinfo {author} {\bibfnamefont {R.~F.}\ \bibnamefont {Frindt}},\ }\bibfield
   {title} {\bibinfo {title} {\emph {The phase transition in {2H}-{TaS\s2} at
  75\,{K}}},\ }\href {https://doi.org/10.1080/14786437408207274} {\bibfield
  {journal} {\bibinfo  {journal} {Philos. Mag.}\ }\textbf {\bibinfo {volume}
  {30}},\ \bibinfo {pages} {1191} (\bibinfo {year} {1974})}\BibitemShut
  {NoStop}%
\bibitem [{\citenamefont {Nagata}\ \emph {et~al.}(1992)\citenamefont {Nagata},
  \citenamefont {Aochi}, \citenamefont {Abe}, \citenamefont {Ebisu},
  \citenamefont {Hagino}, \citenamefont {Seki},\ and\ \citenamefont
  {Tsutsumi}}]{Nagata1992}%
  \BibitemOpen
  \bibfield  {author} {\bibinfo {author} {\bibfnamefont {S.}~\bibnamefont
  {Nagata}}, \bibinfo {author} {\bibfnamefont {T.}~\bibnamefont {Aochi}},
  \bibinfo {author} {\bibfnamefont {T.}~\bibnamefont {Abe}}, \bibinfo {author}
  {\bibfnamefont {S.}~\bibnamefont {Ebisu}}, \bibinfo {author} {\bibfnamefont
  {T.}~\bibnamefont {Hagino}}, \bibinfo {author} {\bibfnamefont
  {Y.}~\bibnamefont {Seki}},\ and\ \bibinfo {author} {\bibfnamefont
  {K.}~\bibnamefont {Tsutsumi}},\ }\bibfield  {title} {\bibinfo {title} {\emph
  {Superconductivity in the layered compound {2H}-{TaS\s2}}},\ }\href
  {https://doi.org/10.1016/0022-3697(92)90242-6} {\bibfield  {journal}
  {\bibinfo  {journal} {J. Phys. Chem. Solids}\ }\textbf {\bibinfo {volume}
  {53}},\ \bibinfo {pages} {1259} (\bibinfo {year} {1992})}\BibitemShut
  {NoStop}%
\bibitem [{\citenamefont {Va\v{n}o}\ \emph {et~al.}(2023)\citenamefont
  {Va\v{n}o}, \citenamefont {Ganguli}, \citenamefont {Amini}, \citenamefont
  {Yan}, \citenamefont {Khosravian}, \citenamefont {Chen}, \citenamefont
  {Kezilebieke}, \citenamefont {Lado},\ and\ \citenamefont
  {Liljeroth}}]{Vano2023}%
  \BibitemOpen
  \bibfield  {author} {\bibinfo {author} {\bibfnamefont {V.}~\bibnamefont
  {Va\v{n}o}}, \bibinfo {author} {\bibfnamefont {S.~C.}\ \bibnamefont
  {Ganguli}}, \bibinfo {author} {\bibfnamefont {M.}~\bibnamefont {Amini}},
  \bibinfo {author} {\bibfnamefont {L.}~\bibnamefont {Yan}}, \bibinfo {author}
  {\bibfnamefont {M.}~\bibnamefont {Khosravian}}, \bibinfo {author}
  {\bibfnamefont {G.}~\bibnamefont {Chen}}, \bibinfo {author} {\bibfnamefont
  {S.}~\bibnamefont {Kezilebieke}}, \bibinfo {author} {\bibfnamefont {J.~L.}\
  \bibnamefont {Lado}},\ and\ \bibinfo {author} {\bibfnamefont
  {P.}~\bibnamefont {Liljeroth}},\ }\bibfield  {title} {\bibinfo {title} {\emph
  {Evidence of nodal superconductivity in monolayer {1H}-{TaS\s2} with hidden
  order fluctuations}},\ }\href {https://doi.org/10.1002/adma.202305409}
  {\bibfield  {journal} {\bibinfo  {journal} {Adv. Mater.}\ }\textbf {\bibinfo
  {volume} {2023}},\ \bibinfo {pages} {2305409} (\bibinfo {year} {2023})},\
  \Eprint {https://arxiv.org/abs/2112.07316} {arXiv:2112.07316}\BibitemShut
  {NoStop}%
\bibitem [{\citenamefont {Navarro-Moratalla}\ \emph {et~al.}(2016)\citenamefont
  {Navarro-Moratalla}, \citenamefont {Island}, \citenamefont {Ma\~nas Valero},
  \citenamefont {Pinilla-Cienfuegos}, \citenamefont {Castellanos-Gomez},
  \citenamefont {Quereda}, \citenamefont {Rubio-Bollinger}, \citenamefont
  {Chirolli}, \citenamefont {Silva-Guill\'en}, \citenamefont {Agra\"it},
  \citenamefont {Steele}, \citenamefont {Guinea}, \citenamefont {van~der
  Zant},\ and\ \citenamefont {Coronado}}]{NavarroMoratalla2016}%
  \BibitemOpen
  \bibfield  {author} {\bibinfo {author} {\bibfnamefont {E.}~\bibnamefont
  {Navarro-Moratalla}}, \bibinfo {author} {\bibfnamefont {J.~O.}\ \bibnamefont
  {Island}}, \bibinfo {author} {\bibfnamefont {S.}~\bibnamefont {Ma\~nas
  Valero}}, \bibinfo {author} {\bibfnamefont {E.}~\bibnamefont
  {Pinilla-Cienfuegos}}, \bibinfo {author} {\bibfnamefont {A.}~\bibnamefont
  {Castellanos-Gomez}}, \bibinfo {author} {\bibfnamefont {J.}~\bibnamefont
  {Quereda}}, \bibinfo {author} {\bibfnamefont {G.}~\bibnamefont
  {Rubio-Bollinger}}, \bibinfo {author} {\bibfnamefont {L.}~\bibnamefont
  {Chirolli}}, \bibinfo {author} {\bibfnamefont {J.~A.}\ \bibnamefont
  {Silva-Guill\'en}}, \bibinfo {author} {\bibfnamefont {N.}~\bibnamefont
  {Agra\"it}}, \bibinfo {author} {\bibfnamefont {G.~A.}\ \bibnamefont
  {Steele}}, \bibinfo {author} {\bibfnamefont {F.}~\bibnamefont {Guinea}},
  \bibinfo {author} {\bibfnamefont {H.~S.~J.}\ \bibnamefont {van~der Zant}},\
  and\ \bibinfo {author} {\bibfnamefont {E.}~\bibnamefont {Coronado}},\
  }\bibfield  {title} {\bibinfo {title} {\emph {Enhanced superconductivity in
  atomically thin {TaS\s2}}},\ }\href {https://doi.org/10.1038/ncomms11043}
  {\bibfield  {journal} {\bibinfo  {journal} {Nat. Commun.}\ }\textbf {\bibinfo
  {volume} {7}},\ \bibinfo {pages} {11043} (\bibinfo {year} {2016})},\ \Eprint
  {https://arxiv.org/abs/1604.05656} {arXiv:1604.05656}\BibitemShut {NoStop}%
\bibitem [{\citenamefont {Yang}\ \emph {et~al.}(2018)\citenamefont {Yang},
  \citenamefont {Fang}, \citenamefont {Fatemi}, \citenamefont {Ruhman},
  \citenamefont {Navarro-Moratalla}, \citenamefont {Watanabe}, \citenamefont
  {Taniguchi}, \citenamefont {Kaxiras},\ and\ \citenamefont
  {Jarillo-Herrero}}]{Yang2018}%
  \BibitemOpen
  \bibfield  {author} {\bibinfo {author} {\bibfnamefont {Y.}~\bibnamefont
  {Yang}}, \bibinfo {author} {\bibfnamefont {S.}~\bibnamefont {Fang}}, \bibinfo
  {author} {\bibfnamefont {V.}~\bibnamefont {Fatemi}}, \bibinfo {author}
  {\bibfnamefont {J.}~\bibnamefont {Ruhman}}, \bibinfo {author} {\bibfnamefont
  {E.}~\bibnamefont {Navarro-Moratalla}}, \bibinfo {author} {\bibfnamefont
  {K.}~\bibnamefont {Watanabe}}, \bibinfo {author} {\bibfnamefont
  {T.}~\bibnamefont {Taniguchi}}, \bibinfo {author} {\bibfnamefont
  {E.}~\bibnamefont {Kaxiras}},\ and\ \bibinfo {author} {\bibfnamefont
  {P.}~\bibnamefont {Jarillo-Herrero}},\ }\bibfield  {title} {\bibinfo {title}
  {\emph {Enhanced superconductivity upon weakening of charge density wave
  transport in {2H}-{TaS\s2} in the two-dimensional limit}},\ }\href
  {https://doi.org/10.1103/PhysRevB.98.035203} {\bibfield  {journal} {\bibinfo
  {journal} {Phys. Rev. B}\ }\textbf {\bibinfo {volume} {98}},\ \bibinfo
  {pages} {035203} (\bibinfo {year} {2018})},\ \Eprint
  {https://arxiv.org/abs/1711.00079} {arXiv:1711.00079}\BibitemShut {NoStop}%
\bibitem [{\citenamefont {Perdew}\ \emph {et~al.}(1996)\citenamefont {Perdew},
  \citenamefont {Burke},\ and\ \citenamefont {Ernzerhof}}]{Perdew1996}%
  \BibitemOpen
  \bibfield  {author} {\bibinfo {author} {\bibfnamefont {J.~P.}\ \bibnamefont
  {Perdew}}, \bibinfo {author} {\bibfnamefont {K.}~\bibnamefont {Burke}},\ and\
  \bibinfo {author} {\bibfnamefont {M.}~\bibnamefont {Ernzerhof}},\ }\bibfield
  {title} {\bibinfo {title} {\emph {Generalized gradient approximation made
  simple}},\ }\href {https://doi.org/10.1103/PhysRevLett.77.3865} {\bibfield
  {journal} {\bibinfo  {journal} {Phys. Rev. Lett.}\ }\textbf {\bibinfo
  {volume} {77}},\ \bibinfo {pages} {3865} (\bibinfo {year}
  {1996})}\BibitemShut {NoStop}%
\bibitem [{\citenamefont {Hamann}(2013)}]{Hamann2013}%
  \BibitemOpen
  \bibfield  {author} {\bibinfo {author} {\bibfnamefont {D.~R.}\ \bibnamefont
  {Hamann}},\ }\bibfield  {title} {\bibinfo {title} {\emph {Optimized
  norm-conserving {Vanderbilt} pseudopotentials}},\ }\href
  {https://doi.org/10.1103/PhysRevB.88.085117} {\bibfield  {journal} {\bibinfo
  {journal} {Phys. Rev. B}\ }\textbf {\bibinfo {volume} {88}},\ \bibinfo
  {pages} {085117} (\bibinfo {year} {2013})},\ \Eprint
  {https://arxiv.org/abs/1306.4707} {arXiv:1306.4707}\BibitemShut {NoStop}%
\bibitem [{\citenamefont {van Setten}\ \emph {et~al.}(2018)\citenamefont {van
  Setten}, \citenamefont {Giantomassi}, \citenamefont {Bousquet}, \citenamefont
  {Verstraete}, \citenamefont {Hamann}, \citenamefont {Gonze},\ and\
  \citenamefont {Rignanese}}]{vanSetten2018}%
  \BibitemOpen
  \bibfield  {author} {\bibinfo {author} {\bibfnamefont {M.~J.}\ \bibnamefont
  {van Setten}}, \bibinfo {author} {\bibfnamefont {M.}~\bibnamefont
  {Giantomassi}}, \bibinfo {author} {\bibfnamefont {E.}~\bibnamefont
  {Bousquet}}, \bibinfo {author} {\bibfnamefont {M.~J.}\ \bibnamefont
  {Verstraete}}, \bibinfo {author} {\bibfnamefont {D.~R.}\ \bibnamefont
  {Hamann}}, \bibinfo {author} {\bibfnamefont {X.}~\bibnamefont {Gonze}},\ and\
  \bibinfo {author} {\bibfnamefont {G.~M.}\ \bibnamefont {Rignanese}},\
  }\bibfield  {title} {\bibinfo {title} {\emph {The \textsc{PseudoDojo}:
  Training and grading a 85 element optimized norm-conserving pseudopotential
  table}},\ }\href {https://doi.org/10.1016/j.cpc.2018.01.012} {\bibfield
  {journal} {\bibinfo  {journal} {Comput. Phys. Commun.}\ }\textbf {\bibinfo
  {volume} {226}},\ \bibinfo {pages} {39} (\bibinfo {year} {2018})},\ \Eprint
  {https://arxiv.org/abs/1710.10138} {arXiv:1710.10138}\BibitemShut {NoStop}%
\bibitem [{\citenamefont {Sohier}\ \emph {et~al.}(2017)\citenamefont {Sohier},
  \citenamefont {Calandra},\ and\ \citenamefont {Mauri}}]{Sohier2017}%
  \BibitemOpen
  \bibfield  {author} {\bibinfo {author} {\bibfnamefont {T.}~\bibnamefont
  {Sohier}}, \bibinfo {author} {\bibfnamefont {M.}~\bibnamefont {Calandra}},\
  and\ \bibinfo {author} {\bibfnamefont {F.}~\bibnamefont {Mauri}},\ }\bibfield
   {title} {\bibinfo {title} {\emph {Density functional perturbation theory for
  gated two-dimensional heterostructures: Theoretical developments and
  application to flexural phonons in graphene}},\ }\href
  {https://doi.org/10.1103/PhysRevB.96.075448} {\bibfield  {journal} {\bibinfo
  {journal} {Phys. Rev. B}\ }\textbf {\bibinfo {volume} {96}},\ \bibinfo
  {pages} {075448} (\bibinfo {year} {2017})},\ \Eprint
  {https://arxiv.org/abs/1705.04973} {arXiv:1705.04973}\BibitemShut {NoStop}%
\bibitem [{\citenamefont {Marzari}\ \emph {et~al.}(1999)\citenamefont
  {Marzari}, \citenamefont {Vanderbilt}, \citenamefont {De~Vita},\ and\
  \citenamefont {Payne}}]{Marzari1999}%
  \BibitemOpen
  \bibfield  {author} {\bibinfo {author} {\bibfnamefont {N.}~\bibnamefont
  {Marzari}}, \bibinfo {author} {\bibfnamefont {D.}~\bibnamefont {Vanderbilt}},
  \bibinfo {author} {\bibfnamefont {A.}~\bibnamefont {De~Vita}},\ and\ \bibinfo
  {author} {\bibfnamefont {M.~C.}\ \bibnamefont {Payne}},\ }\bibfield  {title}
  {\bibinfo {title} {\emph {Thermal contraction and disordering of the
  {Al}(110) surface}},\ }\href {https://doi.org/10.1103/PhysRevLett.82.3296}
  {\bibfield  {journal} {\bibinfo  {journal} {Phys. Rev. Lett.}\ }\textbf
  {\bibinfo {volume} {82}},\ \bibinfo {pages} {3296} (\bibinfo {year}
  {1999})},\ \Eprint {https://arxiv.org/abs/cond-mat/9903147}
  {arXiv:cond-mat/9903147}\BibitemShut {NoStop}%
\bibitem [{\citenamefont {Pickett}\ and\ \citenamefont
  {Gy\H{o}rffy}(1976)}]{Pickett1976}%
  \BibitemOpen
  \bibfield  {author} {\bibinfo {author} {\bibfnamefont {W.~E.}\ \bibnamefont
  {Pickett}}\ and\ \bibinfo {author} {\bibfnamefont {B.~L.}\ \bibnamefont
  {Gy\H{o}rffy}},\ }\bibfield  {title} {\bibinfo {title} {\emph {Theory of the
  lattice dynamics of strong coupling systems in the rigid muffin tin
  approximation}},\ }in\ \href {https://doi.org/10.1007/978-1-4615-8795-8_14}
  {\emph {\bibinfo {booktitle} {Superconductivity in d- and f-Band Metals}}},\
  \bibinfo {editor} {edited by\ \bibinfo {editor} {\bibfnamefont {D.~H.}\
  \bibnamefont {Douglass}}}\ (\bibinfo  {publisher} {Plenum Press},\ \bibinfo
  {address} {New York},\ \bibinfo {year} {1976})\BibitemShut {NoStop}%
\bibitem [{\citenamefont {Gonze}\ \emph {et~al.}(2016)\citenamefont {Gonze}
  \emph {et~al.}}]{Gonze2016}%
  \BibitemOpen
  \bibfield  {author} {\bibinfo {author} {\bibfnamefont {X.}~\bibnamefont
  {Gonze}} \emph {et~al.},\ }\bibfield  {title} {\bibinfo {title} {\emph
  {Recent developments in the \textsc{Abinit} software package}},\ }\href
  {https://doi.org/10.1016/j.cpc.2016.04.003} {\bibfield  {journal} {\bibinfo
  {journal} {Comput. Phys. Commun.}\ }\textbf {\bibinfo {volume} {205}},\
  \bibinfo {pages} {106} (\bibinfo {year} {2016})}\BibitemShut {NoStop}%
\bibitem [{\citenamefont {Gonze}\ \emph {et~al.}(2020)\citenamefont {Gonze}
  \emph {et~al.}}]{Gonze2020}%
  \BibitemOpen
  \bibfield  {author} {\bibinfo {author} {\bibfnamefont {X.}~\bibnamefont
  {Gonze}} \emph {et~al.},\ }\bibfield  {title} {\bibinfo {title} {\emph {The
  \textsc{Abinit} project: Impact, environment and recent developments}},\
  }\href {https://doi.org/10.1016/j.cpc.2019.107042} {\bibfield  {journal}
  {\bibinfo  {journal} {Comput. Phys. Commun.}\ }\textbf {\bibinfo {volume}
  {248}},\ \bibinfo {pages} {107042} (\bibinfo {year} {2020})}\BibitemShut
  {NoStop}%
\bibitem [{\citenamefont {Royo}\ and\ \citenamefont
  {Stengel}(2019)}]{Royo2019}%
  \BibitemOpen
  \bibfield  {author} {\bibinfo {author} {\bibfnamefont {M.}~\bibnamefont
  {Royo}}\ and\ \bibinfo {author} {\bibfnamefont {M.}~\bibnamefont {Stengel}},\
  }\bibfield  {title} {\bibinfo {title} {\emph {First-principles theory of
  spatial dispersion: Dynamical quadrupoles and flexoelectricity}},\ }\href
  {https://doi.org/10.1103/PhysRevX.9.021050} {\bibfield  {journal} {\bibinfo
  {journal} {Phys. Rev. X}\ }\textbf {\bibinfo {volume} {9}},\ \bibinfo {pages}
  {021050} (\bibinfo {year} {2019})},\ \Eprint
  {https://arxiv.org/abs/1812.05935} {arXiv:1812.05935}\BibitemShut {NoStop}%
\bibitem [{Note4()}]{Note4}%
  \BibitemOpen
  \bibinfo {note} {We simultaneously minimize the mean squared error of the
  (complex) dynamical matrix and of the shown coupling (absolute value),
  preserving the symmetries of the quadrupole tensors imposed by the point
  groups of the associated atoms.}\BibitemShut {Stop}%
\bibitem [{Note5()}]{Note5}%
  \BibitemOpen
  \bibinfo {note} {The \ac {DFPT} calculation has been done using $48 \times
  48$ $\protect \mathbf k$~points, but $g(T, 0)$ is only calculated on the
  coarser $12 \times 12$ $\protect \mathbf k$- and $\protect \mathbf q$-point
  meshes for subsequent Fourier interpolation.}\BibitemShut {Stop}%
\bibitem [{\citenamefont {Stefanucci}\ \emph {et~al.}(2023)\citenamefont
  {Stefanucci}, \citenamefont {van Leeuwen},\ and\ \citenamefont
  {Perfetto}}]{Stefanucci2023}%
  \BibitemOpen
  \bibfield  {author} {\bibinfo {author} {\bibfnamefont {G.}~\bibnamefont
  {Stefanucci}}, \bibinfo {author} {\bibfnamefont {R.}~\bibnamefont {van
  Leeuwen}},\ and\ \bibinfo {author} {\bibfnamefont {E.}~\bibnamefont
  {Perfetto}},\ }\bibfield  {title} {\bibinfo {title} {\emph {In and
  out-of-equilibrium ab initio theory of electrons and phonons}},\ }\href
  {https://doi.org/10.1103/PhysRevX.13.031026} {\bibfield  {journal} {\bibinfo
  {journal} {Phys. Rev. X}\ }\textbf {\bibinfo {volume} {13}},\ \bibinfo
  {pages} {031026} (\bibinfo {year} {2023})},\ \Eprint
  {https://arxiv.org/abs/2303.02102} {arXiv:2303.02102}\BibitemShut {NoStop}%
\bibitem [{\citenamefont {Marini}(2023{\natexlab{b}})}]{Marini2023b}%
  \BibitemOpen
  \bibfield  {author} {\bibinfo {author} {\bibfnamefont {A.}~\bibnamefont
  {Marini}},\ }\bibfield  {title} {\bibinfo {title} {\emph {Non-adiabatic
  effects lead to the breakdown of the classical phonon concept}},\ }\Eprint
  {https://arxiv.org/abs/2303.09869} {arXiv:2303.09869} (\bibinfo {year}
  {2023}{\natexlab{b}})\BibitemShut {NoStop}%
\bibitem [{\citenamefont {Monacelli}\ \emph {et~al.}(2021)\citenamefont
  {Monacelli}, \citenamefont {Bianco}, \citenamefont {Cherubini}, \citenamefont
  {Calandra}, \citenamefont {Errea},\ and\ \citenamefont
  {Mauri}}]{Monacelli2021}%
  \BibitemOpen
  \bibfield  {author} {\bibinfo {author} {\bibfnamefont {L.}~\bibnamefont
  {Monacelli}}, \bibinfo {author} {\bibfnamefont {R.}~\bibnamefont {Bianco}},
  \bibinfo {author} {\bibfnamefont {M.}~\bibnamefont {Cherubini}}, \bibinfo
  {author} {\bibfnamefont {M.}~\bibnamefont {Calandra}}, \bibinfo {author}
  {\bibfnamefont {I.}~\bibnamefont {Errea}},\ and\ \bibinfo {author}
  {\bibfnamefont {F.}~\bibnamefont {Mauri}},\ }\bibfield  {title} {\bibinfo
  {title} {\emph {The stochastic self-consistent harmonic approximation:
  Calculating vibrational properties of materials with full quantum and
  anharmonic effects}},\ }\href {https://doi.org/10.1088/1361-648X/ac066b}
  {\bibfield  {journal} {\bibinfo  {journal} {J. Phys. Condens. Matter}\
  }\textbf {\bibinfo {volume} {33}},\ \bibinfo {pages} {363001} (\bibinfo
  {year} {2021})},\ \Eprint {https://arxiv.org/abs/2103.03973}
  {arXiv:2103.03973}\BibitemShut {NoStop}%
\bibitem [{\citenamefont {Garcia-Goiricelaya}\ \emph
  {et~al.}(2020)\citenamefont {Garcia-Goiricelaya}, \citenamefont
  {Lafuente-Bartolome}, \citenamefont {Gurtubay},\ and\ \citenamefont
  {Eiguren}}]{GarciaGoiricelaya2020}%
  \BibitemOpen
  \bibfield  {author} {\bibinfo {author} {\bibfnamefont {P.}~\bibnamefont
  {Garcia-Goiricelaya}}, \bibinfo {author} {\bibfnamefont {J.}~\bibnamefont
  {Lafuente-Bartolome}}, \bibinfo {author} {\bibfnamefont {I.~G.}\ \bibnamefont
  {Gurtubay}},\ and\ \bibinfo {author} {\bibfnamefont {A.}~\bibnamefont
  {Eiguren}},\ }\bibfield  {title} {\bibinfo {title} {\emph {Emergence of large
  nonadiabatic effects induced by the electron-phonon interaction on the
  complex vibrational quasiparticle spectrum of doped monolayer {MoS\s2}}},\
  }\href {https://doi.org/10.1103/PhysRevB.101.054304} {\bibfield  {journal}
  {\bibinfo  {journal} {Phys. Rev. B}\ }\textbf {\bibinfo {volume} {101}},\
  \bibinfo {pages} {054304} (\bibinfo {year} {2020})},\ \Eprint
  {https://arxiv.org/abs/1911.00311} {arXiv:1911.00311}\BibitemShut {NoStop}%
\bibitem [{\citenamefont {Alidoosti}\ \emph {et~al.}(2021)\citenamefont
  {Alidoosti}, \citenamefont {Esfahani},\ and\ \citenamefont
  {Asgari}}]{Alidoosti2021}%
  \BibitemOpen
  \bibfield  {author} {\bibinfo {author} {\bibfnamefont {M.}~\bibnamefont
  {Alidoosti}}, \bibinfo {author} {\bibfnamefont {D.~N.}\ \bibnamefont
  {Esfahani}},\ and\ \bibinfo {author} {\bibfnamefont {R.}~\bibnamefont
  {Asgari}},\ }\bibfield  {title} {\bibinfo {title} {\emph {Charge density wave
  and superconducting phase in monolayer {InSe}}},\ }\href
  {https://doi.org/10.1103/PhysRevB.103.035411} {\bibfield  {journal} {\bibinfo
   {journal} {Phys. Rev. B}\ }\textbf {\bibinfo {volume} {103}},\ \bibinfo
  {pages} {035411} (\bibinfo {year} {2021})},\ \Eprint
  {https://arxiv.org/abs/2007.00818} {arXiv:2007.00818}\BibitemShut {NoStop}%
\bibitem [{\citenamefont {Alidoosti}\ \emph {et~al.}(2022)\citenamefont
  {Alidoosti}, \citenamefont {Esfahani},\ and\ \citenamefont
  {Asgari}}]{Alidoosti2022}%
  \BibitemOpen
  \bibfield  {author} {\bibinfo {author} {\bibfnamefont {M.}~\bibnamefont
  {Alidoosti}}, \bibinfo {author} {\bibfnamefont {D.~N.}\ \bibnamefont
  {Esfahani}},\ and\ \bibinfo {author} {\bibfnamefont {R.}~\bibnamefont
  {Asgari}},\ }\bibfield  {title} {\bibinfo {title} {\emph {Superconducting
  properties of doped blue phosphorene: Effects of non-adiabatic approach}},\
  }\href {https://doi.org/10.1088/2053-1583/ac9069} {\bibfield  {journal}
  {\bibinfo  {journal} {2D Mater.}\ }\textbf {\bibinfo {volume} {9}},\ \bibinfo
  {pages} {045029} (\bibinfo {year} {2022})},\ \Eprint
  {https://arxiv.org/abs/2210.01151} {arXiv:2210.01151}\BibitemShut {NoStop}%
\bibitem [{\citenamefont {Leroux}\ \emph {et~al.}(2012)\citenamefont {Leroux},
  \citenamefont {Le~Tacon}, \citenamefont {Calandra}, \citenamefont {Cario},
  \citenamefont {M\'easson}, \citenamefont {Diener}, \citenamefont
  {Borrissenko}, \citenamefont {Bosak},\ and\ \citenamefont
  {Rodi\`ere}}]{Leroux2012}%
  \BibitemOpen
  \bibfield  {author} {\bibinfo {author} {\bibfnamefont {M.}~\bibnamefont
  {Leroux}}, \bibinfo {author} {\bibfnamefont {M.}~\bibnamefont {Le~Tacon}},
  \bibinfo {author} {\bibfnamefont {M.}~\bibnamefont {Calandra}}, \bibinfo
  {author} {\bibfnamefont {L.}~\bibnamefont {Cario}}, \bibinfo {author}
  {\bibfnamefont {M.-A.}\ \bibnamefont {M\'easson}}, \bibinfo {author}
  {\bibfnamefont {P.}~\bibnamefont {Diener}}, \bibinfo {author} {\bibfnamefont
  {E.}~\bibnamefont {Borrissenko}}, \bibinfo {author} {\bibfnamefont
  {A.}~\bibnamefont {Bosak}},\ and\ \bibinfo {author} {\bibfnamefont
  {P.}~\bibnamefont {Rodi\`ere}},\ }\bibfield  {title} {\bibinfo {title} {\emph
  {Anharmonic suppression of charge density waves in {2H}-{NbS\s2}}},\ }\href
  {https://doi.org/10.1103/PhysRevB.86.155125} {\bibfield  {journal} {\bibinfo
  {journal} {Phys. Rev. B}\ }\textbf {\bibinfo {volume} {86}},\ \bibinfo
  {pages} {155125} (\bibinfo {year} {2012})},\ \Eprint
  {https://arxiv.org/abs/1210.2327} {arXiv:1210.2327}\BibitemShut {NoStop}%
\bibitem [{\citenamefont {Schobert}\ \emph {et~al.}(2023)\citenamefont
  {Schobert}, \citenamefont {Berges}, \citenamefont {van Loon}, \citenamefont
  {Sentef}, \citenamefont {Brener}, \citenamefont {Rossi},\ and\ \citenamefont
  {Wehling}}]{Schobert2023}%
  \BibitemOpen
  \bibfield  {author} {\bibinfo {author} {\bibfnamefont {A.}~\bibnamefont
  {Schobert}}, \bibinfo {author} {\bibfnamefont {J.}~\bibnamefont {Berges}},
  \bibinfo {author} {\bibfnamefont {E.~G. C.~P.}\ \bibnamefont {van Loon}},
  \bibinfo {author} {\bibfnamefont {M.~A.}\ \bibnamefont {Sentef}}, \bibinfo
  {author} {\bibfnamefont {S.}~\bibnamefont {Brener}}, \bibinfo {author}
  {\bibfnamefont {M.}~\bibnamefont {Rossi}},\ and\ \bibinfo {author}
  {\bibfnamefont {T.~O.}\ \bibnamefont {Wehling}},\ }\bibfield  {title}
  {\bibinfo {title} {\emph {Ab initio electron-lattice downfolding: Potential
  energy landscapes, anharmonicity, and molecular dynamics in charge density
  wave materials}},\ }\Eprint {https://arxiv.org/abs/2303.07261}
  {arXiv:2303.07261} (\bibinfo {year} {2023})\BibitemShut {NoStop}%
\bibitem [{\citenamefont {R\"osner}\ \emph {et~al.}(2014)\citenamefont
  {R\"osner}, \citenamefont {Haas},\ and\ \citenamefont
  {Wehling}}]{Rosner2014}%
  \BibitemOpen
  \bibfield  {author} {\bibinfo {author} {\bibfnamefont {M.}~\bibnamefont
  {R\"osner}}, \bibinfo {author} {\bibfnamefont {S.}~\bibnamefont {Haas}},\
  and\ \bibinfo {author} {\bibfnamefont {T.~O.}\ \bibnamefont {Wehling}},\
  }\bibfield  {title} {\bibinfo {title} {\emph {Phase diagram of electron-doped
  dichalcogenides}},\ }\href {https://doi.org/10.1103/PhysRevB.90.245105}
  {\bibfield  {journal} {\bibinfo  {journal} {Phys. Rev. B}\ }\textbf {\bibinfo
  {volume} {90}},\ \bibinfo {pages} {245105} (\bibinfo {year} {2014})},\
  \Eprint {https://arxiv.org/abs/1404.4295} {arXiv:1404.4295}\BibitemShut
  {NoStop}%
\bibitem [{\citenamefont {Bin~Subhan}\ \emph {et~al.}(2021)\citenamefont
  {Bin~Subhan}, \citenamefont {Suleman}, \citenamefont {Moore}, \citenamefont
  {Phu}, \citenamefont {Hoesch}, \citenamefont {Kurebayashi}, \citenamefont
  {Howard},\ and\ \citenamefont {Schofield}}]{BinSubhan2021}%
  \BibitemOpen
  \bibfield  {author} {\bibinfo {author} {\bibfnamefont {M.~K.}\ \bibnamefont
  {Bin~Subhan}}, \bibinfo {author} {\bibfnamefont {A.}~\bibnamefont {Suleman}},
  \bibinfo {author} {\bibfnamefont {G.}~\bibnamefont {Moore}}, \bibinfo
  {author} {\bibfnamefont {P.}~\bibnamefont {Phu}}, \bibinfo {author}
  {\bibfnamefont {M.}~\bibnamefont {Hoesch}}, \bibinfo {author} {\bibfnamefont
  {H.}~\bibnamefont {Kurebayashi}}, \bibinfo {author} {\bibfnamefont {C.~A.}\
  \bibnamefont {Howard}},\ and\ \bibinfo {author} {\bibfnamefont {S.~R.}\
  \bibnamefont {Schofield}},\ }\bibfield  {title} {\bibinfo {title} {\emph
  {Charge density waves in electron-doped molybdenum disulfide}},\ }\href
  {https://doi.org/10.1021/acs.nanolett.1c00677} {\bibfield  {journal}
  {\bibinfo  {journal} {Nano Lett.}\ }\textbf {\bibinfo {volume} {21}},\
  \bibinfo {pages} {5516} (\bibinfo {year} {2021})},\ \Eprint
  {https://arxiv.org/abs/2108.07015} {arXiv:2108.07015}\BibitemShut {NoStop}%
\bibitem [{\citenamefont {Ge}\ and\ \citenamefont {Liu}(2013)}]{Ge2013}%
  \BibitemOpen
  \bibfield  {author} {\bibinfo {author} {\bibfnamefont {Y.}~\bibnamefont
  {Ge}}\ and\ \bibinfo {author} {\bibfnamefont {A.~Y.}\ \bibnamefont {Liu}},\
  }\bibfield  {title} {\bibinfo {title} {\emph {Phonon-mediated
  superconductivity in electron-doped single-layer {MoS\s2}: A first-principles
  prediction}},\ }\href {https://doi.org/10.1103/PhysRevB.87.241408} {\bibfield
   {journal} {\bibinfo  {journal} {Phys. Rev. B}\ }\textbf {\bibinfo {volume}
  {87}},\ \bibinfo {pages} {241408(R)} (\bibinfo {year} {2013})}\BibitemShut
  {NoStop}%
\bibitem [{\citenamefont {Costanzo}\ \emph {et~al.}(2016)\citenamefont
  {Costanzo}, \citenamefont {Jo}, \citenamefont {Berger},\ and\ \citenamefont
  {Morpurgo}}]{Costanzo2016}%
  \BibitemOpen
  \bibfield  {author} {\bibinfo {author} {\bibfnamefont {D.}~\bibnamefont
  {Costanzo}}, \bibinfo {author} {\bibfnamefont {S.}~\bibnamefont {Jo}},
  \bibinfo {author} {\bibfnamefont {H.}~\bibnamefont {Berger}},\ and\ \bibinfo
  {author} {\bibfnamefont {A.~F.}\ \bibnamefont {Morpurgo}},\ }\bibfield
  {title} {\bibinfo {title} {\emph {Gate-induced superconductivity in
  atomically thin {MoS\s2} crystals}},\ }\href
  {https://doi.org/10.1038/nnano.2015.314} {\bibfield  {journal} {\bibinfo
  {journal} {Nat. Nanotechnol.}\ }\textbf {\bibinfo {volume} {11}},\ \bibinfo
  {pages} {339} (\bibinfo {year} {2016})},\ \Eprint
  {https://arxiv.org/abs/1512.03222} {arXiv:1512.03222}\BibitemShut {NoStop}%
\bibitem [{Note6()}]{Note6}%
  \BibitemOpen
  \bibinfo {note} {For n-doped monolayer MoS\protect \textsubscript 2, we use
  ${\epsilon = 3.84}$, ${Z^* _{\protect \textnormal {Mo}} = 1.09 e}$, ${Z^*
  _{\protect \textnormal {S}} = 0.59 e}$, ${Q_{\protect \text {Mo} y y y} =
  2.75~e\protect \,\protect \text {bohr}}$, ${Q_{\protect \text S y y y} =
  -0.13~e\protect \,\protect \text {bohr}}$, ${Q_{\protect \text S z y y} =
  4.77~e\protect \,\protect \text {bohr}}$, and ${L = 3.8~\protect \text
  {bohr}}$.}\BibitemShut {Stop}%
\bibitem [{Note7()}]{Note7}%
  \BibitemOpen
  \bibinfo {note} {It was not possible to neglect the effect of the
  out-of-plane polarizability (cf.\spacefactor \@m {} Sec.~\ref
  {sec:froehlich}) in the bare phonon dispersions.}\BibitemShut {Stop}%
\bibitem [{Note8()}]{Note8}%
  \BibitemOpen
  \bibinfo {note} {Alternative definitions of the bare phonons, which fulfill
  the acoustic sum rule, are possible~\cite {Pickett1976}.}\BibitemShut {Stop}%
\end{thebibliography}%

\end{document}